\documentclass[twocolumn]{aastex631}

\def\approxlt{\ifmmode \rlap{$<$}{}_{{}_{{}_{\textstyle\sim}}} \else%
$\rlap{$<$}{}_{{}_{{}_{\textstyle\sim}}}$\fi}
\def\approxgt{\ifmmode \rlap{$>$}{}_{{}_{{}_{\textstyle\sim}}} \else%
$\rlap{$>$}{}_{{}_{{}_{\textstyle\sim}}}$\fi}

\begin{document}

\title{The kangaroo's first hop: the early fast cooling phase of EP250108a/SN 2025kg}
\shorttitle{The early behaviour of EP250108a/SN 2025kg}
\shortauthors{Eyles-Ferris et al.}

\correspondingauthor{Rob A. J. Eyles-Ferris}
\email{raje1@leicester.ac.uk}

\author[0000-0002-8775-2365]{Rob A. J. Eyles-Ferris}
\affiliation{School of Physics and Astronomy, University of Leicester, University Road, Leicester, LE1 7RH, UK}

\author[0000-0001-5679-0695]{Peter G. Jonker}
\affiliation{Department of Astrophysics/IMAPP, Radboud University, P.O. Box 9010, 6500 GL, Nijmegen, The Netherlands}

\author[0000-0001-7821-9369]{Andrew J. Levan}
\affiliation{Department of Astrophysics/IMAPP, Radboud University, P.O. Box 9010, 6500 GL, Nijmegen, The Netherlands}

\author[0000-0002-7517-326X]{Daniele Bj\o{}rn Malesani}
\affiliation{Cosmic Dawn Center (DAWN), Denmark}
\affiliation{Niels Bohr Institute, University of Copenhagen, Jagtvej 128, Copenhagen, 2200, Denmark}

\author[0000-0003-2700-1030]{Nikhil Sarin}
\affiliation{The Oskar Klein Centre, Department of Physics, Stockholm University, AlbaNova, Stockholm, SE-106 91, Stockholm, Sweden}
\affiliation{Nordita, Stockholm University and KTH Royal Institute of Technology, Hannes Alfvéns väg 12, Stockholm, SE-106 91, Stockholm, Sweden}

\author[0000-0003-2624-0056]{Christopher L. Fryer}
\affiliation{Center for Nonlinear Studies, Los Alamos National Laboratory, Los Alamos, NM 87545 USA}

\author[0000-0002-9267-6213]{Jillian C. Rastinejad}
\affiliation{Center for Interdisciplinary Exploration and Research in Astrophysics (CIERA) and Department of Physics and Astronomy, Northwestern University, Evanston, IL 60208, USA}

\author[0000-0002-2942-3379]{Eric Burns}
\affiliation{Department of Physics and Astronomy, Louisiana State University, Baton Rouge, Louisiana 70803, USA}

\author[0000-0003-3274-6336]{Nial R. Tanvir}
\affiliation{School of Physics and Astronomy, University of Leicester, University Road, Leicester, LE1 7RH, UK}

\author[0000-0002-5128-1899]{Paul T. O'Brien}
\affiliation{School of Physics and Astronomy, University of Leicester, University Road, Leicester, LE1 7RH, UK}

\author[0000-0002-7374-935X]{Wen-fai Fong}
\affiliation{Center for Interdisciplinary Exploration and Research in Astrophysics (CIERA) and Department of Physics and Astronomy, Northwestern University, Evanston, IL 60208, USA}

\author[0000-0002-6134-8946]{Ilya Mandel}
\affiliation{School of Physics and Astronomy, Monash University, Clayton VIC 3800, Australia}
\affiliation{ARC Centre of Excellence for Gravitational-wave Discovery (OzGrav), Melbourne, Australia}

\author[0000-0002-5826-0548]{Benjamin P. Gompertz}
\affiliation{School of Physics and Astronomy, University of Birmingham, Birmingham B15 2TT, UK}
\affiliation{Institute for Gravitational Wave Astronomy, University of Birmingham, Birmingham B15 2TT}

\author[0000-0002-5740-7747]{Charles D. Kilpatrick}
\affiliation{Center for Interdisciplinary Exploration and Research in Astrophysics (CIERA) and Department of Physics and Astronomy, Northwestern University, Evanston, IL 60208, USA}

\author{Steven Bloemen}
\affiliation{Department of Astrophysics/IMAPP, Radboud University, P.O. Box 9010, 6500 GL, Nijmegen, The Netherlands}

\author[0000-0002-7735-5796]{Joe S. Bright}
\affiliation{Astrophysics, Department of Physics, University of Oxford, Keble Road, Oxford OX1 3RH, UK}

\author[0000-0002-0426-3276]{Francesco Carotenuto}
\affiliation{INAF-Osservatorio Astronomico di Roma, Via Frascati 33, I-00078, Monte Porzio Catone (RM), Italy}

\author[0009-0009-1573-8300]{Gregory Corcoran}
\affiliation{School of Physics and Centre for Space Research, University College Dublin, Belfield, Dublin 4, Ireland}

\author[0000-0002-7910-6646]{Laura Cotter}
\affiliation{School of Physics and Centre for Space Research, University College Dublin, Belfield, Dublin 4, Ireland}

\author[0000-0002-4488-726X]{Paul J. Groot}
\affiliation{Department of Astrophysics/IMAPP, Radboud University, P.O. Box 9010, 6500 GL, Nijmegen, The Netherlands}
\affiliation{Department of Astronomy, University of Cape Town, Private Bag X3, Rondebosch 7701, South Africa }
\affiliation{South African Astronomical Observatory, P.O. Box 9, Observatory 7935, South Africa}

\author[0000-0001-9695-8472]{Luca Izzo}
\affiliation{Osservatorio Astronomico di Capodimonte, INAF, Salita Moiariello 16, Napoli, 80131, Italy}
\affiliation{Niels Bohr Institute, University of Copenhagen, DARK, Jagtvej 128, Copenhagen, 2200, Denmark}

\author[0000-0003-1792-2338]{Tanmoy Laskar}
\affiliation{Department of Physics \& Astronomy, University of Utah, Salt Lake City, UT 84112, USA}
\affiliation{Department of Astrophysics/IMAPP, Radboud University, PO Box 9010, 6500 GL Nijmegen, The Netherlands}

\author[0000-0001-5108-0627]{Antonio Martin-Carrillo}
\affiliation{School of Physics and Centre for Space Research, University College Dublin, Belfield, Dublin 4, Ireland}

\author[0000-0002-9408-1563]{Jesse Palmerio}
\affiliation{Universit\'e Paris-Saclay, Universit\'e Paris Cit\'e, CEA, CNRS, AIM, 91191 Gif-sur-Yvette, France}

\author[0000-0003-3193-4714]{Maria E. Ravasio}
\affiliation{Department of Astrophysics/IMAPP, Radboud University, P.O. Box 9010, 6500 GL, Nijmegen, The Netherlands}
\affiliation{INAF – Osservatorio Astronomico di Brera, Via E. Bianchi 46, I-23807 Merate, (LC), Italy}

\author[0000-0002-2626-2872]{Jan van Roestel}
\affiliation{Anton Pannekoek Institute for Astronomy, University of Amsterdam, 1090 GE Amsterdam, The Netherlands}

\author[0000-0002-6950-4587]{Andrea Saccardi}
\affiliation{Universit\'e Paris-Saclay, Universit\'e Paris Cit\'e, CEA, CNRS, AIM, 91191 Gif-sur-Yvette, France}

\author[0000-0001-5803-2038]{Rhaana L. C. Starling}
\affiliation{School of Physics and Astronomy, University of Leicester, University Road, Leicester, LE1 7RH, UK}

\author[0000-0001-9354-2308]{Aishwarya Linesh Thakur}
\affiliation{INAF – Istituto di Astrofisica e Planetologia Spaziali, Via Fosso del Cavaliere 100, 00133 Roma, Italy}

\author[0000-0001-9398-4907]{Susanna D. Vergani}
\affiliation{GEPI, Observatoire de Paris, Universit\'e PSL, CNRS, 5 pace Jules Janssen, F-92190 Meudon, France}
\affiliation{INAF – Osservatorio Astronomico di Brera, Via E. Bianchi 46, I-23807 Merate, (LC), Italy}

\author{Paul M. Vreeswijk}
\affiliation{Department of Astrophysics/IMAPP, Radboud University, P.O. Box 9010, 6500 GL, Nijmegen, The Netherlands}

\author[0000-0002-8686-8737]{Franz E. Bauer}
\affiliation{Instituto de Alta Investigaci{\'{o}}n, Universidad de Tarapac{\'{a}}, Casilla 7D, Arica, Chile}

\author[0000-0001-6278-1576]{Sergio Campana}
\affiliation{INAF – Osservatorio Astronomico di Brera, Via E. Bianchi 46, I-23807 Merate, (LC), Italy}

\author[0009-0000-6374-3221]{Jennifer A. Chac\'on}
\affiliation{Instituto de Astrof\'isica, Facultad de F\'isica, Pontificia Universidad Cat\'olica de Chile, Campus San Joaquín, Av. Vicuña Mackenna 4860, Macul Santiago, Chile, 7820436}
\affiliation{Millennium Institute of Astrophysics, Nuncio Monse\~nor S\'otero Sanz 100, Of 104, Providencia, Santiago, Chile}

\author[0000-0001-9842-6808]{Ashley A. Chrimes}
\affiliation{European Space Agency (ESA), European Space Research and Technology Centre (ESTEC), Keplerlaan 1, 2201 AZ Noordwijk, the Netherlands}
\affiliation{Department of Astrophysics/IMAPP, Radboud University, P.O. Box 9010, 6500 GL, Nijmegen, The Netherlands}

\author[0000-0001-9078-5507]{Stefano Covino}
\affiliation{INAF – Osservatorio Astronomico di Brera, Via E. Bianchi 46, I-23807 Merate, (LC), Italy}
\affiliation{Como Lake centre for AstroPhysics (CLAP), DiSAT, Università dell’Insubria, Via Valleggio 11, 22100 Como, Italy}

\author[0009-0007-6927-7496]{Joyce N. D. van Dalen}
\affiliation{Department of Astrophysics/IMAPP, Radboud University, P.O. Box 9010, 6500 GL, Nijmegen, The Netherlands}

\author[0000-0003-3703-4418]{Valerio D'Elia}
\affiliation{Space Science Data Center (SSDC) - Agenzia Spaziale Italiana (ASI), I-00133 Roma, Italy}

\author[0000-0002-4036-7419]{Massimiliano De Pasquale}
\affiliation{Department of Mathematics and Computer Sciences, Physical Sciences and Earth Sciences, University of Messina, Via F. S. D'Alcontres 31, 98166 Messina, Italy}

\author[0009-0004-9861-8200]{Nusrin Habeeb}
\affiliation{School of Physics and Astronomy, University of Leicester, University Road, Leicester, LE1 7RH, UK}

\author[0000-0002-8028-0991]{Dieter H. Hartmann}
\affiliation{Department of Physics and Astronomy, Clemson University, Clemson, SC 29634-0978, USA}

\author[0009-0005-5404-2745]{Agnes P. C. van Hoof}
\affiliation{Department of Astrophysics/IMAPP, Radboud University, P.O. Box 9010, 6500 GL, Nijmegen, The Netherlands}

\author[0000-0002-9404-5650]{P\'all Jakobsson}
\affiliation{Center for Astrophysics and Cosmology, Science Institute, University of Iceland, Dunhagi 5, 107 Reykijavík, Iceland}

\author[0000-0002-0774-2328]{Yashaswi Julakanti}
\affiliation{School of Physics and Astronomy, University of Leicester, University Road, Leicester, LE1 7RH, UK}

\author[0000-0002-8597-0756]{Giorgos Leloudas}
\affiliation{DTU Space, National Space Institute, Technical University of Denmark, Elektrovej 327, 2800 Kgs. Lyngby, Denmark}

\author[0000-0003-0245-9424]{Daniel Mata S\'anchez}
\affiliation{Instituto de Astrof\'isica de Canarias, E-38205 La Laguna, Tenerife, Spain}
\affiliation{Departamento de Astrofísica, Univ. de La Laguna, E-38206 La Laguna, Tenerife, Spain}

\author[0000-0002-2137-4146]{Christopher J. Nixon}
\affiliation{School of Physics and Astronomy, Sir William Henry Bragg Building, University of Leeds, Woodhouse Ln., Leeds, LS2 9JT, UK}

\author[0000-0003-3114-2733]{Dani\"elle L. A. Pieterse}
\affiliation{Department of Astrophysics/IMAPP, Radboud University, P.O. Box 9010, 6500 GL, Nijmegen, The Netherlands}

\author[0000-0003-3457-9375]{Giovanna Pugliese}
\affiliation{Astronomical Institute Anton Pannekoek, University of Amsterdam, 1090 GE Amsterdam, The Netherlands}

\author[0000-0001-8602-4641]{Jonathan Quirola-V\'asquez}
\affiliation{Department of Astrophysics/IMAPP, Radboud University, P.O. Box 9010, 6500 GL, Nijmegen, The Netherlands}

\author{Ben C. Rayson}
\affiliation{School of Physics and Astronomy, University of Leicester, University Road, Leicester, LE1 7RH, UK}

\author[0000-0002-9393-8078]{Ruben Salvaterra}
\affiliation{INAF—Istituto di Astrofisica Spaziale e Fisica Cosmica di Milano, Via A. Corti 12, 20133 Milano, Italy}

\author[0000-0003-4876-7756]{Ben Schneider}
\affiliation{Aix Marseille University, CNRS, CNES, LAM, Marseille, France}

\author[0000-0002-5297-2683]{Manuel A. P. Torres}
\affiliation{Instituto de Astrof\'isica de Canarias, E-38205 La Laguna, Tenerife, Spain}
\affiliation{Departamento de Astrofísica, Univ. de La Laguna, E-38206 La Laguna, Tenerife, Spain}

\author[0000-0003-3935-7018]{Tayyaba Zafar}
\affiliation{School of Mathematical and Physical Sciences, Macquarie University, NSW 2109, Australia}

\begin{abstract}

Fast X-ray transients (FXTs) are a rare and poorly understood population of events. Previously difficult to detect in real time, the launch of the \textit{Einstein Probe} with its wide field X-ray telescope has led to a rapid expansion in the sample and allowed the exploration of these transients across the electromagnetic spectrum. EP250108a is a recently detected example linked to an optical counterpart, SN 2025kg, or `the kangaroo'. Together with a companion paper \citep{Rastinejad25}, we present our observing campaign and analysis of this event. In this letter, we focus on the early evolution of the optical counterpart over the first six days, including our measurement of the redshift of $z=0.17641$. We compare to other supernovae and fast transients showing similar features, finding significant similarities with SN 2006aj and SN 2020bvc, and show that the source is well-modelled by a rapidly expanding cooling blackbody. We show the observed X-ray and radio properties are consistent with a collapsar-powered jet that is low energy ($\lesssim10^{51}$ erg) and/or fails to break out of the dense material surrounding it. While we examine the possibility that the optical emission emerges from the shock produced as the supernova ejecta expand into a dense shell of circumstellar material, due to our X-ray and radio inferences, we favour a model where it arises from a shocked cocoon resulting from the trapped jet. This makes SN 2025kg one of the few examples of this currently observationally rare event.

\end{abstract}


\section{Introduction} \label{sec:intro}

The \textit{Einstein Probe} \citep[hereafter \textit{EP},][]{Yuan22}, launched just over a year ago, and has already made significant contributions to the field of high energy X-ray astronomy. In particular, it has vastly increased the known sample of fast X-ray transients (FXTs) with its Wide-field X-ray Telescope (\textit{EP}/WXT). 

FXTs are outbursts detected in the soft X-ray regime (typically $<10$ keV), and are characterised by timescales of tens to thousands of seconds. 
While they were found in early sounding rocket experiments, until recently, the meaningful samples have been identified in targeted searches of archival data \citep[e.g.][]{Jonker2013,Glennie2015,Bauer17,Alp20,DeLuca21,QuirolaVasquez22,QuirolaVasquez23}. While there are exceptions to this rule, such as the fortuitous detection of the X-ray flare that accompanied the type Ib supernova SN 2008D \citep[e.g.][]{Mazzali08,Soderberg08,Modjaz09,Malesani09}, these have historically been extremely rare. However, in only a year of operations, \textit{EP}/WXT has detected $\sim$one hundred FXTs and vastly increased the observed sample of these rare transients. These FXTs have been detected and communicated to the community in near real time and allowed in depth analysis of many of their counterparts across the electromagnetic spectrum. 

The properties of the FXTs detected by \textit{EP} are diverse and likely represent a wide range of transient types and progenitor systems. While the nature and origins of many remain mysterious \citep[e.g.][]{Zhang25,OConnor25}, several have been linked to the core collapse of massive stars. Both EP240219a and EP240315a, for instance, were linked to collapsar-driven long gamma-ray bursts \citep[GRBs; e.g.][]{Yin24,Levan24,Liu25}. In these particular cases, the FXT is likely the lower energy counterpart of internal shocks in a powerful relativistic jet produced by the cataclysmic collapse of these stars. Other \textit{EP} FXTs have also been seen to be linked to Type Ic supernovae \citep{vanDalen25,Sun24,Srivastav25} which are known to be produced by similar progenitors as long GRBs and are often observed following such an event \citep[e.g.][]{Woosley99,Woosley06,Georgy09,Modjaz16}.

However, the sources linked to FXTs have often proven to be unusual when compared to the respective populations of these events as observed by other facilities. For example, EP240315a was shown to exhibit soft X-ray emission hundreds of seconds earlier and for far longer than the gamma-ray emission \citep{Liu25}, suggesting significant central engine activity outside the window suggested by the gamma-rays. As another example, SN 2024gsa, the type Ic supernova linked to EP240414a, was accompanied by two distinct thermal emission episodes between the FXT detection and the rise of the supernova suggesting a link to fast blue optical transients \citep[FBOTs;][]{Sun24,vanDalen25,Srivastav25}.

FBOTs are fast evolving and bright at ultraviolet and optical wavelengths. Similarly to the term FXT, FBOT indicates a phenomenological rather than physical description and the observed population is therefore diverse. Several subclasses, such as luminous FBOTS \citep[LFBOTs, e.g.][]{Perley19,Coppejans20,Perley21,Bright22,Ho22,Gutierrez24}, have been identified and the differing natures of these classes implies FBOTs arise from a variety of systems and progenitors. EP240414a/SN 2024gsa implies the existence of systems where weak jets fail to efficiently break out of a dense stellar envelope or circumstellar medium (CSM). Rather than producing a long GRB, the resulting shocked material produces a blue thermal component providing an origin for some FBOTs.

In this letter, we present our observations and analysis of the early evolution of a unique FXT, EP250108a, detected by \textit{EP}/WXT on 8 January 2025 \citep{GCN38861} and which allows us to probe the origin of these enigmatic transients. While this event was reported some 18 hours after the detection and further X-ray observations identified no continued emission \citep{GCN38888}, our rapid optical follow-up identified a counterpart \citep{GCN38878} and prompting multiwavelength observations by the community. EP250108a's optical counterpart was quickly found to be unusual with a spectral energy distribution (SED) consistent with a hot thermal source \citep{GCN38885,GCN38902,GCN38908} characteristic of a fast blue optical transient, tidal disruption event or supernova rather than a power law as is typical for a GRB afterglow. This behaviour is similar to that exhibited by SN 2024gsa, the first \textit{EP} FXT SN, or that of (L)FBOTs. Based on its initial moniker of AT 2025kg, EP250108a's optical counterpart was dubbed `the kangaroo'. The thermal source rapidly faded before starting to rebrighten $\sim$7 days post detection \citep{GCN38972,GCN38983}. Spectroscopic observations showed the rebrightening to be a rising type Ic broad lined (Ic-BL) supernova \citep{GCN38984,GCN38987} and the optical source became SN 2025kg, which we explore in further detail in \citet{Rastinejad25}. Notably, at its redshift of $z=0.17641$ (see below), SN 2025kg is the currently the closest \textit{EP} FXT with a SN connection.

Here we present the early evolution of EP250108a/SN 2025kg covering the first six days (observer frame) post \textit{EP}/WXT detection, hereafter referred to as the fast cooling phase. We also include X-ray and radio data covering the first 20 and 43 days respectively, which we use to inform our analysis of the presence of a jet. The later UV/optical/IR evolution of the source, particularly the Ic-BL supernova, is explored in a companion paper \citep{Rastinejad25}.

EP250108a/SN 2025kg is a valuable opportunity that has been seized by several groups in the community, in particular \citet{Srinivasaragavan25} and \citet{Li25}. We also compare our results to these papers where appropriate.

Throughout this Letter, we adopt a Planck cosmology \citep{Planck20}. At SN 2025kg's redshift of $z=0.17641$, this corresponds to a distance of 880.6 Mpc. Throughout this work, errors are given to 1-$\sigma$.

\section{Observations and data reduction}

Our observing campaign for SN 2025kg has been extensive in terms of both time and wavelength coverage. In this Section, we summarise the observations taken during the fast cooling phase and their reduction.

\subsection{Ultraviolet, optical and NIR}
\label{sec:phot}

\subsubsection{Photometry}

\begin{figure}
\epsscale{1.2}
\plotone{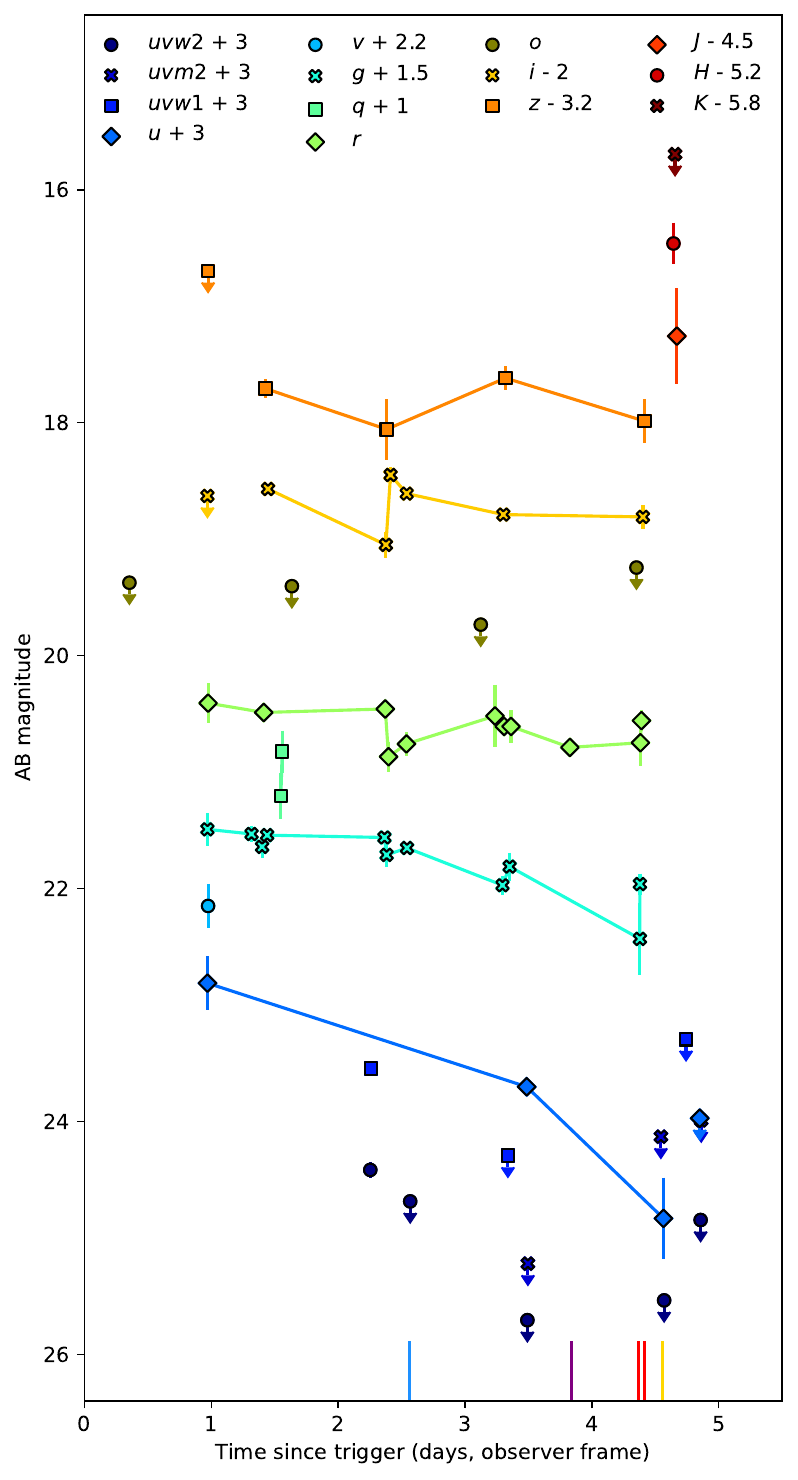}
\caption{The UVOIR light curve of SN 2025kg's fast cooling phase. The photometry has been corrected for Galactic extinction of $A_V = 0.049$ \citep{Schlafly11} using the \citet{Gordon23} Milky Way model. The vertical lines indicate the times of our spectroscopic observations (with colours corresponding to Figure \ref{fig:spectral_sequence}).
\label{fig:uvoir_lc}}
\end{figure}

During the fast cooling phase, we obtained photometric observations with seven telescopes with a total exposure time in excess of 20 ks. We present our assembled photometry for the fast cooling phase of SN 2025kg in Table \ref{tab:photometry}, also including the photometry reported in GCN Circulars, and show the complete light curve in Figure \ref{fig:uvoir_lc}. In our analysis, we correct our photometry for a Milky Way extinction of $A_V = 0.049$ \citep{Schlafly11} using the \texttt{dust\_extinction v1.5} package \citep{Gordon24b} and the \citet[][see also \citet{Gordon09,Fitzpatrick19,Gordon21,Decleir22}]{Gordon23} Milky Way model with $R_V = 3.1$.

\begin{figure*}
\plotone{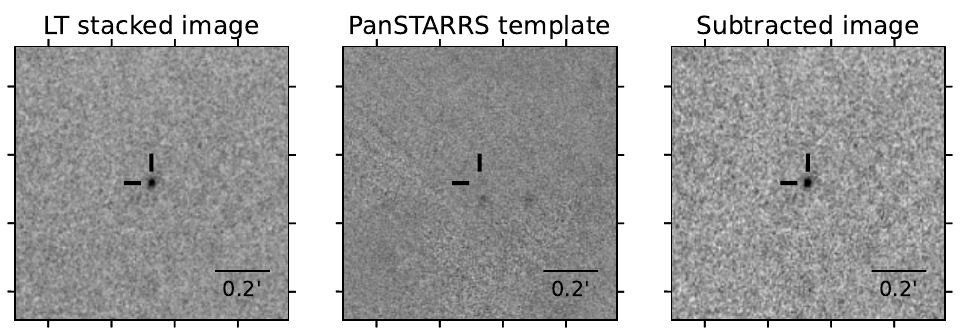}
\caption{The discovery image of SN 2025kg. The stacked \textit{g} LT IO:O band image is shown on the left, the PanSTARRS template is shown in the centre and the subtraction is shown on the right. The images are oriented with North to the top and East to the left and the position of SN 2025kg is marked with the cross hairs.
\label{fig:discovery}}
\end{figure*}

Following the detection of EP250108a, observations were initiated with the 2m Liverpool Telescope \citep[LT,][]{Steele04} using the IO:O instrument (Program IDs PL24B06 \& PL25A25, PI Eyles-Ferris). Six 200~s frames covering the full error region were obtained in the \textit{g} filter $\sim$31.5 hours after the X-ray trigger. These were promptly processed by the automatic pipeline and then by a modified version of the \texttt{photometry-sans-frustration} (\texttt{psf}) package\footnote{\url{https://github.com/mnicholl/photometry-sans-frustration}} \citep{Nicholl23}. The individual frames were aligned, stacked and a template image from the Panoramic Survey Telescope and Rapid Response System (Pan-STARRS or PS1) subtracted using the ZOGY algorithm \citep{Zackay16} as implemented by the \texttt{PyZOGY} package \citep{Guevel21}. A bright source was identified in the subtracted image and its flux measured from the stacked image using PSF photometry calibrated to nearby PanSTARRS stars. This source, later designated SN 2025kg, or the kangaroo, was promptly announced \citep{GCN38878}. We show the discovery image in Figure \ref{fig:discovery}. Additional optical photometry was obtained using LT/IO:O for two further nights during the fast cooling phase, covering the \textit{g}, \textit{r}, \textit{i} and \textit{z} filters. These data were reduced using the same procedure as above.

We further obtained extensive photometric observations from the Nordic Optical Telescope (NOT) using the Alhambra Faint Object Spectrograph and Camera (ALFOSC) instrument (in \textit{griz}; Program ID 70-301, PI Jonker), the Sinistro imager on the Las Cumbres Observatory's  (LCO) South African Astronomical Observatory node 1m telescope (\textit{gr}; Program ID SUPA2024B-004, PI Izzo), BlackGEM (\textit{q}, \citet{Groot24}; Local Transient Survey program) and Gemini-South's FLAMINGOS2 (\textit{JH}; Program ID GS-2024B-Q-105, PI Rastinejad). These data were processed using their respective observatory's pipelines and the flux of SN 2025kg measured using standard methods. Acquisition images from our spectroscopic observations using the Very Large Telescope's (VLT) X-shooter and Gemini-North's GMOS-N were also analysed to supplement our photometry.

To better characterise the properties of the source, particularly at higher energies than accessible through ground based telescopes, we requested three ToO observations by the \textit{Neil Gehrels Swift Observatory} (hereafter \textit{Swift}). These were performed by \textit{Swift} between January 10 and January 13 2025. We acquired the resulting data from the UK \textit{Swift} Science Data Centre\footnote{\url{https://www.swift.ac.uk/index.php}} (UKSSDC) specifically that obtained by the X-ray Telescope (XRT) and Ultraviolet/Optical Telescope (UVOT). Given the blue nature of the transient, we requested observations in the \textit{u}, \textit{w1}, \textit{m2} and \textit{w2} filters covering wavelengths from 3465 \AA~down to 1928 \AA. We used \textsc{uvotproduct v2.9}\footnote{As part of \textsc{HEASoft v6.32} \citep{heasoft}.} to measure SN 2025kg's flux using a circular aperture with a radius of 5\arcsec~and a 3-$\sigma$ detection threshold and converted to AB magnitudes using the standard UVOT zeropoints \citep{Breeveld11}.

To complete our photometry, we used the forced photometry services of the Asteroid Terrestrial-impact Last Alert System\footnote{\url{https://fallingstar-data.com/forcedphot/}} \citep[ATLAS,][]{Tonry18,Smith20,Shingles21} and the Zwicky Transient Facility \citep[ZTF,][]{Masci19, Masci23} to measure the flux at SN 2025kg's position from a few days prior to its detection. For both observatories, we processed the forced photometry into single night epochs following the procedure detailed in the ZTF forced photometry documentation\footnote{\url{https://irsa.ipac.caltech.edu/data/ZTF/docs/ztf_forced_photometry.pdf}} with slight modifications for the ATLAS data. We set a 5-$\sigma$ detection threshold and calculate 3-$\sigma$ upper limits. In both cases, only upper limits were derived and we report these in Table \ref{tab:photometry}. Nevertheless, these upper limits are constraining, particularly a deep upper limit from ZTF $\sim2.3$ days prior to the trigger and an ATLAS upper limit only $\sim 0.4$ days (observer frame) post trigger.

\begin{figure*}
\plotone{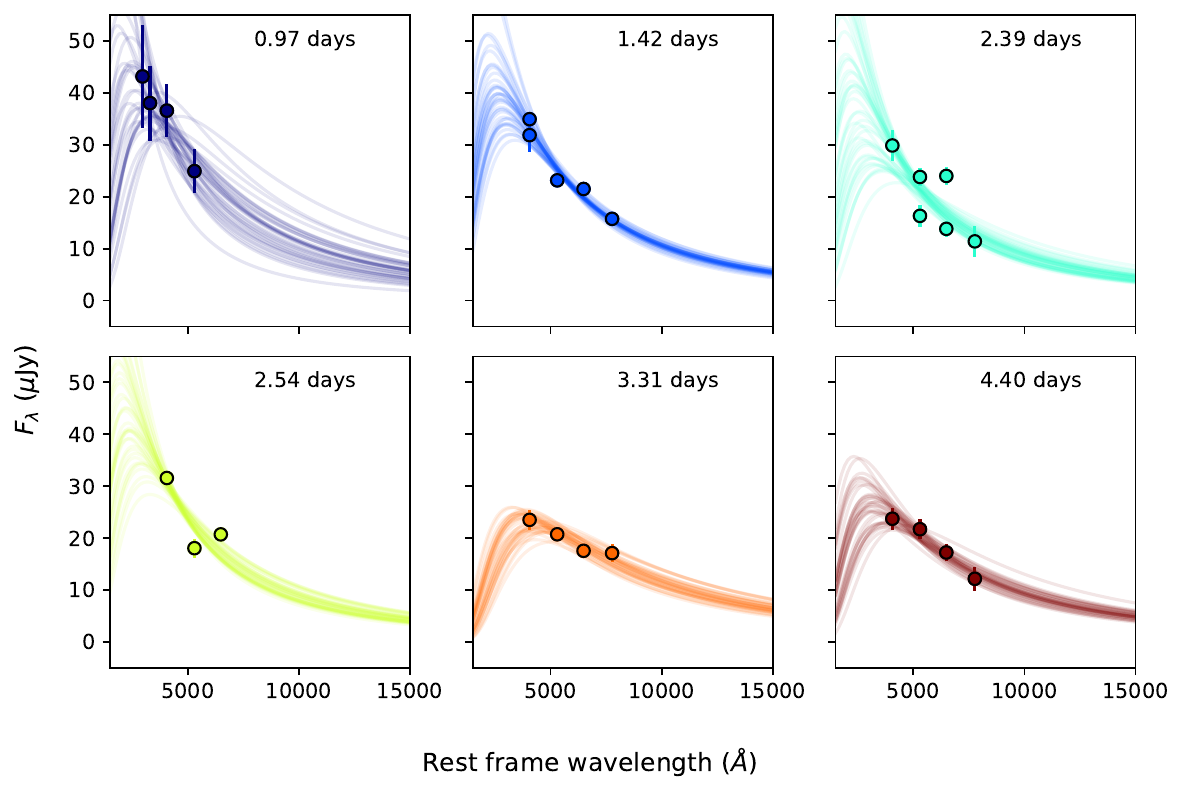}
\caption{The photometric SEDs of SN 2025kg at various epochs (given in the observer frame) fitted with sampled blackbody models. 
\label{fig:bb_phot}}
\end{figure*}

We examine the evolution of the SED of the fast cooling phase by dividing the light curve into epochs and fitting them independently with a blackbody. We limit each epoch to covering up to 0.06 days (observer frame) and derive parameter distributions using Monte Carlo analysis, varying the data within errors and refitting 100 times. We plot the resulting SEDs in Figure \ref{fig:bb_phot} and present the inferred temperatures, radii and luminosities in Table \ref{tab:bb_phot}. We note that we have assumed the SED to be sufficiently accurately represented by a single backbody. However, due to varying optical depth, ejecta temperatures and other factors, the true SED is more likely a superposition of thermal and/or non-thermal emission. While we acknowledge these caveats (and in our advanced modelling below use a more complete description of the SED), this model provides a strong fit to the data and is useful for inferring the bulk properties of the transient and its evolution, guiding further analysis.

\begin{deluxetable}{cccc}
\tablecaption{The blackbody properties inferred from fits to the SED derived from various epochs of our photometry. $\Delta t$ is given in the observer frame. \label{tab:bb_phot}}
\tablehead{\colhead{$\Delta t$ (days)} & \colhead{$T_{\rm bb}$ ($10^4$ K)} & \colhead{$R_{\rm bb}$ ($10^{15}$ cm)} & \colhead{$\log\left( \frac{L_{\rm bb, bol}}{{\rm erg~s}^{-1}} \right)$}}
\startdata
0.97150 & $2.05^{+1.05}_{-0.55}$ & $0.79^{+0.35}_{-0.25}$ & $43.90^{+0.59}_{-0.22}$\\
1.42495 & $1.91^{+0.30}_{-0.22}$ & $0.83\pm0.10$ & $43.83\pm0.14$\\
2.38689 & $2.15^{+0.72}_{-0.48}$ & $0.70\pm0.17$ & $43.88\pm0.28$\\
2.54063 & $2.21^{+0.58}_{-0.34}$ & $0.69\pm0.13$ & $43.91^{+0.22}_{-0.15}$\\
3.30612 & $1.22^{+0.14}_{-0.10}$ & $1.23\pm0.15$ & $43.40\pm0.07$\\
4.37953 & $1.57^{+0.32}_{-0.24}$ & $0.91\pm0.17$ & $43.55\pm0.14$\\
\enddata
\end{deluxetable}

\subsubsection{Spectroscopy}
\label{sec:spec}

\begin{figure*}
\epsscale{0.85}
\plotone{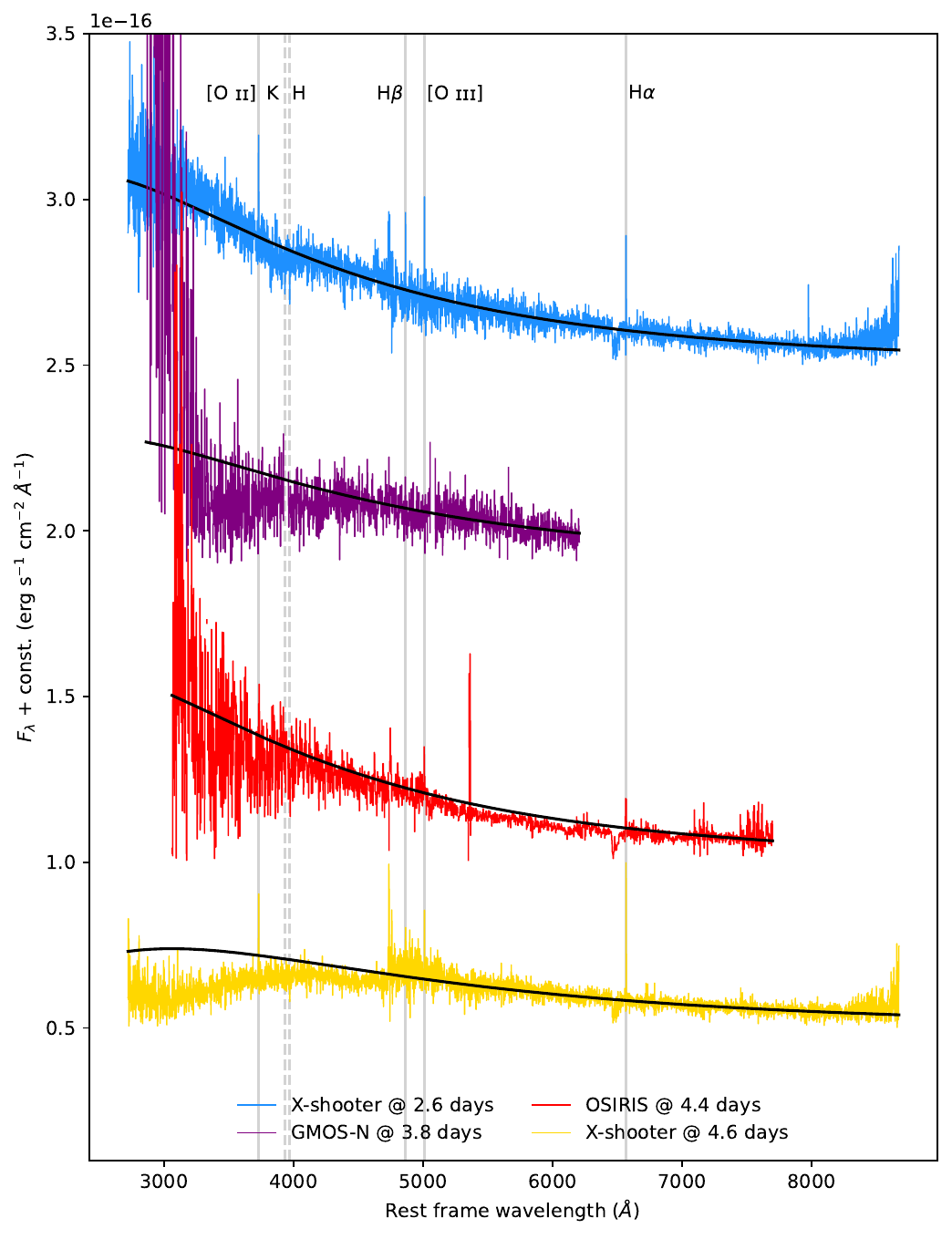}
\caption{The spectral sequence of SN 2025kg's fast cooling phase with times given in the observer frame. Emission and absorption lines typically detected in galaxy spectra are marked in solid and dashed grey lines, respectively. In addition, blackbody fits to each continuum (see text) are plotted in black.
\label{fig:spectral_sequence}}
\end{figure*}

During the fast cooling phase of SN 2025kg, we obtained four spectra from the VLT, Gemini-North and the Gran Telescopio Canarias (GTC). We summarise these observations and our data reduction here. Table~\ref{tab:spectroscopy} presents a log of our spectroscopic observations and we show the full spectral sequence in Figure \ref{fig:spectral_sequence}.

Both VLT spectra were obtained with the  X-shooter instrument \citep{Vernet11}, a multi-wavelength, medium-resolution spectrograph installed at the European Southern Observatory (ESO) VLT UT3/Melipal (Program ID 114.27PZ.001, PI Tanvir). X-shooter covers a total range of $\sim3000 - 25000$ \AA~with a slit width of 0.9 or 1.0\arcsec~depending on the arm. Our VLT spectra were reduced using the standard \textsc{EsoReflex} pipeline \citep{Freudling13}. Here, due to SN 2025kg's blue nature and low redshift, we utilise only the data from the UVB and VIS arms covering $\sim3000 - 10000$ \AA, specifically the 1D spectra reduced in STARE mode. The arms were normalised to each other using their common wavelength area and each spectrum is flux calibrated using the closest photometry in time.

The Gemini-North spectrum was obtained with the GMOS-N instrument (Program ID GN2024B-Q-107, PI Rastinejad) with a slit width of 1\arcsec. The data were reduced using the \texttt{PypeIt} \citep{Prochaska20} package and again flux calibrated using photometry closest in time.

Finally, the OSIRIS+ instrument was used to obtain the spectrum observed by GTC (Program ID GTC1-24ITP, PI Jonker). Data were obtained using both the R1000B and the R1000R grating. A 1\arcsec\, slit width was used for each of the two grating observations and the seeing as measured from the width of the spectroscopic trace is 1.2$\pm$0.1\arcsec. We correct for bias, flats, calibrate in wavelength using arc lamps, and optimally extract the spectra using pyraf and {\sc molly} tasks. Sub-pixel drifts on the position of the [O \textsc{i}]$\lambda$6300 \AA~sky emission line were employed to correct for flexure effects on the wavelength calibration. The spectra were also corrected from the Earth movement using {\sc molly} tasks. Flux calibration was performed using observations of the flux standard G191-B2B at the end of the night. We corrected for slit losses, considering wavelength dependent seeing as well as airmass correction. However, we note that the non-simultaneous observation of the target and the standard introduces uncertainties to the flux calibration due to atmospheric variability. 

To further constrain the properties of SN 2025kg, we fit our spectral continua with the same blackbody model previously applied to the photometric epochs. We use a signal to noise ratio (SNR) cut of ${\rm SNR} > 3$ per bin, fit only to those data and correct for Milky Way extinction using the same model and $A_V=0.049$ mag as our photometric analysis. In the case of the second X-shooter spectrum, due to the increasing opacity from elements formed by nucleosynthesis in the rising supernova, the spectrum is not purely thermal, particularly at blue wavelengths. We therefore apply an additional cut and fit only the data with an observed wavelength $> 5500$ \AA~(4700 \AA~in the rest frame) with the blackbody. Blackbody spectra with parameters set to the median values (i.e., the values reported in Table~\ref{tab:spectroscopy}) are plotted in black over the spectrum for each epoch in Figure~\ref{fig:spectral_sequence} and the inferred properties included in Table~\ref{tab:spectroscopy}. Our results show that, consistent with the photometry, the spectral sequence is broadly consistent with a cooling, expanding blackbody.

\subsubsection{Redshift}

Our spectroscopic observations reveal clear evidence for emission lines from the underlying host galaxy, including lines from [OII], [OIII], H$\beta$ and H$\alpha$ (see Figure~\ref{fig:spectral_sequence}). Measured from our first epoch X-shooter spectrum, these lines provide a systematic redshift for the host galaxy of $z=0.17641 \pm 0.0003$. In addition we also observe absorption lines in the blue arm from Ca H\& K at a similar consistent redshift of $z=0.17637 \pm 0.0002$.  In our cosmology, the luminosity distance is therefore 880.6 Mpc.

\subsection{X-ray}

\begin{figure}
\epsscale{1.18}
\plotone{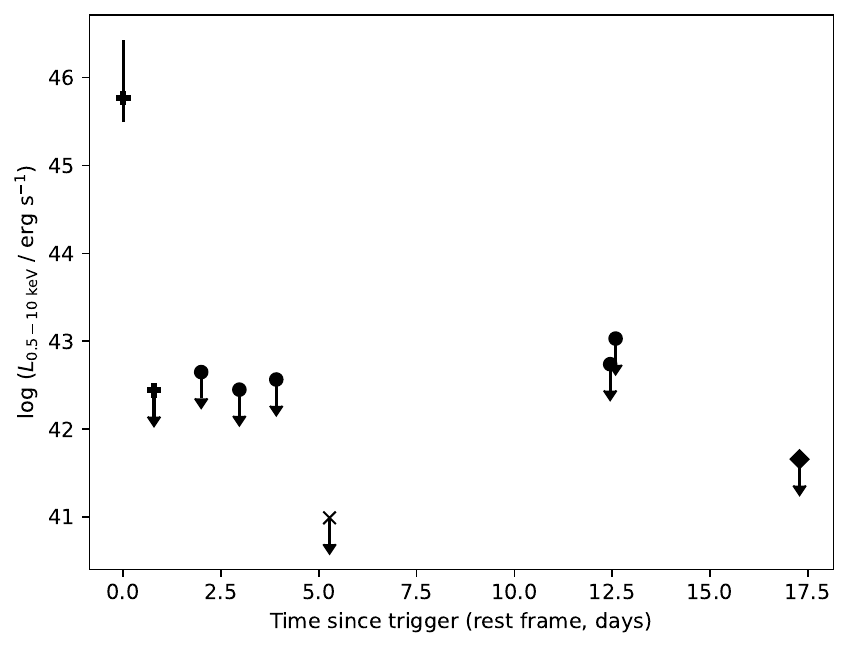}
\caption{The X-ray light curve of EP250108a including data available in GCNs from \textit{EP} \citep[][pluses]{GCN38888,GCN39037} and from our observations with \textit{XMM-Newton} (cross), \textit{Chandra} (diamond) and \textit{Swift} (circles). Note the \textit{EP} detection is the luminosity from 0.5 to 4 keV as given in \citet{GCN39037}.
\label{fig:xray_lc}}
\end{figure}

EP250108a was originally discovered in X-rays and we also performed additional observations to constrain its behaviour at these energies. We summarise our observations and results in Table \ref{tab:xray} and show the X-ray light curve of EP250108a in Figure \ref{fig:xray_lc}, including the initial detection \citep{GCN38861,GCN39037} and upper limit reported by \citet{GCN38888}.

We observed the field of EP250108a using XMM-{\it Newton} starting at 17:24:49 (UTC) on 14 January 2025 (Program ID 096168, PI Jonker). The data were processed with SAS version 20230412\_1735. After filtering for background flares the effective exposure for the pn detector is 30.44~ks, while it is 48 and 47.4~ks for the MOS1 and MOS2 detectors, respectively. However, for the MOS1 and MOS2 detectors we could only choose a source-free background region on an adjacent detector as the use of the Prime Partial Window of the central detector limited the field of view covered by that detector, leaving no room to estimate the background effectively. Given the larger uncertainties associated with estimating the background on another detector, and because of the higher pn sensitivity, we only consider the pn detector measurements for our upper limit calculations. From visual inspection of the image made from the filtered pn data we conclude we do not detect a source at the position of the optical transient. We calculated the number of source counts needed to be detected in a circle of 10\arcsec\, radius in order to be 3-$\sigma$ above the expected background in such a region in the energy range 0.5-10 keV. We used this number of 64 counts to calculate a 95\% confidence upper limit on the 0.5-10 keV unabsorbed source flux of $\approxlt 1\times 10^{-15}$ erg cm$^{-2}$ s$^{-1}$ assuming a power law spectrum with index 3.03 as observed by \citet{GCN39037}.

We obtained further observations with the {\em Chandra} X-ray Observatory (DDT Program ID 30747, PI Jonker). An observation was obtained starting at 20:35:15 (UTC) on 28 January 2025 for 10.851 ks. The source was placed at the default location on the ACIS-S3 chip which was operated in the {\sc very faint} mode. 

The \textit{Chandra} data were processed using {\sc ciao} v4.16. We detect no photons in a circle of 1\arcsec\, radius centred on the source localisation. Using the method of \cite{Kraft91} we obtain a 90\% confidence upper limit of 2.3 counts. Again assuming the spectrum observed by \citet{GCN39037}, this corresponds to a 0.5--10 keV unabsorbed source flux upper limit of $5\times 10^{-15}$ erg s$^{-1}$ cm$^{-2}$.  

We also investigate the X-ray data acquired by \textit{Swift} in our ToO observations, plus two additional observations also undertaken by \textit{Swift}. No source was identified in any single observation. The Living \textit{Swift}-XRT Point Source Catalogue Upper Limit Server\footnote{\url{https://www.swift.ac.uk/LSXPS/ulserv.php}} \citep{Evans23} automatically derives 3-$\sigma$ upper limits which range from 0.002 to 0.009 count s$^{-1}$. Using \textsc{PIMMS v4.15} and the spectrum reported in \citet{GCN39037}, this indicates unabsorbed 0.5 - 10 keV flux upper limits of $3.1\times10^{-14}$ to $1.2\times10^{-13}$ erg cm$^{-2}$ s$^{-1}$. Along with the \textit{XMM-Newton} and \textit{Chandra} data, we convert these to luminosities and include them in Table \ref{tab:xray}.

\subsection{Radio}

We observed the position of EP250108a with the MeerKAT radio telescope \citep{Camilo2018, Jonas2018}, as part of program SCI-20241101-FC-01 (PI Carotenuto) and summarise our observations in Table \ref{tab:radio}. We conducted {\bf four} observations log-spaced in time, each with the same total on-source time of 42 minutes. The first observation started on 13 January 2025 at 17:57 UTC (5.2 days after the first X-ray detection). The second and third observations were performed on 30 January 2025 at 13:41 UTC (22.0 days after the first X-ray detection) and 20 February 2025 at 15:55 UTC (43.2 days after the first X-ray detection), respectively. {\bf A fourth observation was performed on 3 April 2025 at 09:24 UTC (84.9 days after the first X-ray detection).}

We observed at a central frequency of 3.06\,GHz (S-band, S4), with a total bandwidth of 875\,MHz. PKS~J1939--6342 and PKS~J0409–1757 were used as flux and complex gain calibrators, respectively. The data were reduced with the \texttt{OxKAT} pipeline \citep{oxkat}, which performs standard flagging, calibration and imaging using \texttt{tricolour} \citep{Hugo_2022}, \texttt{CASA} \citep{CASA_team_2022} and \texttt{WSCLEAN} \citep{Offringa_wsclean}, respectively. In the imaging step, we adopted a Briggs weighting scheme with a $-0.3$ robust parameter, yielding a $2.8\arcsec \times 2.8\arcsec$ beam and a $8 \, \mu$Jy\,beam$^{-1}$ rms noise in the target field.

We do not detect radio emission at the position of the optical counterpart of EP250108a, and we place a 3$\sigma$ upper limit on the flux density of the target at 24, 24, 27, and {\bf 26} $\mu$Jy\,beam$^{-1}$ for the first, second, third, and fourth observation, respectively.

\section{SN 2025kg and other transients with early thermal features}

\textit{EP}'s unprecedented observational capabilities are likely to uncover a significant population of events similar to EP250108a/SN 2025kg in the near future. For instance, a source with similar optical properties, EP250304a \citep{GCN39580,GCN39583,GCN39584,GCN39585,GCN39587}, was detected during the preparation of this Letter. However, there are also a number of previously detected core-collapse supernovae with similar cooling phases. Here we compare the observed properties of SN 2025kg to a selection of these supernovae including SN 2006aj \citep{Campana06,Mirabal06,Sollerman06,Ferrero06}, SN 2008D \citep{Mazzali08,Soderberg08,Modjaz09,Malesani09}, SN 2010bh \citep{Cano11,Oliveras12,Bufano12}, SN 2017iuk \citep{DElia18,Izzo19}, SN 2020bvc \citep{Izzo20,Ho20,Rho21}, and SN 2024gsa \citep{vanDalen25, Sun24, Srivastav25}. Of these, SN 2024gsa is particularly notable as the optical counterpart to EP240414a, an FXT also detected by \textit{EP}. There are also several supernovae that show rapid rises, possibly due to early contributions from shock heating or cocoon emission, such as iPTF16asu \citep{Whitesides17, Wang19} and SN 2018gep \citep{Ho19}. However, these sources lack a distinct early phase common to the rest of our sample and we therefore do not compare directly here. All of the above sources are Ic-BL supernovae (i.e. the same type as SN 2025kg) with the exception of SN 2008D, which is a Type Ib \citep{Malesani09} and therefore likely has a progenitor with a more intact helium envelope. We also compare to GRB 101225A \citep{Thone11,Campana11,Levan14}, an ultra-long GRB with an early blue component before becoming more consistent with a typical GRB afterglow. In \citet{Rastinejad25}, we extend the comparison to the properties of the supernovae themselves i.e. at later times than examined here.

\begin{figure*}
\epsscale{1.1}
\plotone{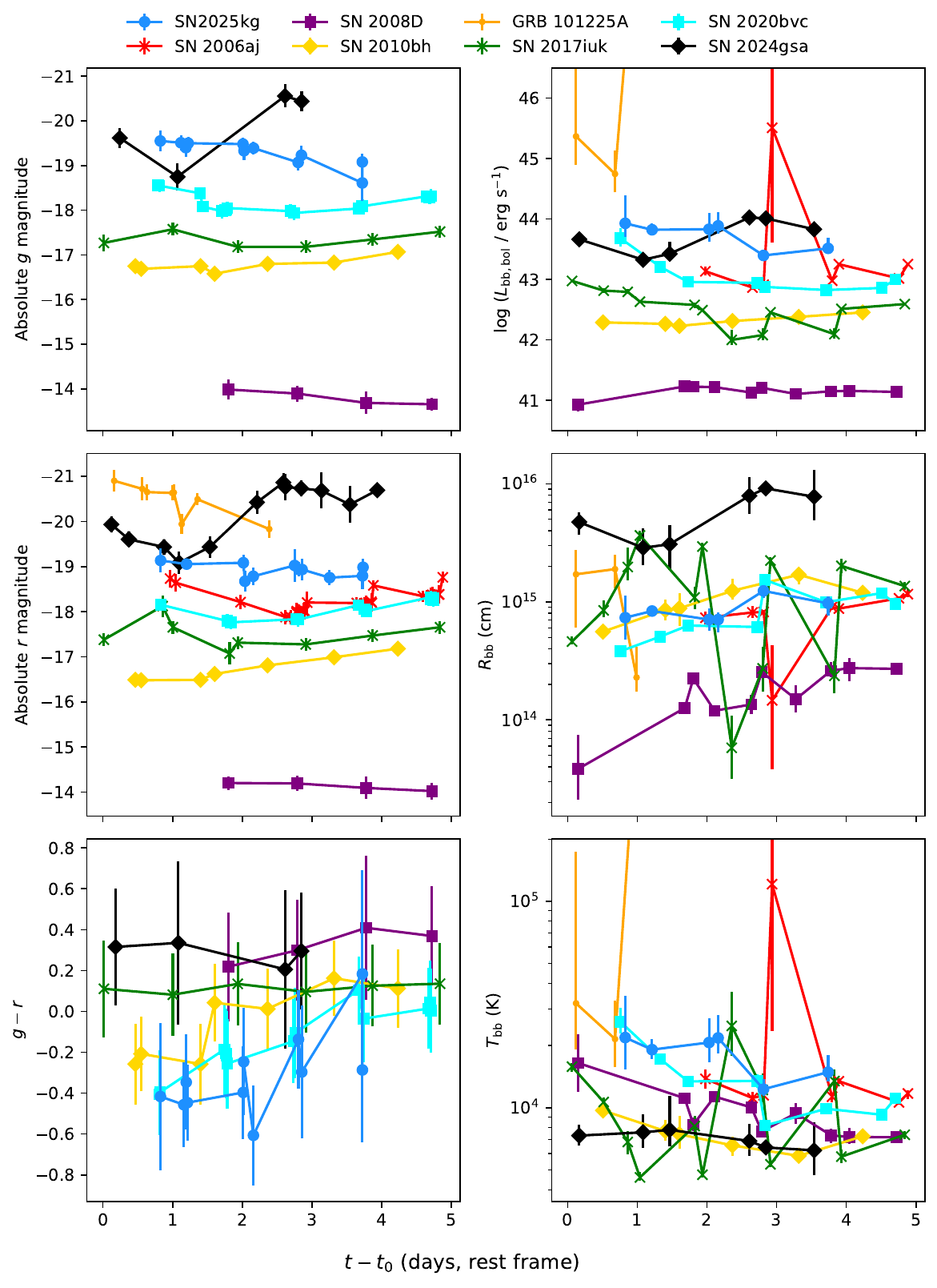}
\caption{The photometric and blackbody properties of SN 2025kg's fast cooling phase compared to the properties of supernovae with similar early features. We include GRB 101225A which also displays a resemblant feature. In the case of SN 2006aj, we have used the Bessell \textit{R} filter as a proxy for the \textit{r} filter.
\label{fig:sn_comp}}
\end{figure*}

In Figure \ref{fig:sn_comp}, we plot the absolute magnitude and colour evolution of these supernovae. In addition, we also fit blackbodies to individual epochs of these supernovae's early components using the same procedure as detailed in Section \ref{sec:phot} allowing us to directly compare the temperature, radius and luminosity evolution.

We find that SN 2025kg most closely resembles SN 2006aj and SN 2020bvc. Its blackbody radius, in particular, is in very strong agreement with the early phases of these supernovae although SN 2025kg is somewhat hotter and therefore higher luminosity. There is greater discrepancy with SN 2008D, SN 2010bh, GRB 101225A, SN 2017iuk and SN 2024gsa, although there is difficulty in comparing the latter as there are likely to be overlapping emission components during the early phase \citep{vanDalen25,Srivastav25,Hamidani25,Zheng25}. Even accounting for this, however, SN 2024gsa appears to show more rapid radius and luminosity evolution than SN 2025kg but it and both SN 2010bh and SN 2017iuk are somewhat cooler. GRB 101225A's early optical counterpart is much more luminous but its temperature and radius evolution are somewhat poorly constrained by our methods here. The more in depth modelling of \citet{Thone11} indicates a significantly higher temperature but smaller radius, pointing towards a different origin than SN2025kg's early fast cooling phase.

Finally, while SN 2008D does show a similar slope in its temperature evolution (notably \citet{Soderberg08} show $T\propto t^{-0.5}$ consistent with our result in Section \ref{sec:phot_modelling}), it is much cooler, less luminous and has a smaller radius. This is consistent with a separate origin for the early phase of SN 2008D as a  shock breakout from the helium envelope of a Ib progenitor \citep[e.g.]{Soderberg08} rather than a cocoon or CSM interaction as suggested by our analyses in this work.

\begin{figure}
\epsscale{1.18}
\plotone{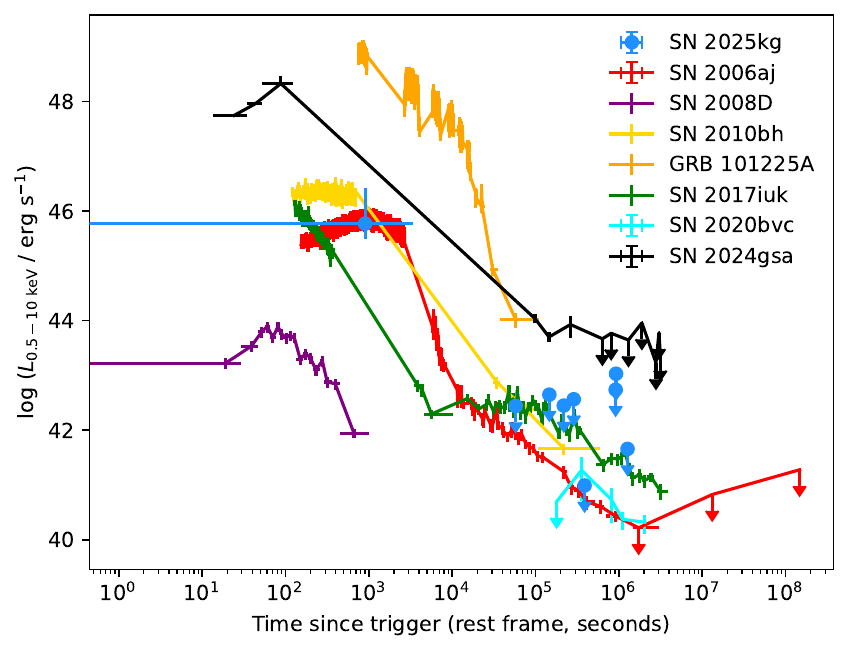}
\caption{The X-ray light curve of EP250108a compared to our sample of FXT/GRB supernovae. EP250108a is shown in blue.
\label{fig:xray_comp}}
\end{figure}

In Figure \ref{fig:xray_comp}, we compare the X-ray light curves of our sample to EP250108a/SN 2025kg, taken either from the references above or acquired from the UKSSDC. Again we find a strong agreement with SN 2006aj and SN 2020bvc - in particular SN2006aj's light curve is entirely consistent with the observed data and limits from EP250108a. We also find a strong agreement with SN 2010bh/XRF 100316D and while SN 2017iuk/GRB 171205A's light curve is not particularly well constrained at early times and hints towards a significantly higher luminosity, it is also broadly compatible. As expected, GRB 101225A and SN 2024gsa are significantly more luminous suggesting either a different origin, e.g. afterglow emission or continued central engine activity in GRB 101225A. Another possibility, which we explore in depth in Section \ref{sec:nature}, is that while all the X-ray emission in the above cases is powered by some kind of relativistic jet, SN 2006aj, SN 2020bvc and SN 2025kg's jets are either significantly less powerful or are unable to break out of their progenitor's outer envelope or the surrounding CSM. Such a jet is deemed to be `failed' \citep[e.g.][]{Nakar15, Nakar2017, Piro2018,Hamidani25b} and the resulting high energy emission is much fainter. The material accelerated by the jet may instead form a rapidly expanding cocoon which produces the bulk of the observed emisison. SN 2008D is again much fainter as expected from its obviously different origin.

At radio wavelengths and including the limits from \citet{GCN38970} and \citet{GCN38998}, we find SN 2025kg is again compatible with SN 2020bvc \citep{Ho20} and SN 2006aj's radio counterparts \citep{Soderberg06}. This is consistent with a picture where all of these events either had similar weak or failed jets or, as suggested elsewhere \citep[e.g.][]{Cobb06,Toma07,Emery19}, were observed at least somewhat off-axis. In contrast, SN 2017iuk and SN 2024gsa are significantly brighter in radio than SN 2025kg, again suggesting that if they had emerged from a similar progenitor system, a powerful jet successfully broke out. We are unable to compare to SN 2010bh due to a lack of data.

To summarise, we find SN 2025kg to be similar to SN 2006aj and SN 2020bvc suggesting both that a similar progenitor powered all these events and that the environment surrounding the explosion was similar. In particular, it is likely that if there were jets, they were either relatively low energy and/or failed to break out. There are also some similarities to SN 2017iuk and SN 2024gsa but these particular events were more likely to have successful jets. In particular, SN 2024gsa is significantly more luminous at both X-ray and radio energies. SN 2017iuk, on the other hand, is roughly compatible in X-rays with SN 2025kg but its radio is much brighter, suggesting this could be a case of a somewhat off-axis but successful jet. The resemblance between SN 2010bh and SN 2017iuk also suggests a similar nature for the former. The additional optical component in SN 2024gsa may arise due to the presence of both cocoon and dense-CSM emission \citep{vanDalen25} or could be the result of a more successful jet \citep{Hamidani25,Zheng25}. On the other hand, both GRB 101225A and SN 2008D show significant differences to SN 2025kg, however, and are likely to be the result of substantially different systems. 

As explored in \citet{Rastinejad25}, the supernovae themselves in SN 2006aj, SN 2020bvc, SN 2024gsa and SN 2025kg also show distinct similarities that further tie these events together and suggest similar origins.

\section{The nature of EP250108a/SN 2025kg}\label{sec:nature}

\subsection{Energetics}
\label{sec:energetics}

To explore the origins of EP250108a/SN 2025kg, we first evaluate the observed energetics. At the luminosity distance of EP250108a, the flux reported by \citet{GCN39037} equates to a luminosity of $5.9^{+20.9}_{-2.8}\times10^{45}$ erg s$^{-1}$ and the 2200 s assumed duration of the FXT implies a total isotropic equivalent energy of $1.3^{+4.6}_{-0.6}\times10^{49}$ erg in the 0.5 to 4 keV band. \citet{GCN39146} also place constraints on higher energy emission using the Gamma-ray Burst Monitor (GBM) instrument on the \textit{Fermi} satellite. In the 10 to 1000 keV band, any emission is limited to an isotropic equivalent  luminosity $\lesssim2.4\times10^{48}$ erg s$^{-1}$.

The energy is therefore $\sim$two orders of magnitude lower than the bulk of the long GRB population \citep[e.g.][]{Nava12}. If we implicitly assume that the jet did break out but did not produce gamma-rays in our observable window \citep{GCN39146}, we can constrain the luminosity of such a jet. Only bursts with peak isotropic-equivalent gamma-ray luminosities under $\sim10^{49}$ erg s$^{-1}$ are viable, leaving only low-luminosity GRBs \citep[e.g.][]{Liang07,Virgili09}. These GRBs are significantly lower luminosity than the majority of the population and often have softer prompt emission \citep{Patel23}. The reasons for the stark differences between these sources and other GRBs are not entirely clear but one possibility is intrinsically weaker or possibly failed jets \citep[e.g.][]{Nakar15,Senno16} as suggested by our comparison to similar transients. 

\subsection{Photometric modelling}

\label{sec:phot_modelling}

To fully explore the evolution of SN 2025kg's fast cooling phase, we first fit the full light curve using a simple cooling, expanding blackbody model. While this is primarily a phenomenological model, we can infer the apparent temperature and luminosity evolution. Assuming the photospheric radius broadly coincides with the expanding shock or cocoon and that the resulting emission is thermally dominated\footnote{Or is at least quasi-thermal.}, this model also provides some initial constraints on the expansion velocity. As noted above, we are simplifying the presumably complex superposition of various emission components that make up the true SED. However, this procedure is useful for the constraints that can be fed into our other modelling.

We assume the photospheric radius varies as
\begin{equation}
    R_{\rm ph} (t) = R_0 + \int_{t_0}^t v(t) dt,
\end{equation}
where $R_0$ is the radius of the photosphere at $t_0 = 0.5$ days (rest frame) after the \textit{EP}/WXT trigger time, and $v(t)$ is the expansion velocity,
\begin{equation}
    v (t) = v_0 \left(\frac{t + \tau_d}{\tau_d}\right)^{-\alpha_d},
\end{equation}
where $v_0$ is the expansion velocity at $t_0 = 0.5$ days (rest frame) after trigger time, $\tau_d$ is a deceleration timescale and $\alpha_d$ is positive. Similarly, the temperature evolution is assumed to vary as
\begin{equation}
    T (t) = T_0 \left(\frac{t + \tau_c}{\tau_c}\right)^{-\alpha_c},
    \label{eq:bb_T}
\end{equation}
where $T_0$ is the temperature at $t_0 = 0.5$ rest frame days after the \textit{EP}/WXT trigger time, $\tau_c$ is a cooling timescale and $\alpha_c$ is positive. We do not include any additional extinction component intrinsic to the host. However, if we do fit for such a component using \texttt{dust\_extinction} and the \citet{Gordon24a} averaged model for the Small Magellanic Cloud, we find $A_{V,{\rm host}}\sim0.2$ mag and our other results to be broadly consistent within errors.

We fit this model to SN 2025kg's light curve using the Markov Chain Monte Carlo (MCMC) algorithm implemented in \texttt{emcee v3.1.6} \citep{ForemanMackey13} with 32 walkers. We use 5000 iterations, discarding the first 500 as burn-in, and include an additional parameter $f$ representing the fractional underestimation of our errors\footnote{$f$ enters into our likelihood function as an additional factor of $F_{\nu,\,{\rm model}}e^{\log f}$ added in quadrature to the measured errors.}. We summarise our priors in Table \ref{tab:bb_fitted_params}.

\begin{figure*}
\epsscale{0.75}
\plotone{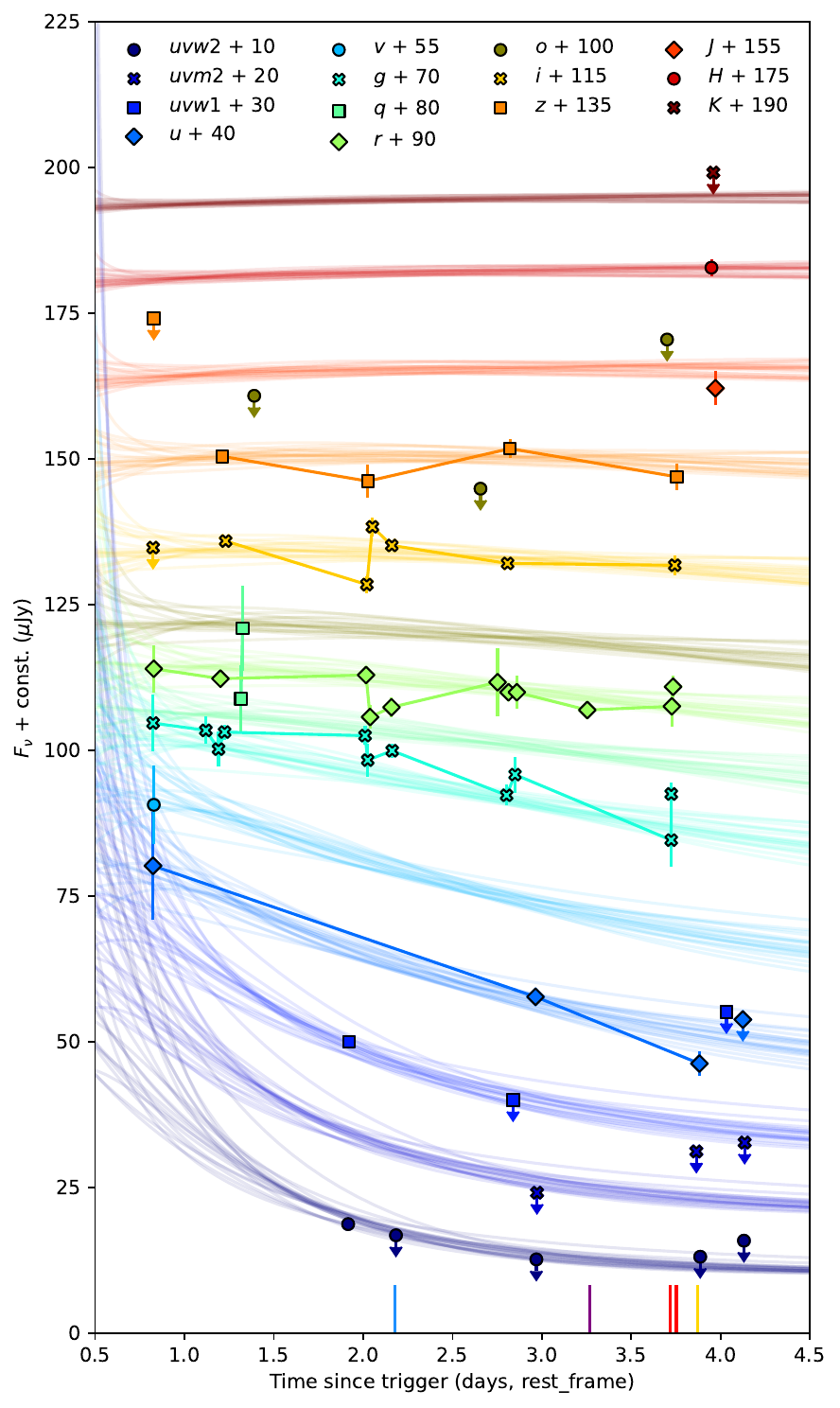}
\caption{The UVOIR light curve of SN 2025kg's fast cooling phase fitted with our cooling, expanding blackbody model. We show traces from 20 randomly selected fits in our MCMC chain. The vertical lines indicate the times of our spectroscopic observations (with colours corresponding to Figure \ref{fig:spectral_sequence}).
\label{fig:bb_fittedlc}}
\end{figure*}

\begin{figure*}
\epsscale{}
\plotone{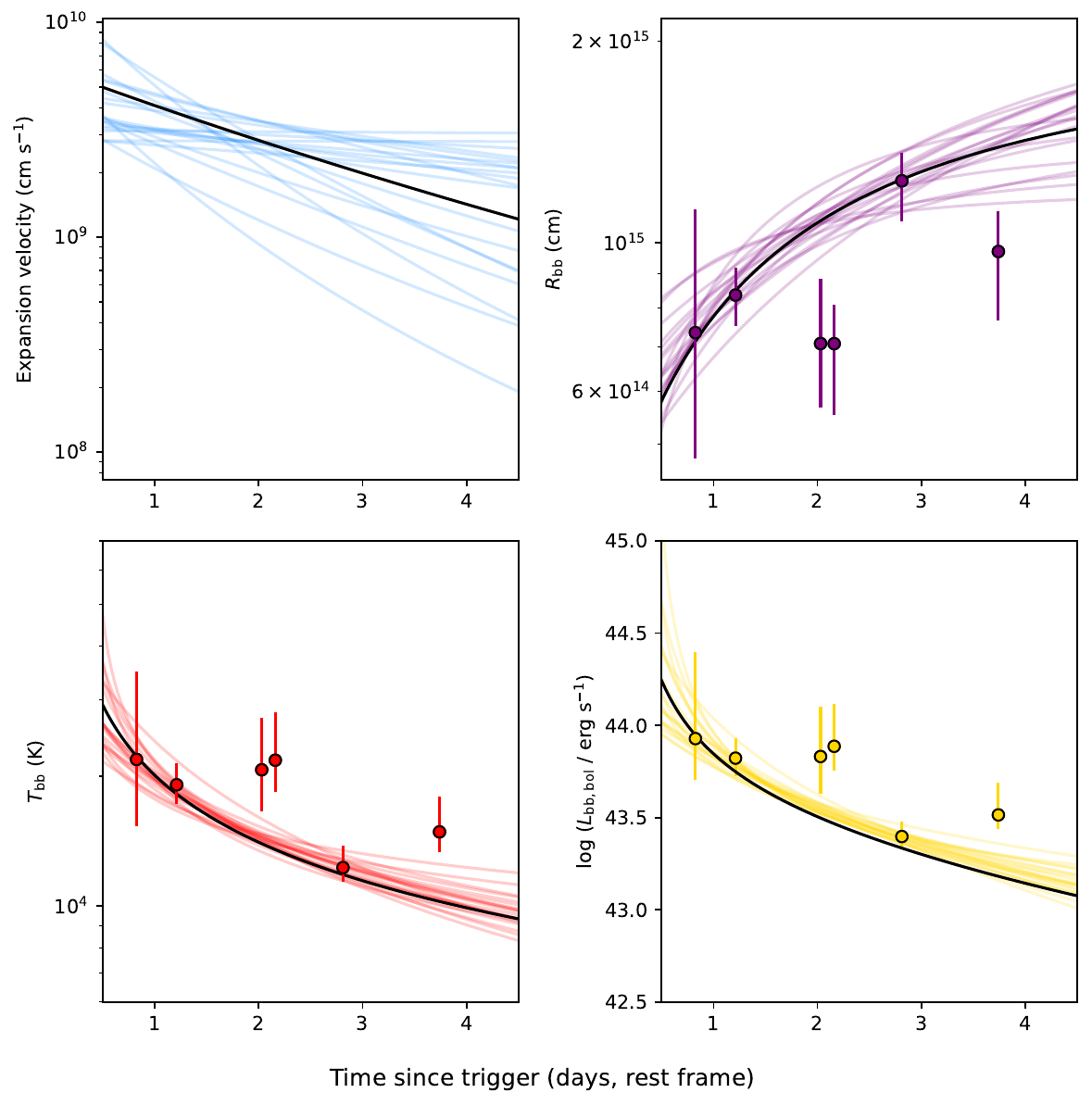}
\caption{The velocity, radius, temperature and luminosity evolution of SN 2025kg's fast cooling phase inferred from our cooling, expanding blackbody model with traces from the same fits as in Figure \ref{fig:bb_fittedlc}. The black lines in each panel indicate the evolution derived when using the median of each parameter (i.e. the values reported in Table \ref{tab:bb_fitted_params}). We also plot the properties inferred from our blackbody fits to our photometric epochs (circles) and spectroscopic continua (crosses).
\label{fig:bb_vRTL}}
\end{figure*}

As shown in Figure \ref{fig:bb_fittedlc}, we find this model is a reasonable representation of the light curve in the fast cooling phase and give our inferred parameters in Table \ref{tab:bb_fitted_params} and show the corner plot in Figure \ref{fig:bb_corner}. The inferred temperature, radius and bolometric luminosity evolution of the fast cooling phase are shown in Figure \ref{fig:bb_vRTL} and are in good agreement with the values inferred from blackbody fits to the photometric epochs and our spectroscopic observations (see above). We note that the model slightly underpredicts the \textit{g} band and overpredicts the UV bands (\textit{u}, \textit{uvw1}, \textit{uvm2} and \textit{uvw2}). This is likely due to us neglecting the effect of increasing opacity as heavy elements are formed in the rising supernova, an effect modelled in detail below, and possibly due to underestimating the deceleration of the photosphere at late times (see top right panel of Figure \ref{fig:bb_vRTL}).

Our results also indicate that the initial expansion velocity of SN 2025kg was at least mildly relativistic - the inferred $R_0$ would require an average expansion velocity over the first 0.5 rest frame days of $\sim1.3\times10^{10}$ cm s$^{-1}$ or $\sim0.45 c$. However, this is inconsistent with a simple extrapolation of the inferred velocity evolution. This suggests the early expansion velocity is much faster before rapidly decelerating to match the value inferred at 0.5 days. To constrain this further, we repeat our fit with $v_0$, $R_0$ and $T_0$ as the expansion velocity, radius and temperature at the trigger time. We fix $R_0 = 10^{12}$ cm, i.e. the largest radius expected for the presumed Wolf-Rayet progenitor and find $v_0 \sim 0.6c$. This is maintained for $\sim$eight hours before a rapid deceleration. We note, however, that this fit is an extrapolation of later data and the velocity and temperature evolution are significantly more complex than we have assumed.

We can however compare the \textit{EP}/WXT data with the prediction from this model of a soft, thermal X-ray source at the trigger time. Such a source is inconsistent with the harder observed spectrum of the transient \citep{GCN39037,Li25}, indicating a different origin for the X-ray and early optical emission. We explore this further in Sections \ref{sec:origin_of_xray} and \ref{sec:optical}.

\begin{deluxetable}{ccc}
\tablecaption{The priors and inferred values from our fit to SN 2025kg's fast cooling phase light curve with our cooling, expanding blackbody model. \label{tab:bb_fitted_params}}
\tablehead{\colhead{Parameter} & \colhead{Prior} & \colhead{Fitted value}}
\startdata
$R_0$ & $>1 \times 10^{12}$ cm & $5.81^{+1.46}_{-1.06} \times 10^{14}$ cm \\
$v_0$ & $0.0 < v_0 <c$ & $5.00^{+2.58}_{-1.81} \times 10^9$ cm s$^{-1}$ \\
$\tau_d$ & $>0.0$ s & $1.32^{+3.44}_{-0.87} \times 10^6$ s \\
$\alpha_d$ & $0.0 < \beta < 10.0$ & $6.10^{+2.85}_{-3.28}$ \\
$T_0$ & $>0.0$ K & $2.92^{+1.25}_{-0.59} \times 10^4$ K \\
$\tau_c$ & $>0.0$ s & $3.79^{+6.21}_{-2.46} \times 10^4$ s \\
$\alpha_c$ & $0.0 < \alpha < 1.0$ & $0.49^{+0.21}_{-0.13}$ \\
$f$ & $-10.0 < \log f < 1.0$ & $-2.00^{+0.18}_{-0.19}$ \\
\enddata
\end{deluxetable}

\subsection{Origin of the X-ray emission}
\label{sec:origin_of_xray}

Having established that the same thermal transient is unlikely to generate both the X-ray and optical emission, we now investigate the source of the FXT itself. The luminous X-ray signal can be explained through Bremsstrahlung emission arising from a large mass of material moving at a high velocity. From the massive star progenitor of EP250108a, this could be from either shock heating in the supernova blastwave or the propagation of a failed jet through the star or clumpy stellar wind. Here, we use the framework developed in Fryer et al. (in prep) to explore these possibilities.

The fastest ejecta is described by a distribution of velocities and the combined emission from this distribution produces the observed X-rays. For the energy distribution of high-velocity ejecta, we use a simple power-law:
\begin{eqnarray}
E(\Gamma \beta) & \propto  & (\Gamma \beta)^{-p}~{\rm~for~\Gamma < \Gamma_{\rm max}} \\ \nonumber
 & = & 0 \hskip 0.35 in {\rm \, otherwise}
\end{eqnarray}
where $\Gamma$ is the Lorentz factor, $\beta$ is the velocity divided by the speed of light and $p$ describes the velocity distribution. We explore two different engine scenarios known to power massive star explosions. The first is a convective engine where the core collapse results in a hot proto-neutron star \citep[e.g.][]{Woosley93,Herant94} and is expected in the vast majority of core collapse supernovae. The neutrino luminosity of the remnant drives a blastwave which generates a relativistic component ($\Gamma_{\rm max} ~\sim 2$) from shock acceleration as it exits the star. The second is a collapsar engine where the star explodes through a combination of disk winds and collimated jets. This engine was invoked to explain long GRBs and their associated Ic-BL supernovae \citep{MacFadyen99,Woosley06} and may also explain other Ic-BLs \citep{Fryer24}. In this case, the jet fails to efficiently break out of the star's outer envelope or CSM and the resulting ejecta is a semi-relativistic cocoon ($\Gamma_{\rm max} \sim 5-10$) of a baryon-loaded jet ejecta.  For this failed jet model, we assume the velocity distribution is relatively flat ($p=0-0.5$). For the convective engine, we use steeper slopes ($p \sim 2-4$) to match shock acceleration as the blastwave breaks out of the star~\citep{2001ApJ...551..946T}.

\begin{figure*}
\epsscale{1.1}
\plotone{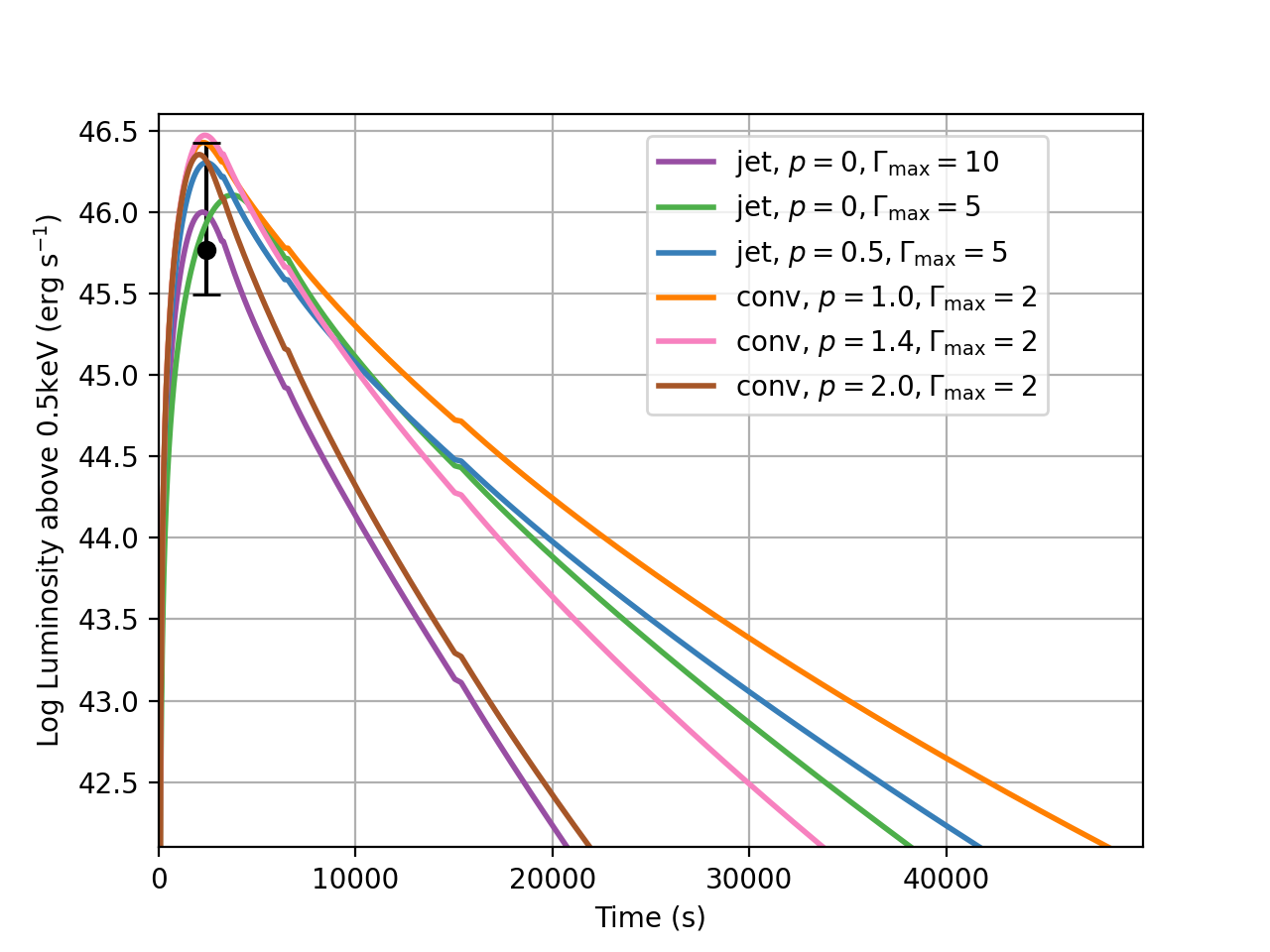}
\caption{X-ray luminosities as a function of time from the high-velocity, high-velocity ejecta. Here we assume two basic models:  a convective engine where the high-velocity ejecta is produced as the shock propagates out of the stellar edge, and a jet-driven model where the high-velocity is dictated by the failed jet + cocoon ejecta. These models are differentiated by the distribution (mass as a function of ejecta velocity) and the maximum Lorentz factor of the ejecta.}
\label{fig:model_xray}
\end{figure*}

We find that we can reproduce the peak luminosity, half-peak duration and the rapid decay with both convective and jet-driven models under certain parameters. We show several such models in Figure~\ref{fig:model_xray} and can use the parameters inferred from this modelling to place lower limits on the mass moving at high velocities.  For our jet driven model, matching the observed X-ray data requires roughly $10^{-4}-10^{-3}\,M_\odot$ of ejecta with velocities above $0.1c$.  This is consistent with recent calculations of the breakout of the jet-driven cocoon~\citep{2024arXiv240815973G}. For our convective engine model with a shallow power law ($p=2$), we are able to fit the data only if we assume the high velocity ejecta ($\beta>0.1$) exceeds 1\,M$_\odot$.  We can reduce this requirement by more than an order of magnitude (down to 0.04\,M$_\odot$) by flattening the velocity distribution ($p \sim 1$).  For the convective engine, the energy is limited to two to three times the energy at the launch of the explosion. For current SN progenitor models, the total explosion energy is limited to $\sim 1-3 \times 10^{51} {\rm \, erg}$~\citep{2001ApJ...554..548F}. Although long-lived engines can increase the total energy by a factor of $\sim 2$, most convective-driven engine simulations have comparable energies to the explosion energy. These models are expected to have less than $0.001-0.01\,M_\odot$ of ejecta at the required high ($\beta>0.1$) velocities~\citep{2018ApJ...856...63F}. We can therefore only produce X-ray signals that match the observations in an extreme case where we both assume supernova energies exceeding current models and push the limits of the mass/energy distribution with ejecta velocity.  

\begin{figure*}
\epsscale{1.1}
\plotone{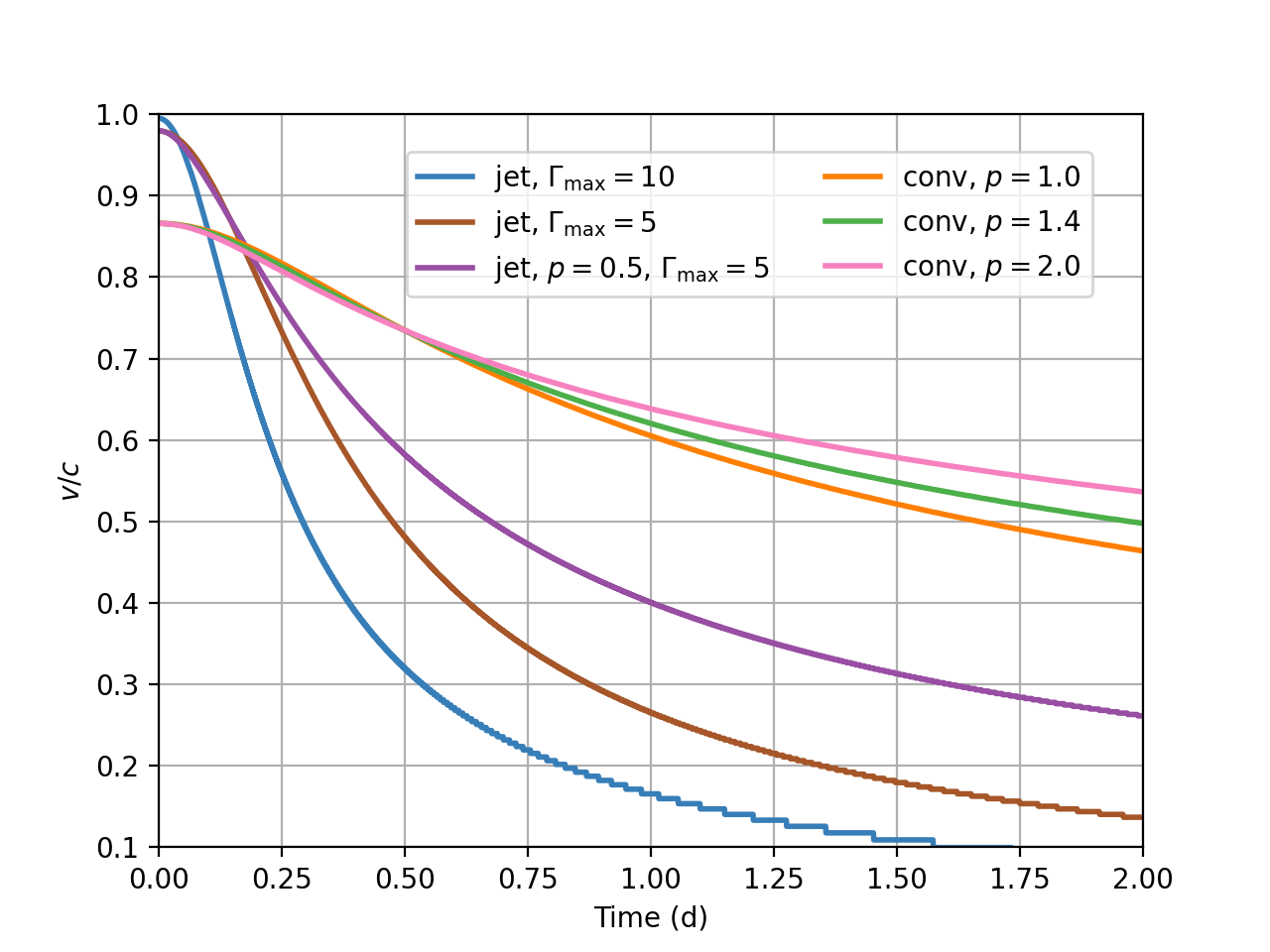}
\caption{Model photospheric velocity of the shock breakout/interaction models from Figure~\ref{fig:model_xray}.  The high mass in the convective-engine models causes them to decelerate slowly, arguing for a high-velocity photosphere in the first 2 days. The lower mass in the jet-driven explosions leads to rapid deceleration of the photosphere.}
\label{fig:model_velvt}
\end{figure*}

We can further constrain the shock interaction conditions by following the evolution of our fast-moving ejecta in the first few days and, in particular, the velocity of the photosphere.  The progenitors of Ic supernovae can eject material from their outer envelope as they approach core collapse in the form of stellar winds \citep[e.g.][]{Dessart20}. As the ejecta propagates through this stellar wind, it will sweep up mass and decelerate.  At the same time, as it becomes optically thin, the photosphere recedes relative to the outflowing ejecta, thus probing slower-moving inner material.  To calculate the photospheric velocity, we assume the mass of the circumstellar medium is set by a wind profile, the enclosed mass out to radial extent $R$ is:
\begin{equation}
M_{\rm wind, enclosed}(R) = (\dot{M}_{\rm wind}/v_{\rm wind}) R
\end{equation}
where $\dot{M}_{\rm wind}$ is the wind mass loss rate, $v_{\rm wind} \approx 1000 {\rm \, km \,s^{-1}}$, and $R$ is the radial extent.  As discussed below, our optical light curves require strong shock interactions at early times (either a shell or a strong wind).  For this calculation, we assume a strong wind: $\dot{M}_{\rm wind} = 10^{-4} \, {\rm M_\odot \, yr^{-1}}$.  We approximate the expansion velocity by applying momentum conservation, decelerating the fastest moving material and moving inward as the high-velocity blastwave propagates through the circumstellar medium.  This rapid deceleration (and deposition of this high-velocity material) is required in our SN calculations to explain the optical emission (see Section~\ref{sec:phot_modelling}).  By also including a simple gray opacity, we can calculate the position of the photosphere (optical depth $~\sim 1$) and the velocity at that photosphere.  Figure~\ref{fig:model_velvt} shows this photospheric velocity evolution with time for our best-fit X-ray models (both jet-driven and convective engines) from Figure~\ref{fig:model_xray}.  Although the photospheric velocity in our convective-engine model is initially slower, the larger high-velocity ejecta mass means that it decelerates slowly with time and therefore the velocity is too high to match the data at a day or so (see Section \ref{sec:phot_modelling}).  In comparison, the low-mass jet-driven models do decelerate sufficiently quickly. We thus conclude this event is most likely powered by a collapsar engine, with the explosion driven by an accretion disk and jet. The rapid drop-off in the X-ray luminosity at late-times provides a final constraint on the jet-driven models. A fraction of these models have X-ray luminosities in excess of $10^{43} {\rm \, erg \, s^{-1}}$ at $10^5 {\rm \, s}$, inconsistent with the observed limits, and we therefore exclude them. We next explore whether we can constrain whether the jet successfully broke through the star/CSM.

\subsection{Constraints on jet/ejecta behaviour}

Having shown how a jet driven engine can produce the X-ray emission observed by \textit{EP}, we further examine the constraints on its behaviour from other observations. In addition to the energy limitation from the lack of gamma-ray detections (see Section \ref{sec:energetics}), our radio observations also provide some constraints on the presence of any other fast-moving ejecta.

\begin{figure*}
\epsscale{1}
\plotone{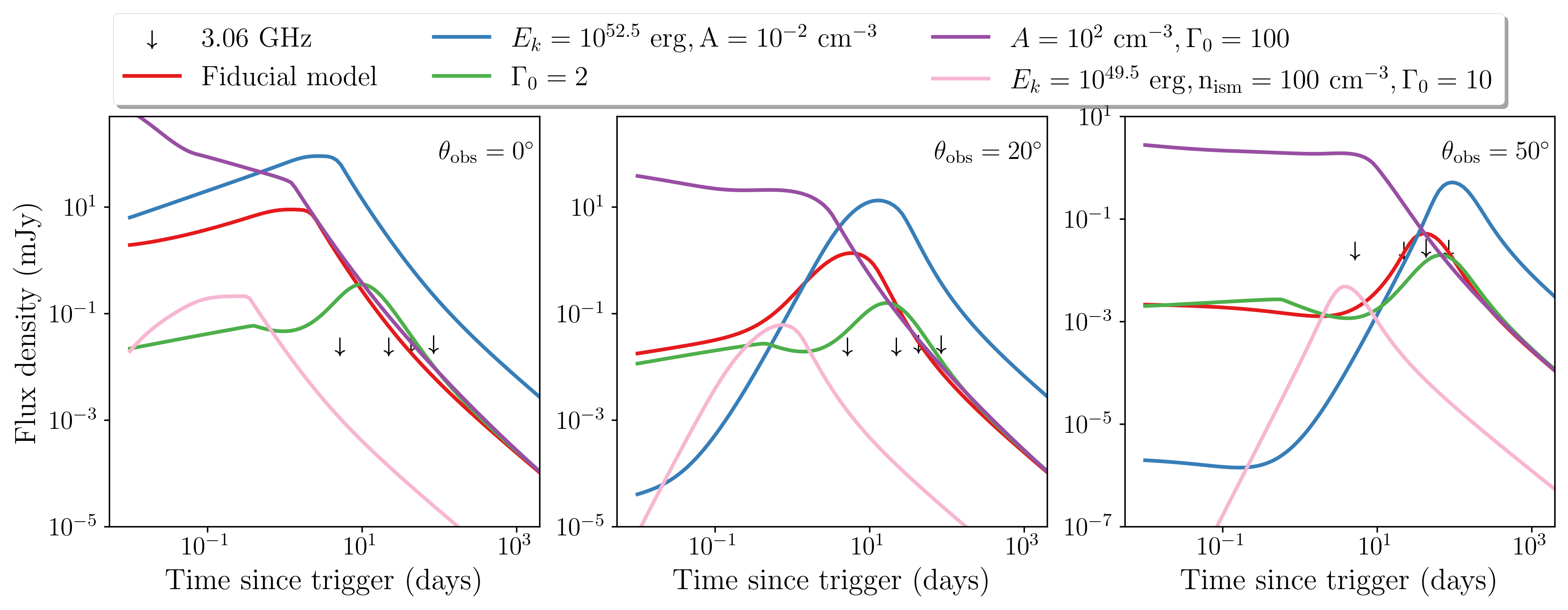}
\caption{Radio afterglows at $3.06$~GHz following a `top hat' jet for different assumptions over the `fiducial model' indicated by the red curves. The three panels correspond to different observer viewing angles while the arrows indicate upper limits from our radio observations. \label{fig:radio_jet_constraints}}
\end{figure*}

In Fig.~\ref{fig:radio_jet_constraints}, we show the predicted radio light curves for different parameters for jets viewed on-axis and $20^{\circ}$ and $50^{\circ}$ off-axis in first, middle, and last panels, respectively. All models assume a `tophat' jet following {\sc jetsimpy}~\citep{Wang2024} using {\sc Redback}~\citep{sarin_redback}. We note that these models do not include synchrotron self-absorption. However, this, or using a structured jet instead of a tophat, does not appreciably impact our interpretation, as the dominant uncertainty is from the parameters themselves. The black arrows indicate upper limits from our radio observations at $3.06$~GHz, while the different colours indicate different assumptions about the jet parameters. The `fiducial' model shown in red shows the radio afterglow at $3.06$~GHz of a $10^{51.5}$~erg kinetic energy jet with an initial Lorentz factor of $1000$ and $\theta_j=10^{\circ}$, travelling into a wind-medium with ISM density following $n_{\rm ism} = A \left(\frac{r}{10^{17}\rm{cm}}\right)^{-2}$ where $A=1~{\rm cm}^{-3}$, i.e., the density at a radius of $10^{17}$cm, with typical values for microphysical parameters $p=2.17$, $\epsilon_{\rm e} = 0.1$, and $\epsilon_{B} = 0.01$. Different colours indicate different assumptions from this fiducial model, with the pink curve representing light curves from a constant density ISM with no wind. Our radio observations can confidently rule out jets with kinetic energies $\gtrsim 10^{51}$~ergs in a wind-like medium, even for far-off-axis scenarios, with the only viable solution where such a jet has a low initial Lorentz factor ($\sim 2$) and also observed at least $50^{\circ}$ off-axis (green curves in last panel). Our non-detections cannot rule out on- or off-axis jets weaker than $10^{50}$~ergs propagating into a low-density ISM.

The radio limits from \citet{Srinivasaragavan25} and \citet{Li25} are also broadly consistent with our results. \citeauthor{Srinivasaragavan25} also model possible afterglows but assume a constant density medium as opposed to the wind medium we assume here. They also conclude a low energy on-axis jet is plausible, but suggest that higher energy off-axis jets may also be possible and would be expected to peak at late times.

While the above only considers jets, our results are broadly applicable to any quasi-spherical outflow such as the fast-ejecta from the supernova as implied by our X-ray modelling above. We do note, however, that our constraints are inconsistent with the slowly coasting jet suggested by \citet{Li25} to explain the early optical emission.

\subsection{Origin of the early optical emission}
\label{sec:optical}

Our blackbody analyses show that we can broadly match the optical emission with a fast-moving, cooling blackbody. Here, we explore more physically-motivated models. The broad similarity of our light curves to past GRBs and supernovae e.g. SN2020bvc, provides some hints towards the appropriate models, i.e. some form of shock breakout and cooling from supernova ejecta expanding out into a dense CSM. This could be the envelope of the star, eruptive mass-loss in the years leading up to the supernova, or winds from the progenitor. 

We explore these different scenarios further by fitting the data for $t \leq 6.5$d with models for shock cooling~\citep{piro21} and dense-CSM shell shock cooling~\citep{margalit22} implemented in {\sc Redback}~\citep{sarin_redback} using the {\sc Pymultinest}~\citep{pymultinest} via {\sc Bilby}~\citep{ashton19}. We fit assuming a Gaussian likelihood with a systematic error added in quadrature to the statistical errors on all data points i.e.,
$\sigma^2_{i} = \sigma_{i, {\rm data}}^2 + \sigma_{\rm sys}^2$, where the first term is the original error on our data points and $\sigma_{\rm sys} = 0.15$ is a systematic uncertainty to capture any discrepancies caused by differences in photometric reduction or filter transmission curves, and broad uniform priors. We also assume the contribution from the supernova to be relatively negligible at this time as there is little evidence of plateauing or rising in the light curve. In addition, the uncertainties in the models themselves are notably larger than the possible supernova contribution. We note the key difference between the models we investigate here is the assumed profile of the CSM, with the \citet{piro21} model assuming a broken power law density profile aimed to resemble homologous expansion of material following shock breakout from the star~\citep[e.g.,][]{matznermckee99}, while the later assumes a CSM shell located at initial radius, $R_0$ of width $\Delta R_{0}$ with a sharp drop in density at larger radii, with further differences in the treatment of radiative diffusion~\citep{margalit22}.

Both models capture the evolution of the light curve, however, the estimated parameters from the shock cooling model~\citep{piro21} are inconsistent with expectations and our analysis of the X-ray emission, such as the overall energetics and velocities. In particular, our inference with this model suggests $\geq 10^{52}$~erg of energy and a total CSM mass of $\geq 6M_{\odot}$. We note that this is significantly larger than that found by \citet{Srinivasaragavan25}, likely due to them including an additional supernova contribution in their fit. In contrast, the dense-CSM model provides more reasonable estimates consistent with the X-ray, with a CSM mass of $0.2-0.9 M_{\odot}$, with a kinetic energy of $\lesssim 10^{51}$~erg with $R_{0} \sim 7\times10^{14}$~cm and a shell width of $\sim 0.3 R_{0}$. Our results suggest that if the light curve is powered by some form of CSM interaction, the CSM has an abrupt drop in density to adequately describe the light curve and be consistent with inferences from the X-ray observations. 
We note that while the model formally assumes the presence of a shell at $R_{0}$, the model is also compatible with the assumption of a wind, provided the wind is truncated at $r > R_{0}$. This provides some clues into the progenitor and perhaps a binary companion (as the CSM need not be from the progenitor of EP250108a), requiring either eruptive mass-loss in the years leading up to the supernova that could place a dense-CSM shell at $R_{0}\sim 10^{4}R_{\odot}$ or a wind profile that terminates at such a radius. 

\begin{figure*}
\epsscale{0.75}
\plotone{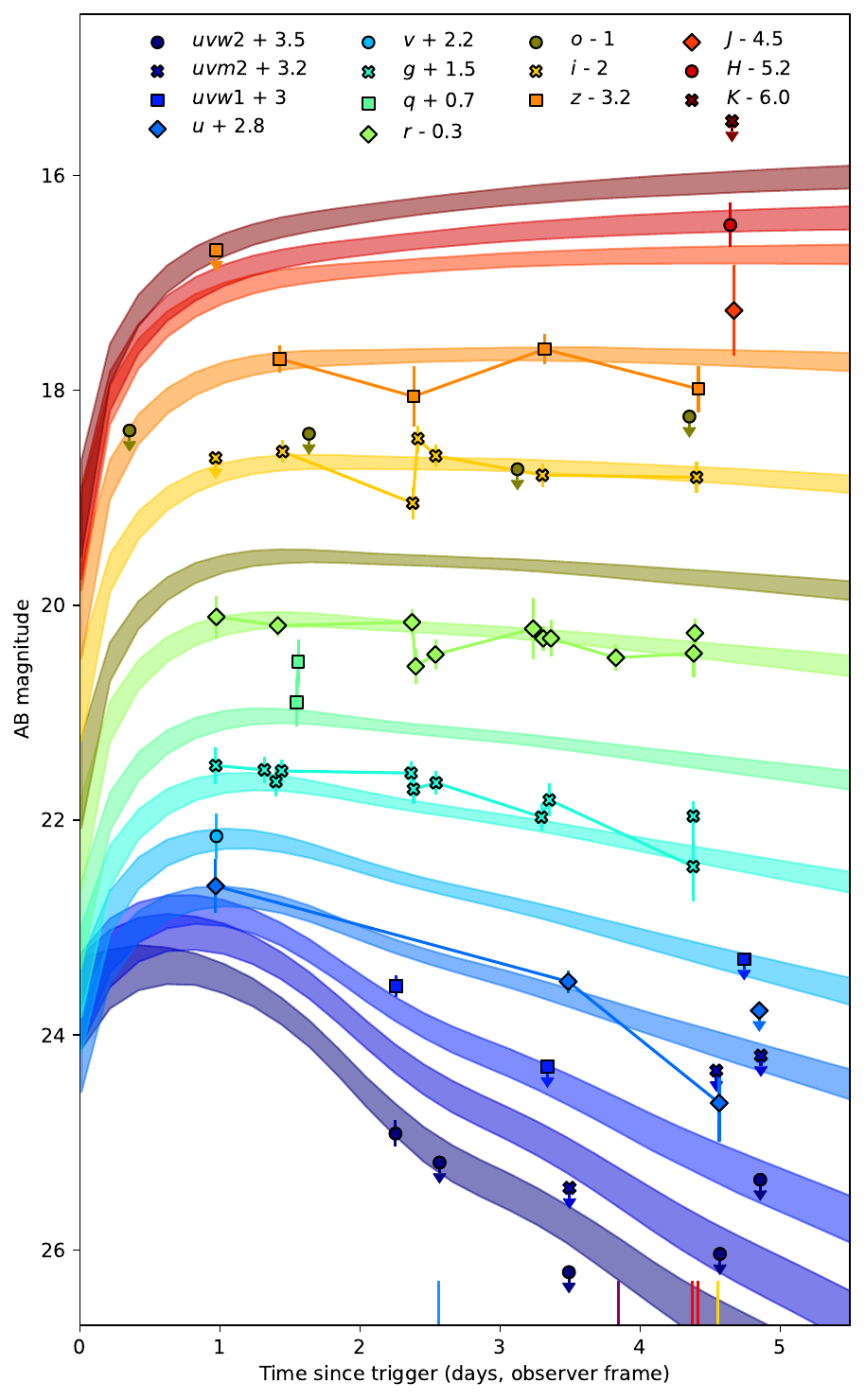}
\caption{The UVOIR light curve of SN 2025kg's fast cooling phase fitted with a shocked cocoon model. The shaded regions indicate the 68\% credible interval of the posterior predictive distribution for each band. The vertical lines indicate the times of our spectroscopic observations (with colours corresponding to Figure \ref{fig:spectral_sequence}).
\label{fig:lc_cocoon_fit}}
\end{figure*}

Motivated by our X-ray analyses, which suggest that jet-driven models are most likely, we also explore a shocked cocoon/failed GRB-jet model. We broadly follow the models for shocked cocoon emission outlined~\citet{Nakar2017} and~\citet{Piro2018}. In particular, we assume a GRB jet of negligible mass propagates out through ejecta of total mass $M$, with a mass profile which follows a power-law distribution i.e.,
\begin{equation}
m(>v) = M \left(\frac{v}{v_0}\right)^{-(s+1)}, 
\end{equation}
where $v_0$ is the minimum velocity of the ejecta, and $s$ is the power-law exponent of the energy distribution with respect to velocity. The energy provided by this interaction is $E \propto \frac{mvR}{t}$ where $t$ is time, while the luminosity is $\approx E(t)/t_{\rm diff}$ where $t_{\rm diff}$ is the characteristic diffusion timescale. We fit this model using {\sc Redback} with the same likelihood and data treatment as described above, further assuming that the shocked ejecta is confined to some angle $\theta_{\rm cocoon}$. Our fits to the light curve are shown in Fig.~\ref{fig:lc_cocoon_fit}. We infer that the cocoon contains $\sim 0.04-0.15M_{\odot}$ of shocked ejecta, confined to $\sim 20-25^{\circ}$, with a kinetic energy of $\sim 4-45\times10^{50}$~erg, with a shock radius of $\sim 1.9-3.5R_{\odot}$. This model provides a good fit to the data, and our parameters are broadly in agreement with inferences from the X-ray observations and the constraints from the radio observations above. They are also consistent with the results of \citet{Srinivasaragavan25} who fit a similar model. While we cannot definitively rule out whether any jet successfully broke out, the inferred energetics from our modelling above, the radio and X-rays constraints suggest that if EP250108a/SN 2025kg was jet-driven, either no jet broke out or was weaker than $\approx 10^{51}$~erg. We note that here, we only considered the thermal emission from the shocked cocoon. Such shocked cocoons are also expected to produce non-thermal emission as discussed earlier in Sec.~\ref{sec:nature}. However, this is extremely sensitive to assumptions about microphysical parameters and the quantity of fast ejecta, which our modelling here does not constrain.

\section{Conclusions}

In this Letter, we have detailed our observations and analysis of the fast cooling phase of EP250108a/SN 2025kg, comprising the first $\sim6$ days of data. We have also used X-ray and radio data from later times to further investigate the transient. The remaining data from our campaign and its analysis are presented in \citet{Rastinejad25} which examines the properties of the supernova itself in much greater detail. We summarise our conclusions below:

\begin{itemize}
    \item SN 2025kg's early evolution shows distinct similarity to similar behaviour in the broad line Ic SN 2006aj and SN 2020bvc suggesting these events arise from similar progenitor systems. There are also similarities to early features in other broad line Ic supernovae SN 2017iuk and possibly SN 2024gsa. In our companion paper, \citet{Rastinejad25}, we show the properties of the supernovae themselves in SN 2006aj, SN 2020bvc and SN 2024gsa are also similar.
    \item We find that the optical transient is consistent with an expanding, cooling blackbody similar to several other examples in the GRB-SN population, in particular SN 2006aj and SN 2020bvc. We simulate the photometry using this model and find, by extrapolation, an initial expansion velocity of order $\sim0.4-0.6c$ that later declines to $\sim0.1-0.2c$ by about 0.5 days (rest frame).
    \item The X-ray emission likely arises from a jet driven engine. To achieve the deceleration and rapid fading required from the observed data, a convection engine would require both extreme supernova energies and masses. However, the low masses required by a collapsar jet engine allow this rapid deceleration and decline in luminosity. The limits set by our radio observations and \textit{Fermi} gamma-ray observations indicate the jet is low energy and/or failed to successfully break out, assuming tophat jets in a wind-medium. We cannot, at this point, constrain whether we are viewing the jet off-axis.
    \item These results suggest that the optical transient is driven by either shock cooling as supernova ejecta expand into a dense shell of CSM or the jet inferred from the X-rays propagates through the ejecta to produce a shocked cocoon. Both models are consistent with the data and support the suggestion that the jet was low energy or failed to successfully break out. Our conclusions are consistent with the findings of \citet{Srinivasaragavan25} and \citet{Li25}. In \citet{Rastinejad25}, we further discuss the progenitor system and rates of these failed/weak jet powered FXT-SNe and show that they are likely higher than similar GRB-SNe.
\end{itemize}

Despite similarities to previous supernova, SN 2025kg is an extraordinary source in terms of both physics and the quality of the data produced by both us and other groups \citep{Srinivasaragavan25,Li25}. \textit{Einstein Probe} and its unique ability to identify FXTs in real time will allow many more of these rare events to be uncovered and the exact physics underpinning them to be explored.

\section*{Acknowledgments}

We are deeply grateful to Tom Marsh for developing the `molly' software, one of his many contributions to advancing the field of compact objects.

We thank Genevieve Schroeder and Gokul Prem Srinivasaragavan for useful discussion.

We also thank the anonymous referee for their constructive comments to improve our original manuscript.

The Liverpool Telescope is operated on the island of La Palma by Liverpool John Moores University in the Spanish Observatorio del Roque de los Muchachos of the Instituto de Astrofisica de Canarias with financial support from the UK Science and Technology Facilities Council.

The data presented here were obtained in part with ALFOSC, which is provided by the Instituto de Astrofisica de Andalucia (IAA) under a joint agreement with the University of Copenhagen and NOT.

This work is partly based on observations collected at the European Southern Observatory under ESO programme 114.27PZ.001.

This work is partly based on observations made with the Gran Telescopio Canarias (GTC), installed at the Spanish Observatorio del Roque de los Muchachos of the Instituto de Astrofísica de Canarias, on the island of La Palma, under program GTC1-24ITP. These data were obtained with the instrument OSIRIS, built by a Consortium led by the Instituto de Astrofísica de Canarias in collaboration with the Instituto de Astronomía of the Universidad Autónoma de México. OSIRIS was funded by GRANTECAN and the National Plan of Astronomy and Astrophysics of the Spanish Government.

This work is partly based on observations obtained at the international Gemini Observatory, a program of NSF NOIRLab, which is managed by the Association of Universities for Research in Astronomy (AURA) under a cooperative agreement with the U.S. National Science Foundation on behalf of the Gemini Observatory partnership: the U.S. National Science Foundation (United States), National Research Council (Canada), Agencia Nacional de Investigaci\'{o}n y Desarrollo (Chile), Ministerio de Ciencia, Tecnolog\'{i}a e Innovaci\'{o}n (Argentina), Minist\'{e}rio da Ci\^{e}ncia, Tecnologia, Inova\c{c}\~{o}es e Comunica\c{c}\~{o}es (Brazil), and Korea Astronomy and Space Science Institute (Republic of Korea), under programs GN-2024B-Q-107 and GS-2024B-Q-105. 

This work made use of data supplied by the UK Swift Science Data Centre at the University of Leicester.

Data for this paper has in part been obtained under the International Time Programme of the CCI (International Scientific Committee of the Observatorios de Canarias of the IAC) with the NOT and GTC operated on the island of La Palma by the Roque de los Muchachos.

This work makes use of observations using the Sinistro imager on the LCOGT 1m telescope at the South African Astronomical Observatory.

The MeerKAT telescope is operated by the South African Radio Astronomy Observatory, which is a facility of the National Research Foundation, an agency of the Department of Science and Innovation. This work has made use of the “MPIfR S-band receiver system” designed, constructed and maintained by funding of the MPI f\"ur Radioastronomy and the Max Planck Society.

Based on observations with the BlackGEM telescope array. The BlackGEM telescope array is built and run by a consortium consisting of Radboud University, the Netherlands Research School for Astronomy (NOVA), and KU Leuven with additional support from Armagh Observatory and Planetarium, Durham University, Hamburg Observatory, Hebrew University, Las Cumbres Observatory, Tel Aviv University, Texas Tech University, Technical University of Denmark, University of California Davis, the University of Barcelona, the University of Manchester, University of Potsdam, the University of Valparaiso, the University of Warwick, and Weizmann Institute of science. BlackGEM is hosted and supported by ESO at La Silla.

PGJ, MER \& JNDvD are funded by the European Union (ERC, Starstruck, 101095973). Views and opinions expressed are however those of the authors only and do not necessarily reflect those of the European Union or the European Research Council Executive Agency. Neither the European Union nor the granting authority can be held responsible for them.
NS acknowledges support from the Knut and Alice Wallenberg Foundation through the "Gravity Meets Light" project and the research environment grant ``Gravitational Radiation and Electromagnetic Astrophysical Transients'' (GREAT) funded by the Swedish Research Council (VR) under Dnr 2016-06012.
The work by CLF was supported by the US Department of Energy through the Los Alamos National Laboratory. Los Alamos National Laboratory is operated by Triad National Security, LLC, for the National Nuclear Security Administration of U.S.\ Department of Energy (Contract No.\ 89233218CNA000001).  
PTO acknowledges support from UKRI under grant ST/W000857/1.
BPG acknowledges support from STFC grant No. ST/Y002253/1 and The Leverhulme Trust grant No. RPG-2024-117.
AS acknowledges support by a postdoctoral fellowship from the CNES.
RLCS acknowledges support from Leverhulme Trust grant RPG-2023-240.
GL is supported by a research grant (VIL60862) from VILLUM FONDEN.
DMS and MAPT acknowledge support by the Spanish Ministry of Science via the Plan de Generacion de conocimiento PID2020-120323GB-I00. DMS also acknowledges support via a Ramon y Cajal Fellowship RYC2023-044941.
CJN acknowledges support from the Science and Technology Facilities Council (grant No. ST/Y000544/1) and from the Leverhulme Trust (grant No. RPG-2021-380).

For the purpose of open access, the author has applied a Creative Commons Attribution (CC BY) licence to the Author Accepted Manuscript version arising from this submission.

\vspace{5mm}
\facilities{Gemini:Gillett (GMOS-N), Gemini:South (FLAMINGOS2), GTC(OSIRIS), LCOGT (SINISTRO), Liverpool:2m (IO:O), NOT (ALFOSC), Swift (XRT and UVOT), VLT:Melipal (X-shooter)}

\software{astropy \citep{astropy13,astropy18,astropy22}, dust\_extinction \citep{Gordon24b}, emcee \citep{ForemanMackey13}, EsoReflex \citep{Freudling13}, HEASoft \citep{heasoft}, photometry-sans-frustration \citep{Nicholl23}, PIMMS, PyZOGY \citep{Guevel21}, {\sc Redback} \citep{sarin_redback}}

\appendix

\section{Observing logs}

\startlongtable
\begin{deluxetable*}{ccccccc}
\tabletypesize{\scriptsize}
\tablecaption{The assembled UV, optical and NIR photometry from our observations and other sources. $\Delta t$ is given in the observer frame and magnitudes are as measured and are not corrected for Galactic extinction. \label{tab:photometry}}
\tablehead{\colhead{Date (UT)} & \colhead{$\Delta t$ (days)} & \colhead{Instrument} & \colhead{Filter} & \colhead{Exposure time (s)} & \colhead{AB magnitude} & \colhead{Reference}}
\startdata
2025 Jan 10.77415 & 2.25299 & \textit{Swift}/UVOT & \textit{uvw2} & 1007.4 & $21.55 \pm 0.07$ & Revision of \citet{GCN38909}. \\
2025 Jan 11.08888 & 2.56773 & \textit{Swift}/UVOT & \textit{uvw2} & 1080.2 & $>21.82$ & This work. \\
2025 Jan 12.01286 & 3.49171 & \textit{Swift}/UVOT & \textit{uvw2} & 2053.3 & $>22.84$ & This work. \\
2025 Jan 13.09120 & 4.57005 & \textit{Swift}/UVOT & \textit{uvw2} & 1667.2 & $>22.67$ & This work. \\
2025 Jan 13.37897 & 4.85782 & \textit{Swift}/UVOT & \textit{uvw2} & 597.2 & $>21.98$ & This work. \\
\tableline
2025 Jan 12.01700 & 3.49584 & \textit{Swift}/UVOT & \textit{uvm2} & 917.4 & $>22.37$ & This work. \\
2025 Jan 13.06583 & 4.54467 & \textit{Swift}/UVOT & \textit{uvm2} & 215.0 & $>21.28$ & This work. \\
2025 Jan 13.38372 & 4.86257 & \textit{Swift}/UVOT & \textit{uvm2} & 198.8 & $>21.14$ & This work. \\
\tableline
2025 Jan 10.77958 & 2.25842 & \textit{Swift}/UVOT & \textit{uvw1} & 843.9 & $20.65 \pm 0.02$ & Revision of \citet{GCN38909}. \\
2025 Jan 11.85942 & 3.33827 & \textit{Swift}/UVOT & \textit{uvw1} & 251.2 & $>21.40$ & This work. \\
2025 Jan 13.26392 & 4.74277 & \textit{Swift}/UVOT & \textit{uvw1} & 67.6 & $>20.40$ & This work. \\
\tableline
2025 Jan 09.49019 & 0.969 & MEPHISTO & \textit{u}\tablenotemark{\scriptsize a} & $2\times180$ & $19.89 \pm 0.23$ & \citet{GCN38914}. \\
2025 Jan 12.00747 & 3.48631 & \textit{Swift}/UVOT & \textit{u} & 2892.0 & $20.78 \pm 0.03$ & This work. \\
2025 Jan 13.08614 & 4.56499 & \textit{Swift}/UVOT & \textit{u} & 2744.4 & $21.91 \pm 0.35$ & This work. \\
2025 Jan 13.37186 & 4.85071 & \textit{Swift}/UVOT & \textit{u} & 597.2 & $>21.05$ & This work. \\
\tableline
2025 Jan 09.41519 & 0.974 & MEPHISTO & \textit{v} & $2\times180$ & $20.02 \pm 0.19$ & \citet{GCN38914}. \\
\tableline
2025 Jan 06.20545 & -2.31571 & ZTF & \textit{g} & 30 & $>21.54$ & This work. \\
2025 Jan 09.49020 & 0.969 & MEPHISTO & \textit{g} & $6\times50$ & $20.05 \pm 0.14$ & \citet{GCN38914}. \\
2025 Jan 09.83781 & 1.31665 & LT/IO:O & \textit{g} & $1\times200$ & $20.09 \pm 0.07$ & Revision of \citet{GCN38878}. \\
2025 Jan 9.920945 & $\sim1.4$ & LT/IO:O & \textit{g} & --- & $20.20 \pm 0.10$ & \citet{GCN38907}. \\
2025 Jan 9.96101 & 1.43985 & NOT/ALFOSC & \textit{g} & $2\times300$ & $20.10 \pm 0.04$ & Revision of \citet{GCN38885}. \\
2025 Jan 10.19027 & 1.66912 & ZTF & \textit{g} & 30 & $>19.00$ & This work. \\
2025 Jan 10.88571 & 2.36454 & NOT/ALFOSC & \textit{g} & $2\times300$ & $20.12 \pm 0.05$ & \citet{GCN38902}. \\
2025 Jan 10.90305 & 2.38189 & LT/IO:O & \textit{g} & $6\times200$ & $20.27 \pm 0.10$ & This work. \\
2025 Jan 11.06419 & 2.54302 & VLT/X-shooter & \textit{g} & $3\times40$ & $20.21 \pm 0.04$ & This work. \\
2025 Jan 11.81677 & 3.29560 & NOT/ALFOSC & \textit{g} & $2\times300$ & $20.53 \pm 0.08$ & This work. \\
2025 Jan 11.87469 & 3.35160 & LCO/Sinistro & \textit{g} & $3\times300$ & $20.37 \pm 0.12$ & Revision of \citet{GCN38912}. \\
2025 Jan 12.16815 & 3.64700 & ZTF & \textit{g} & 30 & $>19.01$ & This work. \\
2025 Jan 12.90017 & 4.37901 & LCO/Sinistro & \textit{g} & 3x300 & $20.99 \pm 0.31$ & This work.\\
2025 Jan 12.90049 & 4.37933 & LT/IO:O & \textit{g} & $6\times150$ & $20.52 \pm 0.09$ & This work. \\
2025 Jan 13.18974 & 4.66859 & ZTF & \textit{g} & 30 & $>19.63$ & This work. \\
\tableline
2025 Jan 10.06904 & 1.54788 & BlackGEM & \textit{q} & 60 & $20.25 \pm 0.20$ & This work. \\
2025 Jan 10.07963 & 1.55847 & BlackGEM & \textit{q} & 60 & $19.87 \pm 0.18$ & This work. \\
\tableline
2025 Jan 09.41519 & 0.974 & MEPHISTO & \textit{r} & $6\times50$ & $20.45 \pm 0.17$ & \citet{GCN38914}. \\
2025 Jan 9.93392 & 1.41277 & NOT/ALFOSC & \textit{r} & $3\times300$ & $20.53 \pm 0.02$ & Revision of \citet{GCN38885}. \\
2025 Jan 10.89215 & 2.37098 & NOT/ALFOSC & \textit{r} & $2\times180$ & $20.50 \pm 0.07$ & \citet{GCN38902}. \\
2025 Jan 10.91841 & 2.39725 & LT/IO:O & \textit{r} & $6\times200$ & $20.91 \pm 0.13$ & This work. \\
2025 Jan 11.05948 & 2.53831 & VLT/X-shooter & \textit{r} & $3\times30$ & $20.80 \pm 0.10$ & This work. \\
2025 Jan 11.75736 & 3.2362 & SAO RAS Zeis-1000 & \textit{r} & $8\times300$ & $20.56 \pm 0.27$ & \citet{GCN38925}. \\
2025 Jan 11.82939 & 3.30822 & NOT/ALFOSC & \textit{r} & $3\times180$ & $20.65 \pm 0.06$ & This work. \\
2025 Jan 11.88624 & 3.363 & LCO/Sinistro & \textit{r} & $3\times300$ & $20.65 \pm 0.14$ & Revision of \citet{GCN38912}. \\
2025 Jan 12.23224 & 3.71109 & ZTF & \textit{r} & 30 & $>19.58$ & This work. \\
2025 Jan 12.34830 & 3.82715 & Gemini-North/GMOS-N & \textit{r} & 50 & $20.83 \pm 0.07$ & This work. \\
2025 Jan 12.90408 & 4.38292 & LCO/Sinistro & \textit{r} & 3x300 & $20.79 \pm 0.20$ & This work. \\
2025 Jan 12.91241 & 4.39125 & LT/IO:O & \textit{r} & $6\times150$ & $20.60 \pm 0.09$ & This work. \\
2025 Jan 13.21237 & 4.69121 & ZTF & \textit{r} & 30 & $>19.06$ & This work. \\
\tableline
2025 Jan 08.87477 & 0.35362 & ATLAS & \textit{o} & $3\times30$ & $>19.41$ & This work. \\
2025 Jan 10.15593 & 1.63478 & ATLAS & \textit{o} & $4\times30$ & $>19.44$ & This work. \\
2025 Jan 11.64527 & 3.12411 & ATLAS & \textit{o} & $6\times30$ & $>19.77$ & This work. \\
2025 Jan 12.87307 & 4.35191 & ATLAS & \textit{o} & $4\times30$ & $>19.28$ & This work. \\
2025 Jan 14.12614 & 5.60499 & ATLAS & \textit{o} & $3\times30$ & $>19.10$ & This work. \\
\tableline
2025 Jan 09.49020 & 0.969 & MEPHISTO & \textit{i} & $4\times79$ & $>20.66$ & \citet{GCN38914}. \\
2025 Jan 9.96796 & 1.44681 & NOT/ALFOSC & \textit{i} & $2\times180$ & $20.60 \pm 0.03$ & Revision of \citet{GCN38885}. \\
2025 Jan 10.89718 & 2.37601 & NOT/ALFOSC & \textit{i} & $2\times180$ & $21.08 \pm 0.11$ & \citet{GCN38902}. \\
2025 Jan 10.93375 & 2.41259 & LT/IO:O & \textit{i} & $6\times200$ & $20.48 \pm 0.07$ & This work. \\
2025 Jan 11.06172 & 2.54056 & VLT/X-shooter & \textit{i} & $3\times60$ & $20.64 \pm 0.04$ & This work. \\
2025 Jan 11.82321 & 3.30204 & NOT/ALFOSC & \textit{i} & $2\times180$ & $20.82 \pm 0.05$ & This work. \\
2025 Jan 12.92426 & 4.4031 & LT/IO:O & \textit{i} & $6\times150$ & $20.84 \pm 0.10$ & This work. \\
\tableline
2025 Jan 09.49519 & 0.974 & MEPHISTO & \textit{z} & $4\times79$ & $>19.92$ & \citet{GCN38914}. \\
2025 Jan 9.94647 & 1.4253 & NOT/ALFOSC & \textit{z} & $5\times200$ & $20.93 \pm 0.08$ & Revision of \citet{GCN38885}. \\
2025 Jan 10.90377 & 2.3826 & NOT/ALFOSC & \textit{z} & $3\times200$ & $21.28 \pm 0.26$ & \citet{GCN38902}. \\
2025 Jan 11.83977 & 3.3186 & NOT/ALFOSC & \textit{z} & $5\times200$ & $20.84 \pm 0.10$ & This work. \\
2025 Jan 12.93758 & 4.41642 & LT/IO:O & \textit{z} & $6\times200$ & $21.21 \pm 0.19$ & This work. \\
\tableline
2025-01-13 4:35:25 & 4.67009 & Gemini-South/F2 & \textit{J} & $3\times30$ & $21.77 \pm 0.41$ & This work. \\
\tableline
2025-01-13 3:57:32 & 4.64378 & Gemini-South/F2 & \textit{H} & $25\times15$ & $21.67 \pm 0.18$ & This work. \\
\tableline
2025-01-13 4:15:23 & 4.65618 & Gemini-South/F2 & \textit{K} & $24\times15$ & $>21.50$ & This work. \\
\enddata
\tablenotetext{a}{The effective wavelengths of the MEPHISTO \textit{u} and \textit{Swift}/UVOT \textit{u} filters are separated by approximately 35\AA, in this work we assume them to have the same effective wavelength of 3483 \AA.}
\end{deluxetable*}

\begin{deluxetable*}{ccccccc}[b]
\tablecaption{The log of our spectroscopic observations. $\Delta t$ is given in the observer frame. \label{tab:spectroscopy}}
\tablehead{\colhead{Date (UT)} & \colhead{$\Delta t$ (days)} & \colhead{Instrument} & \colhead{Exposure time (s)} & \colhead{$T_{\rm bb}$ ($10^4$ K)} & \colhead{$R_{\rm bb}$ ($10^{15}$ cm)} & \colhead{$\log\left( \frac{L_{\rm bb, bol}}{{\rm erg~s}^{-1}} \right)$}}
\startdata
2025 Jan 11.08112 & 2.56000 & VLT/X-shooter & $4\times600$ & $1.23\pm0.01$ & $1.13\pm0.01$ & $43.32\pm0.01$ \\
2025 Jan 12.36373 & 3.84257 & Gemini-North/GMOS-N & $3\times400$ & $1.17\pm0.04$ & $1.04\pm0.05$ & $43.17\pm0.02$ \\
2025 Jan 12.89132 & 4.37016 & GTC/OSIRIS+/R1000R\tablenotemark{\scriptsize{a}} & $3\times1200$ & $1.24\pm0.02$ & $1.12\pm0.02$ & $43.33\pm0.01$ \\
2025 Jan 12.93447 & 4.41331 & GTC/OSIRIS+/R1000B\tablenotemark{\scriptsize{a}} & $3\times1200$ & $1.24\pm0.02$ & $1.12\pm0.02$ & $43.33\pm0.01$ \\
2025 Jan 13.07514 & 4.55398 & VLT/X-shooter & $4\times600$ & $0.96\pm0.01$ & $1.34\pm0.02$ & $43.04\pm0.01$ \\
\enddata
\tablenotetext{a}{Note that we combine the data from the two GTC/OSIRIS+ observations to produce a single spectrum covering the full wavelength range of both gratings.}
\end{deluxetable*}

\begin{deluxetable*}{ccccc}
\tablecaption{The log of our X-ray observations and additional public data observed by \textit{Swift}. $\Delta t$ is given in the observer frame. \label{tab:xray}}
\tablehead{\colhead{Date (UT)} & \colhead{$\Delta t$ (days)} & \colhead{Instrument} & \colhead{Exposure time (ks)} & \colhead{$\log\left( \frac{L_{\rm 0.5-10~keV}}{{\rm erg~s}^{-1}} \right)$}}
\startdata
2025 Jan 10.86591 & 2.34475 & \textit{Swift}/XRT & 2.989 & $<42.65$\\
2025 Jan 12.01331 & 3.49215 & \textit{Swift}/XRT & 4.919 & $<42.45$\\
2025 Jan 13.12290 & 4.60175 & \textit{Swift}/XRT & 5.139 & $<42.57$\\
2025 Jan 14.72557 & 6.20441 & \textit{XMM-Newton}/pn & 30.44 & $<40.99$\\
2025 Jan 23.16258 & 14.64143 & \textit{Swift}/XRT & 1.59 & $<42.74$\\
2025 Jan 23.32540 & 14.80424 & \textit{Swift}/XRT & 1.248 & $<43.03$\\
2025 Jan 28.85781 & 20.33665 & \textit{Chandra}/ACIS-S3 & 10.851 & $<41.66$\\
\enddata
\end{deluxetable*}

\begin{deluxetable*}{cccccc}
\tablecaption{The log of our radio observations. $\Delta t$ is given in the observer frame. \label{tab:radio}}
\tablehead{\colhead{Date (UT)} & \colhead{$\Delta t$ (days)} & \colhead{Telescope} & \colhead{Central frequency (GHz)} & \colhead{Integration time (minutes)} & \colhead{$F_\nu$ ($\mu$Jy\,beam$^{-1}$)}}
\startdata
2025 Jan 13.74792 & 5.22676 & MeerKAT & 3.06 & 42 & 24 \\
2025 Jan 30.57014 & 22.04898 & MeerKAT & 3.06 & 42 & 24 \\
2025 Feb 20.66319 & 43.14203 & MeerKAT & 3.06 & 42 & 27 \\
2025 Apr 3.39167 & 84.87051 & MeerKAT &3.06 & 42 & 26 \\ 
\enddata
\end{deluxetable*}

\section{Corner plot}

\begin{figure*}
\epsscale{1.18}
\plotone{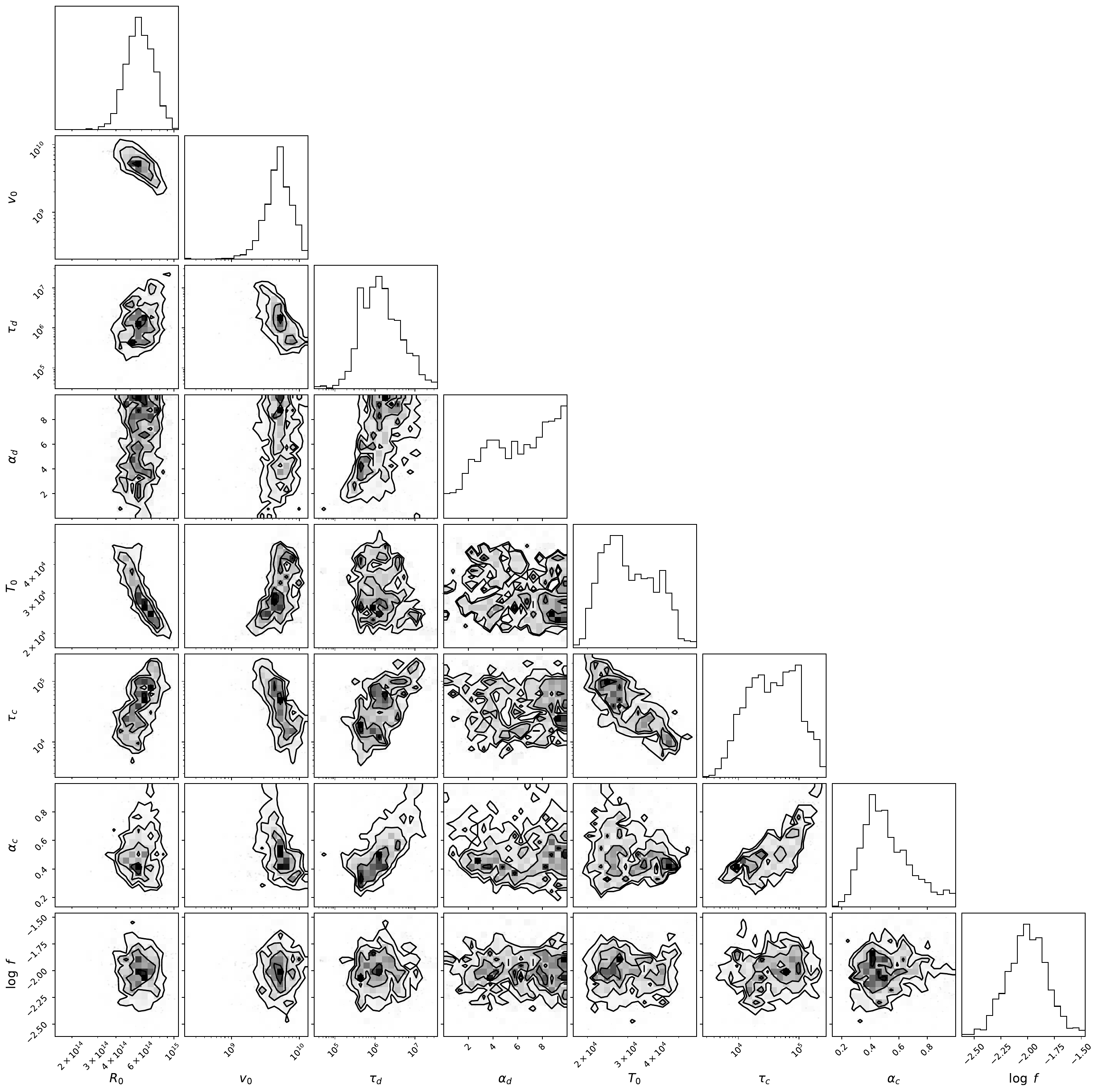}
\caption{A corner plot showing the parameter covariance in our fit to SN 2025kg's fast cooling phase with a cooling, expanding blackbody.
\label{fig:bb_corner}}
\end{figure*}

\section{SN 2025kg in relation to luminous fast blue optical transients}

Another class of transients that also display a blue colour with rapid temperature and luminosity evolution are LFBOTs. Initially, SN 2025kg was suggested to be a member of this class \citep{GCN38908} while a link has also been suggested in the case of the FXT supernova SN 2024gsa \citep{vanDalen25}. Here we briefly consider the fast cooling phase's properties in relation to observed LFBOTs. 

\begin{figure*}
\epsscale{1.18}
\plotone{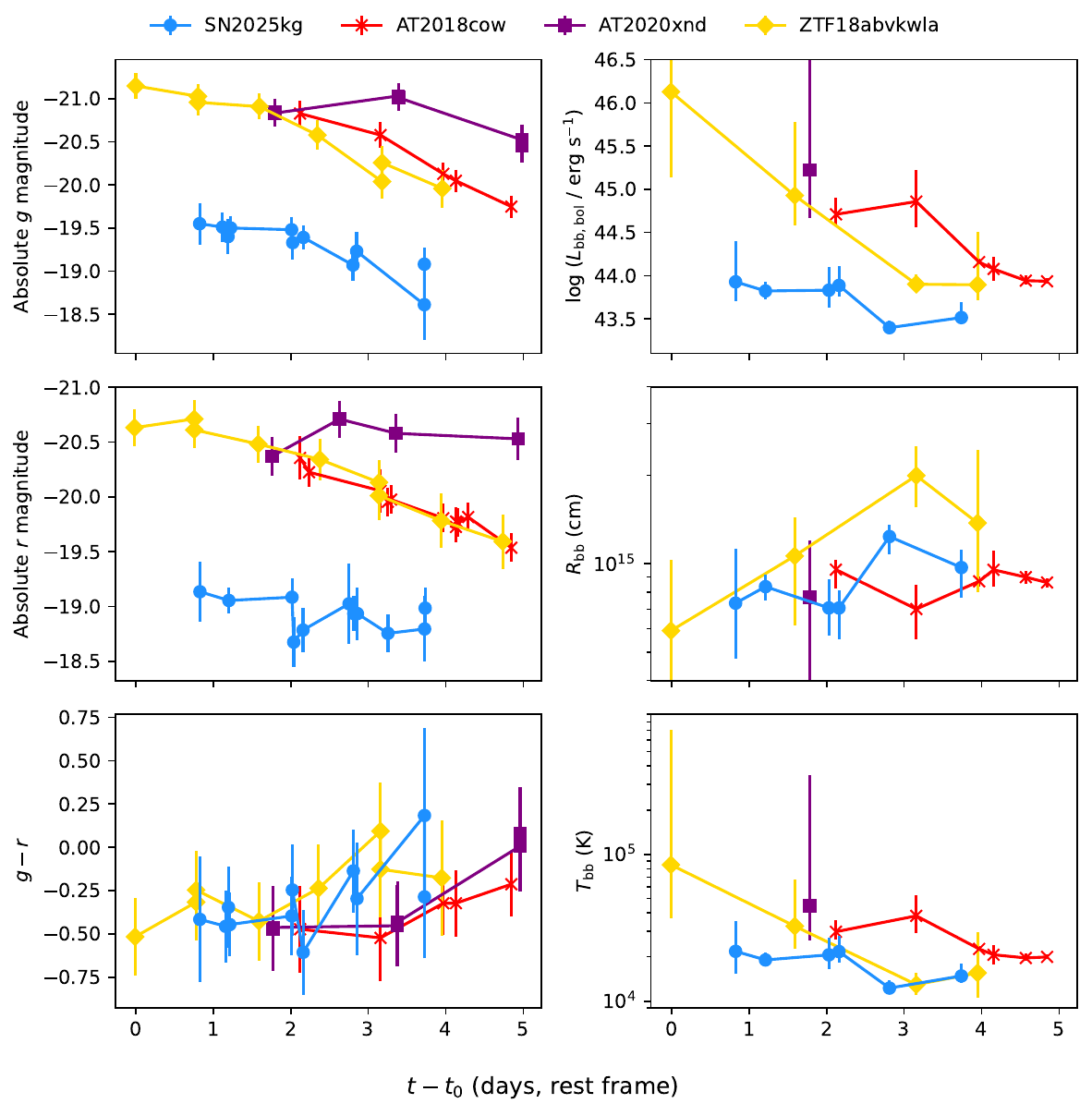}
\caption{The photometric and blackbody properties of SN 2025kg's fast cooling phase compared to the early properties of the LFBOTs AT2018cow, ZTF18abvkwla and AT2020xnd. Note that we have taken $t_0 = 58374.41$ MJD for ZTF18abvkwla i.e. the peak time in the \textit{g} band light curve.}
\label{fig:fbot_comp}
\end{figure*}

In Figure \ref{fig:fbot_comp}, we compare the fast cooling phase of SN 2025kg to early data from AT2018cow \citep{Prentice18,Margutti19,Perley19}, ZTF18abvkwla \citep{Ho20b} and AT2020xnd/ZTF20acigmel \citep{Perley21,Bright22,Ho22}. Additional examples include AT2020mrf \citep{Yao22} and AT2022tsd \citep{Ho22b,Ho22c,Matthews23}. Similarly to our supernovae comparison above, blackbodies are fitted to individual epochs of the catalogued photometry using the procedure in Section \ref{sec:phot}.

From Figure \ref{fig:fbot_comp}, it is clear that from the optical properties alone, SN 2025kg is very different to behaviour observed in the LFBOT population. In particular, it is less luminous and cooler than both the prototypical AT2018cow and AT2020xnd. The blackbody radius of SN 2025kg is also somewhat larger and continuing to expand. While Figure \ref{fig:fbot_comp} only shows the earliest times, both LFBOTs' photospheres were shown to recede. Extending the comparison to other wavelengths, the rapid fading of EP250108a is also inconsistent with the X-ray behaviour of both of these events which display emission over tens of days at luminosities incompatible with the upper limits we derive above. At this time cannot rule out similar behaviour to that seen in another LFBOT CSS161010 \citep{Coppejans20,Gutierrez24} which displayed lower luminosity X-ray emission at a $\sim100$ days post-outburst. Similarly, our current radio limits for SN 2025kg are compatible with the behaviour seen in the LFBOT sample which are still rising at a comparable time.

To conclude, SN 2025kg is inconsistent with many properties of LFBOTs at optical and X-ray wavelengths and it is unlikely that the same progenitor system is responsible for producing both classes of transient.

\bibliography{main}{}

\begin{thebibliography}{}
\expandafter\ifx\csname natexlab\endcsname\relax\def\natexlab#1{#1}\fi
\providecommand{\url}[1]{\href{#1}{#1}}
\providecommand{\dodoi}[1]{doi:~\href{http://doi.org/#1}{\nolinkurl{#1}}}
\providecommand{\doeprint}[1]{\href{http://ascl.net/#1}{\nolinkurl{http://ascl.net/#1}}}
\providecommand{\doarXiv}[1]{\href{https://arxiv.org/abs/#1}{\nolinkurl{https://arxiv.org/abs/#1}}}

\bibitem[{{Alp} \& {Larsson}(2020)}]{Alp20}
{Alp}, D., \& {Larsson}, J. 2020, \apj, 896, 39, \dodoi{10.3847/1538-4357/ab91ba}

\bibitem[{{An} {et~al.}(2025){An}, {Liu}, {Geng}, {Wu}, {Carotenuto}, \& {Einstein Probe radio follow-up Team}}]{GCN38998}
{An}, T., {Liu}, Y., {Geng}, J., {et~al.} 2025, GRB Coordinates Network, 38998, 1

\bibitem[{{Ashton} {et~al.}(2019){Ashton}, {H{\"u}bner}, {Lasky}, {Talbot}, {Ackley}, {Biscoveanu}, {Chu}, {Divakarla}, {Easter}, {Goncharov}, {Hernandez Vivanco}, {Harms}, {Lower}, {Meadors}, {Melchor}, {Payne}, {Pitkin}, {Powell}, {Sarin}, {Smith}, \& {Thrane}}]{ashton19}
{Ashton}, G., {H{\"u}bner}, M., {Lasky}, P.~D., {et~al.} 2019, \apjs, 241, 27, \dodoi{10.3847/1538-4365/ab06fc}

\bibitem[{{Astropy Collaboration} {et~al.}(2013){Astropy Collaboration}, {Robitaille}, {Tollerud}, {Greenfield}, {Droettboom}, {Bray}, {Aldcroft}, {Davis}, {Ginsburg}, {Price-Whelan}, {Kerzendorf}, {Conley}, {Crighton}, {Barbary}, {Muna}, {Ferguson}, {Grollier}, {Parikh}, {Nair}, {Unther}, {Deil}, {Woillez}, {Conseil}, {Kramer}, {Turner}, {Singer}, {Fox}, {Weaver}, {Zabalza}, {Edwards}, {Azalee Bostroem}, {Burke}, {Casey}, {Crawford}, {Dencheva}, {Ely}, {Jenness}, {Labrie}, {Lim}, {Pierfederici}, {Pontzen}, {Ptak}, {Refsdal}, {Servillat}, \& {Streicher}}]{astropy13}
{Astropy Collaboration}, {Robitaille}, T.~P., {Tollerud}, E.~J., {et~al.} 2013, \aap, 558, A33, \dodoi{10.1051/0004-6361/201322068}

\bibitem[{{Astropy Collaboration} {et~al.}(2018){Astropy Collaboration}, {Price-Whelan}, {Sip{\H{o}}cz}, {G{\"u}nther}, {Lim}, {Crawford}, {Conseil}, {Shupe}, {Craig}, {Dencheva}, {Ginsburg}, {VanderPlas}, {Bradley}, {P{\'e}rez-Su{\'a}rez}, {de Val-Borro}, {Aldcroft}, {Cruz}, {Robitaille}, {Tollerud}, {Ardelean}, {Babej}, {Bach}, {Bachetti}, {Bakanov}, {Bamford}, {Barentsen}, {Barmby}, {Baumbach}, {Berry}, {Biscani}, {Boquien}, {Bostroem}, {Bouma}, {Brammer}, {Bray}, {Breytenbach}, {Buddelmeijer}, {Burke}, {Calderone}, {Cano Rodr{\'\i}guez}, {Cara}, {Cardoso}, {Cheedella}, {Copin}, {Corrales}, {Crichton}, {D'Avella}, {Deil}, {Depagne}, {Dietrich}, {Donath}, {Droettboom}, {Earl}, {Erben}, {Fabbro}, {Ferreira}, {Finethy}, {Fox}, {Garrison}, {Gibbons}, {Goldstein}, {Gommers}, {Greco}, {Greenfield}, {Groener}, {Grollier}, {Hagen}, {Hirst}, {Homeier}, {Horton}, {Hosseinzadeh}, {Hu}, {Hunkeler}, {Ivezi{\'c}}, {Jain}, {Jenness}, {Kanarek}, {Kendrew}, {Kern}, {Kerzendorf}, {Khvalko}, {King}, {Kirkby}, {Kulkarni},
  {Kumar}, {Lee}, {Lenz}, {Littlefair}, {Ma}, {Macleod}, {Mastropietro}, {McCully}, {Montagnac}, {Morris}, {Mueller}, {Mumford}, {Muna}, {Murphy}, {Nelson}, {Nguyen}, {Ninan}, {N{\"o}the}, {Ogaz}, {Oh}, {Parejko}, {Parley}, {Pascual}, {Patil}, {Patil}, {Plunkett}, {Prochaska}, {Rastogi}, {Reddy Janga}, {Sabater}, {Sakurikar}, {Seifert}, {Sherbert}, {Sherwood-Taylor}, {Shih}, {Sick}, {Silbiger}, {Singanamalla}, {Singer}, {Sladen}, {Sooley}, {Sornarajah}, {Streicher}, {Teuben}, {Thomas}, {Tremblay}, {Turner}, {Terr{\'o}n}, {van Kerkwijk}, {de la Vega}, {Watkins}, {Weaver}, {Whitmore}, {Woillez}, {Zabalza}, \& {Astropy Contributors}}]{astropy18}
{Astropy Collaboration}, {Price-Whelan}, A.~M., {Sip{\H{o}}cz}, B.~M., {et~al.} 2018, \aj, 156, 123, \dodoi{10.3847/1538-3881/aabc4f}

\bibitem[{{Astropy Collaboration} {et~al.}(2022){Astropy Collaboration}, {Price-Whelan}, {Lim}, {Earl}, {Starkman}, {Bradley}, {Shupe}, {Patil}, {Corrales}, {Brasseur}, {N{\"o}the}, {Donath}, {Tollerud}, {Morris}, {Ginsburg}, {Vaher}, {Weaver}, {Tocknell}, {Jamieson}, {van Kerkwijk}, {Robitaille}, {Merry}, {Bachetti}, {G{\"u}nther}, {Aldcroft}, {Alvarado-Montes}, {Archibald}, {B{\'o}di}, {Bapat}, {Barentsen}, {Baz{\'a}n}, {Biswas}, {Boquien}, {Burke}, {Cara}, {Cara}, {Conroy}, {Conseil}, {Craig}, {Cross}, {Cruz}, {D'Eugenio}, {Dencheva}, {Devillepoix}, {Dietrich}, {Eigenbrot}, {Erben}, {Ferreira}, {Foreman-Mackey}, {Fox}, {Freij}, {Garg}, {Geda}, {Glattly}, {Gondhalekar}, {Gordon}, {Grant}, {Greenfield}, {Groener}, {Guest}, {Gurovich}, {Handberg}, {Hart}, {Hatfield-Dodds}, {Homeier}, {Hosseinzadeh}, {Jenness}, {Jones}, {Joseph}, {Kalmbach}, {Karamehmetoglu}, {Ka{\l}uszy{\'n}ski}, {Kelley}, {Kern}, {Kerzendorf}, {Koch}, {Kulumani}, {Lee}, {Ly}, {Ma}, {MacBride}, {Maljaars}, {Muna}, {Murphy}, {Norman},
  {O'Steen}, {Oman}, {Pacifici}, {Pascual}, {Pascual-Granado}, {Patil}, {Perren}, {Pickering}, {Rastogi}, {Roulston}, {Ryan}, {Rykoff}, {Sabater}, {Sakurikar}, {Salgado}, {Sanghi}, {Saunders}, {Savchenko}, {Schwardt}, {Seifert-Eckert}, {Shih}, {Jain}, {Shukla}, {Sick}, {Simpson}, {Singanamalla}, {Singer}, {Singhal}, {Sinha}, {Sip{\H{o}}cz}, {Spitler}, {Stansby}, {Streicher}, {{\v{S}}umak}, {Swinbank}, {Taranu}, {Tewary}, {Tremblay}, {de Val-Borro}, {Van Kooten}, {Vasovi{\'c}}, {Verma}, {de Miranda Cardoso}, {Williams}, {Wilson}, {Winkel}, {Wood-Vasey}, {Xue}, {Yoachim}, {Zhang}, {Zonca}, \& {Astropy Project Contributors}}]{astropy22}
{Astropy Collaboration}, {Price-Whelan}, A.~M., {Lim}, P.~L., {et~al.} 2022, \apj, 935, 167, \dodoi{10.3847/1538-4357/ac7c74}

\bibitem[{{Bauer} {et~al.}(2017){Bauer}, {Treister}, {Schawinski}, {Schulze}, {Luo}, {Alexander}, {Brandt}, {Comastri}, {Forster}, {Gilli}, {Kann}, {Maeda}, {Nomoto}, {Paolillo}, {Ranalli}, {Schneider}, {Shemmer}, {Tanaka}, {Tolstov}, {Tominaga}, {Tozzi}, {Vignali}, {Wang}, {Xue}, \& {Yang}}]{Bauer17}
{Bauer}, F.~E., {Treister}, E., {Schawinski}, K., {et~al.} 2017, \mnras, 467, 4841, \dodoi{10.1093/mnras/stx417}

\bibitem[{{Breeveld} {et~al.}(2011){Breeveld}, {Landsman}, {Holland}, {Roming}, {Kuin}, \& {Page}}]{Breeveld11}
{Breeveld}, A.~A., {Landsman}, W., {Holland}, S.~T., {et~al.} 2011, in American Institute of Physics Conference Series, Vol. 1358, Gamma Ray Bursts 2010, ed. J.~E. {McEnery}, J.~L. {Racusin}, \& N.~{Gehrels} (AIP), 373--376, \dodoi{10.1063/1.3621807}

\bibitem[{{Bright} {et~al.}(2022){Bright}, {Margutti}, {Matthews}, {Brethauer}, {Coppejans}, {Wieringa}, {Metzger}, {DeMarchi}, {Laskar}, {Romero}, {Alexander}, {Horesh}, {Migliori}, {Chornock}, {Berger}, {Bietenholz}, {Devlin}, {Dicker}, {Jacobson-Gal{\'a}n}, {Mason}, {Milisavljevic}, {Motta}, {Mroczkowski}, {Ramirez-Ruiz}, {Rhodes}, {Sarazin}, {Sfaradi}, \& {Sievers}}]{Bright22}
{Bright}, J.~S., {Margutti}, R., {Matthews}, D., {et~al.} 2022, \apj, 926, 112, \dodoi{10.3847/1538-4357/ac4506}

\bibitem[{{Buchner} {et~al.}(2014){Buchner}, {Georgakakis}, {Nandra}, {Hsu}, {Rangel}, {Brightman}, {Merloni}, {Salvato}, {Donley}, \& {Kocevski}}]{pymultinest}
{Buchner}, J., {Georgakakis}, A., {Nandra}, K., {et~al.} 2014, \aap, 564, A125, \dodoi{10.1051/0004-6361/201322971}

\bibitem[{{Bufano} {et~al.}(2012){Bufano}, {Pian}, {Sollerman}, {Benetti}, {Pignata}, {Valenti}, {Covino}, {D'Avanzo}, {Malesani}, {Cappellaro}, {Della Valle}, {Fynbo}, {Hjorth}, {Mazzali}, {Reichart}, {Starling}, {Turatto}, {Vergani}, {Wiersema}, {Amati}, {Bersier}, {Campana}, {Cano}, {Castro-Tirado}, {Chincarini}, {D'Elia}, {de Ugarte Postigo}, {Deng}, {Ferrero}, {Filippenko}, {Goldoni}, {Gorosabel}, {Greiner}, {Hammer}, {Jakobsson}, {Kaper}, {Kawabata}, {Klose}, {Levan}, {Maeda}, {Masetti}, {Milvang-Jensen}, {Mirabel}, {M{\o}ller}, {Nomoto}, {Palazzi}, {Piranomonte}, {Salvaterra}, {Stratta}, {Tagliaferri}, {Tanaka}, {Tanvir}, \& {Wijers}}]{Bufano12}
{Bufano}, F., {Pian}, E., {Sollerman}, J., {et~al.} 2012, \apj, 753, 67, \dodoi{10.1088/0004-637X/753/1/67}

\bibitem[{{Camilo} {et~al.}(2018){Camilo}, {Scholz}, {Serylak}, {Buchner}, {Merryfield}, {Kaspi}, {Archibald}, {Bailes}, {Jameson}, {van Straten}, {Sarkissian}, {Reynolds}, {Johnston}, {Hobbs}, {Abbott}, {Adam}, {Adams}, {Alberts}, {Andreas}, {Asad}, {Baker}, {Baloyi}, {Bauermeister}, {Baxana}, {Bennett}, {Bernardi}, {Booisen}, {Booth}, {Botha}, {Boyana}, {Brederode}, {Burger}, {Cheetham}, {Conradie}, {Conradie}, {Davidson}, {De Bruin}, {de Swardt}, {de Villiers}, {de Villiers}, {de Villiers}, {de Villiers}, {De Waal}, {Dikgale}, {du Toit}, {du Toit}, {Esterhuyse}, {Fanaroff}, {Fataar}, {Foley}, {Foster}, {Fourie}, {Gamatham}, {Gatsi}, {Geschke}, {Goedhart}, {Grobler}, {Gumede}, {Hlakola}, {Hokwana}, {Hoorn}, {Horn}, {Horrell}, {Hugo}, {Isaacson}, {Jacobs}, {Jansen van Rensburg}, {Jonas}, {Jordaan}, {Joubert}, {Joubert}, {J{\'o}zsa}, {Julie}, {Julius}, {Kapp}, {Karastergiou}, {Karels}, {Kariseb}, {Karuppusamy}, {Kasper}, {Knox-Davies}, {Koch}, {Kotz{\'e}}, {Krebs}, {Kriek}, {Kriel}, {Kusel}, {Lamoor},
  {Lehmensiek}, {Liebenberg}, {Liebenberg}, {Lord}, {Lunsky}, {Mabombo}, {Macdonald}, {Macfarlane}, {Madisa}, {Mafhungo}, {Magnus}, {Magozore}, {Mahgoub}, {Main}, {Makhathini}, {Malan}, {Malgas}, {Manley}, {Manzini}, {Marais}, {Marais}, {Marais}, {Maree}, {Martens}, {Matshawule}, {Matthysen}, {Mauch}, {McNally}, {Merry}, {Millenaar}, {Mjikelo}, {Mkhabela}, {Mnyand u}, {Moeng}, {Mokone}, {Monama}, {Montshiwa}, {Moss}, {Mphego}, {New}, {Ngcebetsha}, {Ngoasheng}, {Niehaus}, {Ntuli}, {Nzama}, {Obies}, {Obrocka}, {Ockards}, {Olyn}, {Oozeer}, {Otto}, {Padayachee}, {Passmoor}, {Patel}, {Paula}, {Peens-Hough}, {Pholoholo}, {Prozesky}, {Rakoma}, {Ramaila}, {Rammala}, {Ramudzuli}, {Rasivhaga}, {Ratcliffe}, {Reader}, {Renil}, {Richter}, {Robyntjies}, {Rosekrans}, {Rust}, {Salie}, {Sambu}, {Schollar}, {Schwardt}, {Seranyane}, {Sethosa}, {Sharpe}, {Siebrits}, {Sirothia}, {Slabber}, {Smirnov}, {Smith}, {Sofeya}, {Songqumase}, {Spann}, {Stappers}, {Steyn}, {Steyn}, {Strong}, {Struthers}, {Stuart}, {Sunnylall}, {Swart},
  {Taljaard}, {Tasse}, {Taylor}, {Theron}, {Thondikulam}, {Thorat}, {Tiplady}, {Toruvanda}, {van Aardt}, {van Balla}, {van den Heever}, {van der Byl}, {van der Merwe}, {van der Merwe}, {van Niekerk}, {van Rooyen}, {van Staden}, {van Tonder}, {van Wyk}, {Wait}, {Walker}, {Wallace}, {Welz}, {Williams}, {Xaia}, {Young}, \& {Zitha}}]{Camilo2018}
{Camilo}, F., {Scholz}, P., {Serylak}, M., {et~al.} 2018, \apj, 856, 180, \dodoi{10.3847/1538-4357/aab35a}

\bibitem[{{Campana} {et~al.}(2006){Campana}, {Mangano}, {Blustin}, {Brown}, {Burrows}, {Chincarini}, {Cummings}, {Cusumano}, {Della Valle}, {Malesani}, {M{\'e}sz{\'a}ros}, {Nousek}, {Page}, {Sakamoto}, {Waxman}, {Zhang}, {Dai}, {Gehrels}, {Immler}, {Marshall}, {Mason}, {Moretti}, {O'Brien}, {Osborne}, {Page}, {Romano}, {Roming}, {Tagliaferri}, {Cominsky}, {Giommi}, {Godet}, {Kennea}, {Krimm}, {Angelini}, {Barthelmy}, {Boyd}, {Palmer}, {Wells}, \& {White}}]{Campana06}
{Campana}, S., {Mangano}, V., {Blustin}, A.~J., {et~al.} 2006, \nat, 442, 1008, \dodoi{10.1038/nature04892}

\bibitem[{{Campana} {et~al.}(2011){Campana}, {Lodato}, {D'Avanzo}, {Panagia}, {Rossi}, {Della Valle}, {Tagliaferri}, {Antonelli}, {Covino}, {Ghirlanda}, {Ghisellini}, {Melandri}, {Pian}, {Salvaterra}, {Cusumano}, {D'Elia}, {Fugazza}, {Palazzi}, {Sbarufatti}, \& {Vergani}}]{Campana11}
{Campana}, S., {Lodato}, G., {D'Avanzo}, P., {et~al.} 2011, \nat, 480, 69, \dodoi{10.1038/nature10592}

\bibitem[{{Cano} {et~al.}(2011){Cano}, {Bersier}, {Guidorzi}, {Kobayashi}, {Levan}, {Tanvir}, {Wiersema}, {D'Avanzo}, {Fruchter}, {Garnavich}, {Gomboc}, {Gorosabel}, {Kasen}, {Kopa{\v{c}}}, {Margutti}, {Mazzali}, {Melandri}, {Mundell}, {Nugent}, {Pian}, {Smith}, {Steele}, {Wijers}, \& {Woosley}}]{Cano11}
{Cano}, Z., {Bersier}, D., {Guidorzi}, C., {et~al.} 2011, \apj, 740, 41, \dodoi{10.1088/0004-637X/740/1/41}

\bibitem[{{CASA Team} {et~al.}(2022){CASA Team}, {Bean}, {Bhatnagar}, {Castro}, {Donovan Meyer}, {Emonts}, {Garcia}, {Garwood}, {Golap}, {Gonzalez Villalba}, {Harris}, {Hayashi}, {Hoskins}, {Hsieh}, {Jagannathan}, {Kawasaki}, {Keimpema}, {Kettenis}, {Lopez}, {Marvil}, {Masters}, {McNichols}, {Mehringer}, {Miel}, {Moellenbrock}, {Montesino}, {Nakazato}, {Ott}, {Petry}, {Pokorny}, {Raba}, {Rau}, {Schiebel}, {Schweighart}, {Sekhar}, {Shimada}, {Small}, {Steeb}, {Sugimoto}, {Suoranta}, {Tsutsumi}, {van Bemmel}, {Verkouter}, {Wells}, {Xiong}, {Szomoru}, {Griffith}, {Glendenning}, \& {Kern}}]{CASA_team_2022}
{CASA Team}, {Bean}, B., {Bhatnagar}, S., {et~al.} 2022, \pasp, 134, 114501, \dodoi{10.1088/1538-3873/ac9642}

\bibitem[{{Chen} {et~al.}(2025){Chen}, {Wang}, {Zhang}, {Dai}, {Liu}, \& {Einstein Probe Team}}]{GCN39580}
{Chen}, W., {Wang}, W.~X., {Zhang}, Y.~J., {et~al.} 2025, GRB Coordinates Network, 39580, 1

\bibitem[{{Cobb} {et~al.}(2006){Cobb}, {Bailyn}, {van Dokkum}, \& {Natarajan}}]{Cobb06}
{Cobb}, B.~E., {Bailyn}, C.~D., {van Dokkum}, P.~G., \& {Natarajan}, P. 2006, \apjl, 645, L113, \dodoi{10.1086/506271}

\bibitem[{{Coppejans} {et~al.}(2020){Coppejans}, {Margutti}, {Terreran}, {Nayana}, {Coughlin}, {Laskar}, {Alexander}, {Bietenholz}, {Caprioli}, {Chandra}, {Drout}, {Frederiks}, {Frohmaier}, {Hurley}, {Kochanek}, {MacLeod}, {Meisner}, {Nugent}, {Ridnaia}, {Sand}, {Svinkin}, {Ward}, {Yang}, {Baldeschi}, {Chilingarian}, {Dong}, {Esquivia}, {Fong}, {Guidorzi}, {Lundqvist}, {Milisavljevic}, {Paterson}, {Reichart}, {Shappee}, {Stroh}, {Valenti}, {Zauderer}, \& {Zhang}}]{Coppejans20}
{Coppejans}, D.~L., {Margutti}, R., {Terreran}, G., {et~al.} 2020, \apjl, 895, L23, \dodoi{10.3847/2041-8213/ab8cc7}

\bibitem[{{De Luca} {et~al.}(2021){De Luca}, {Salvaterra}, {Belfiore}, {Carpano}, {D'Agostino}, {Haberl}, {Israel}, {Law-Green}, {Lisini}, {Marelli}, {Novara}, {Read}, {Rodriguez-Castillo}, {Rosen}, {Salvetti}, {Tiengo}, {Vianello}, {Watson}, {Delvaux}, {Dickens}, {Esposito}, {Greiner}, {H{\"a}mmerle}, {Kreikenbohm}, {Kreykenbohm}, {Oertel}, {Pizzocaro}, {Pye}, {Sandrelli}, {Stelzer}, {Wilms}, \& {Zagaria}}]{DeLuca21}
{De Luca}, A., {Salvaterra}, R., {Belfiore}, A., {et~al.} 2021, \aap, 650, A167, \dodoi{10.1051/0004-6361/202039783}

\bibitem[{{Decleir} {et~al.}(2022){Decleir}, {Gordon}, {Andrews}, {Clayton}, {Cushing}, {Misselt}, {Pendleton}, {Rayner}, {Vacca}, \& {Whittet}}]{Decleir22}
{Decleir}, M., {Gordon}, K.~D., {Andrews}, J.~E., {et~al.} 2022, \apj, 930, 15, \dodoi{10.3847/1538-4357/ac5dbe}

\bibitem[{{D'Elia} {et~al.}(2018){D'Elia}, {Campana}, {D'A{\`\i}}, {De Pasquale}, {Emery}, {Frederiks}, {Lien}, {Melandri}, {Page}, {Starling}, {Burrows}, {Breeveld}, {Oates}, {O'Brien}, {Osborne}, {Siegel}, {Tagliaferri}, {Brown}, {Cenko}, {Svinkin}, {Tohuvavohu}, \& {Tsvetkova}}]{DElia18}
{D'Elia}, V., {Campana}, S., {D'A{\`\i}}, A., {et~al.} 2018, \aap, 619, A66, \dodoi{10.1051/0004-6361/201833847}

\bibitem[{{Dessart} {et~al.}(2020){Dessart}, {Yoon}, {Aguilera-Dena}, \& {Langer}}]{Dessart20}
{Dessart}, L., {Yoon}, S.-C., {Aguilera-Dena}, D.~R., \& {Langer}, N. 2020, \aap, 642, A106, \dodoi{10.1051/0004-6361/202038763}

\bibitem[{{Emery} {et~al.}(2019){Emery}, {Page}, {Breeveld}, {Brown}, {Kuin}, {Oates}, \& {De Pasquale}}]{Emery19}
{Emery}, S.~W.~K., {Page}, M.~J., {Breeveld}, A.~A., {et~al.} 2019, \mnras, 484, 5484, \dodoi{10.1093/mnras/stz373}

\bibitem[{{Evans} {et~al.}(2023){Evans}, {Page}, {Beardmore}, {Eyles-Ferris}, {Osborne}, {Campana}, {Kennea}, \& {Cenko}}]{Evans23}
{Evans}, P.~A., {Page}, K.~L., {Beardmore}, A.~P., {et~al.} 2023, \mnras, 518, 174, \dodoi{10.1093/mnras/stac2937}

\bibitem[{{Eyles-Ferris}(2025)}]{GCN38878}
{Eyles-Ferris}, R.~A.~J. 2025, GRB Coordinates Network, 38878, 1

\bibitem[{{Eyles-Ferris} {et~al.}(2025){Eyles-Ferris}, {Malesani}, {O'Brien}, {Jonker}, {Levan}, {van Dalen}, {Kumar}, {Gompertz}, \& {Tanvir}}]{GCN38983}
{Eyles-Ferris}, R.~A.~J., {Malesani}, D.~B., {O'Brien}, P.~T., {et~al.} 2025, GRB Coordinates Network, 38983, 1

\bibitem[{{Ferrero} {et~al.}(2006){Ferrero}, {Kann}, {Zeh}, {Klose}, {Pian}, {Palazzi}, {Masetti}, {Hartmann}, {Sollerman}, {Deng}, {Filippenko}, {Greiner}, {Hughes}, {Mazzali}, {Li}, {Rol}, {Smith}, \& {Tanvir}}]{Ferrero06}
{Ferrero}, P., {Kann}, D.~A., {Zeh}, A., {et~al.} 2006, \aap, 457, 857, \dodoi{10.1051/0004-6361:20065530}

\bibitem[{{Fitzpatrick} {et~al.}(2019){Fitzpatrick}, {Massa}, {Gordon}, {Bohlin}, \& {Clayton}}]{Fitzpatrick19}
{Fitzpatrick}, E.~L., {Massa}, D., {Gordon}, K.~D., {Bohlin}, R., \& {Clayton}, G.~C. 2019, \apj, 886, 108, \dodoi{10.3847/1538-4357/ab4c3a}

\bibitem[{{Foreman-Mackey} {et~al.}(2013){Foreman-Mackey}, {Hogg}, {Lang}, \& {Goodman}}]{ForemanMackey13}
{Foreman-Mackey}, D., {Hogg}, D.~W., {Lang}, D., \& {Goodman}, J. 2013, \pasp, 125, 306, \dodoi{10.1086/670067}

\bibitem[{{Freudling} {et~al.}(2013){Freudling}, {Romaniello}, {Bramich}, {Ballester}, {Forchi}, {Garc{\'{\i}}a-Dabl{\'o}}, {Moehler}, \& {Neeser}}]{Freudling13}
{Freudling}, W., {Romaniello}, M., {Bramich}, D.~M., {et~al.} 2013, \aap, 559, A96, \dodoi{10.1051/0004-6361/201322494}

\bibitem[{{Fryer} {et~al.}(2018){Fryer}, {Andrews}, {Even}, {Heger}, \& {Safi-Harb}}]{2018ApJ...856...63F}
{Fryer}, C.~L., {Andrews}, S., {Even}, W., {Heger}, A., \& {Safi-Harb}, S. 2018, \apj, 856, 63, \dodoi{10.3847/1538-4357/aaaf6f}

\bibitem[{{Fryer} {et~al.}(2024){Fryer}, {Burns}, {Ho}, {Corsi}, {Lien}, {Perley}, {Vail}, \& {Villar}}]{Fryer24}
{Fryer}, C.~L., {Burns}, E., {Ho}, A. Y.~Q., {et~al.} 2024, arXiv e-prints, arXiv:2410.10378, \dodoi{10.48550/arXiv.2410.10378}

\bibitem[{{Fryer} \& {Kalogera}(2001)}]{2001ApJ...554..548F}
{Fryer}, C.~L., \& {Kalogera}, V. 2001, \apj, 554, 548, \dodoi{10.1086/321359}

\bibitem[{{Georgy} {et~al.}(2009){Georgy}, {Meynet}, {Walder}, {Folini}, \& {Maeder}}]{Georgy09}
{Georgy}, C., {Meynet}, G., {Walder}, R., {Folini}, D., \& {Maeder}, A. 2009, \aap, 502, 611, \dodoi{10.1051/0004-6361/200811339}

\bibitem[{{Glennie} {et~al.}(2015){Glennie}, {Jonker}, {Fender}, {Nagayama}, \& {Pretorius}}]{Glennie2015}
{Glennie}, A., {Jonker}, P.~G., {Fender}, R.~P., {Nagayama}, T., \& {Pretorius}, M.~L. 2015, \mnras, 450, 3765, \dodoi{10.1093/mnras/stv801}

\bibitem[{{Gordon}(2024)}]{Gordon24b}
{Gordon}, K. 2024, The Journal of Open Source Software, 9, 7023, \dodoi{10.21105/joss.07023}

\bibitem[{{Gordon} {et~al.}(2009){Gordon}, {Cartledge}, \& {Clayton}}]{Gordon09}
{Gordon}, K.~D., {Cartledge}, S., \& {Clayton}, G.~C. 2009, \apj, 705, 1320, \dodoi{10.1088/0004-637X/705/2/1320}

\bibitem[{{Gordon} {et~al.}(2023){Gordon}, {Clayton}, {Decleir}, {Fitzpatrick}, {Massa}, {Misselt}, \& {Tollerud}}]{Gordon23}
{Gordon}, K.~D., {Clayton}, G.~C., {Decleir}, M., {et~al.} 2023, \apj, 950, 86, \dodoi{10.3847/1538-4357/accb59}

\bibitem[{{Gordon} {et~al.}(2021){Gordon}, {Misselt}, {Bouwman}, {Clayton}, {Decleir}, {Hines}, {Pendleton}, {Rieke}, {Smith}, \& {Whittet}}]{Gordon21}
{Gordon}, K.~D., {Misselt}, K.~A., {Bouwman}, J., {et~al.} 2021, \apj, 916, 33, \dodoi{10.3847/1538-4357/ac00b7}

\bibitem[{{Gordon} {et~al.}(2024){Gordon}, {Fitzpatrick}, {Massa}, {Bohlin}, {Chastenet}, {Murray}, {Clayton}, {Lennon}, {Misselt}, \& {Sandstrom}}]{Gordon24a}
{Gordon}, K.~D., {Fitzpatrick}, E.~L., {Massa}, D., {et~al.} 2024, \apj, 970, 51, \dodoi{10.3847/1538-4357/ad4be1}

\bibitem[{{Groot} {et~al.}(2024){Groot}, {Bloemen}, {Vreeswijk}, {van Roestel}, {Jonker}, {Nelemans}, {Klein-Wolt}, {Lepoole}, {Pieterse}, {Rodenhuis}, {Boland}, {Haverkorn}, {Aerts}, {Bakker}, {Balster}, {Bekema}, {Dijkstra}, {Dolron}, {Elswijk}, {van Elteren}, {Engels}, {Fokker}, {de Haan}, {Hahn}, {ter Horst}, {Lesman}, {Kragt}, {Morren}, {Nillissen}, {Pessemier}, {Raskin}, {de Rijke}, {Scheers}, {Schuil}, {Timmer}, {Antunes Amaral}, {Arancibia-Rojas}, {Arcavi}, {Blagorodnova}, {Biswas}, {Breton}, {Dawson}, {Dayal}, {De Wet}, {Duffy}, {Faris}, {Fausnaugh}, {Gal-Yam}, {Geier}, {Horesh}, {Johnston}, {Katusiime}, {Kelley}, {Kosakowski}, {Kupfer}, {Leloudas}, {Levan}, {Modiano}, {Mogawana}, {Munday}, {Paice}, {Patat}, {Pelisoli}, {Ramsay}, {Ranaivomanana}, {Ruiz-Carmona}, {Schaffenroth}, {Scaringi}, {Stoppa}, {Street}, {Tranin}, {Uzundag}, {Valenti}, {Veresvarska}, {Vuc̆kovi{\'c}}, {Wichern}, {Wijers}, {Wijnands}, \& {Zimmerman}}]{Groot24}
{Groot}, P.~J., {Bloemen}, S., {Vreeswijk}, P.~M., {et~al.} 2024, \pasp, 136, 115003, \dodoi{10.1088/1538-3873/ad8b6a}

\bibitem[{Guevel {et~al.}(2021)Guevel, Hosseinzadeh, Bostroem, \& Burke}]{Guevel21}
Guevel, D., Hosseinzadeh, G., Bostroem, A., \& Burke, C.~J. 2021, dguevel/PyZOGY: v0.0.2, v0.0.2,  Zenodo, \dodoi{10.5281/zenodo.4570234}

\bibitem[{{Guti{\'e}rrez} {et~al.}(2024{\natexlab{a}}){Guti{\'e}rrez}, {Mattila}, {Lundqvist}, {Dessart}, {Gonz{\'a}lez-Gait{\'a}n}, {Jonker}, {Dong}, {Coppejans}, {Chen}, {Charalampopoulos}, {Elias-Rosa}, {Reynolds}, {Kochanek}, {Fraser}, {Pastorello}, {Gromadzki}, {Neustadt}, {Benetti}, {Kankare}, {Kangas}, {Kotak}, {Stritzinger}, {Wevers}, {Zhang}, {Bersier}, {Bose}, {Buckley}, {Dastidar}, {Gangopadhyay}, {Hamanowicz}, {Kollmeier}, {Mao}, {Misra}, {Potter}, {Prieto}, {Romero-Colmenero}, {Singh}, {Somero}, {Terreran}, {Vaisanen}, \& {Wyrzykowski}}]{Gutierrez24}
{Guti{\'e}rrez}, C.~P., {Mattila}, S., {Lundqvist}, P., {et~al.} 2024{\natexlab{a}}, \apj, 977, 162, \dodoi{10.3847/1538-4357/ad89a5}

\bibitem[{{Guti{\'e}rrez} {et~al.}(2024{\natexlab{b}}){Guti{\'e}rrez}, {Bhattacharya}, {Radice}, {Murase}, \& {Bernuzzi}}]{2024arXiv240815973G}
{Guti{\'e}rrez}, E.~M., {Bhattacharya}, M., {Radice}, D., {Murase}, K., \& {Bernuzzi}, S. 2024{\natexlab{b}}, arXiv e-prints, arXiv:2408.15973, \dodoi{10.48550/arXiv.2408.15973}

\bibitem[{{Hamidani} {et~al.}(2025{\natexlab{a}}){Hamidani}, {Ioka}, {Kashiyama}, \& {Tanaka}}]{Hamidani25b}
{Hamidani}, H., {Ioka}, K., {Kashiyama}, K., \& {Tanaka}, M. 2025{\natexlab{a}}, arXiv e-prints, arXiv:2503.16242, \dodoi{10.48550/arXiv.2503.16242}

\bibitem[{{Hamidani} {et~al.}(2025{\natexlab{b}}){Hamidani}, {Sato}, {Kashiyama}, {Tanaka}, {Ioka}, \& {Kimura}}]{Hamidani25}
{Hamidani}, H., {Sato}, Y., {Kashiyama}, K., {et~al.} 2025{\natexlab{b}}, arXiv e-prints, arXiv:2503.16243, \dodoi{10.48550/arXiv.2503.16243}

\bibitem[{{Herant} {et~al.}(1994){Herant}, {Benz}, {Hix}, {Fryer}, \& {Colgate}}]{Herant94}
{Herant}, M., {Benz}, W., {Hix}, W.~R., {Fryer}, C.~L., \& {Colgate}, S.~A. 1994, \apj, 435, 339, \dodoi{10.1086/174817}

\bibitem[{{Heywood}(2020)}]{oxkat}
{Heywood}, I. 2020, {oxkat: Semi-automated imaging of MeerKAT observations}, Astrophysics Source Code Library, record ascl:2009.003

\bibitem[{{Ho} {et~al.}(2022{\natexlab{a}}){Ho}, {Perley}, {Chen}, {Schulze}, {Sollerman}, \& {Gal-Yam}}]{Ho22c}
{Ho}, A.~Y.~Q., {Perley}, D.~A., {Chen}, P., {et~al.} 2022{\natexlab{a}}, Transient Name Server AstroNote, 267, 1

\bibitem[{{Ho} {et~al.}(2022{\natexlab{b}}){Ho}, {Perley}, {Filippenko}, {Zheng}, {Brink}, {Li}, \& {Wang}}]{Ho22b}
{Ho}, A.~Y.~Q., {Perley}, D.~A., {Filippenko}, A.~V., {et~al.} 2022{\natexlab{b}}, Transient Name Server AstroNote, 199, 1

\bibitem[{{Ho} {et~al.}(2019){Ho}, {Goldstein}, {Schulze}, {Khatami}, {Perley}, {Ergon}, {Gal-Yam}, {Corsi}, {Andreoni}, {Barbarino}, {Bellm}, {Blagorodnova}, {Bright}, {Burns}, {Cenko}, {Cunningham}, {De}, {Dekany}, {Dugas}, {Fender}, {Fransson}, {Fremling}, {Goldstein}, {Graham}, {Hale}, {Horesh}, {Hung}, {Kasliwal}, {Kuin}, {Kulkarni}, {Kupfer}, {Lunnan}, {Masci}, {Ngeow}, {Nugent}, {Ofek}, {Patterson}, {Petitpas}, {Rusholme}, {Sai}, {Sfaradi}, {Shupe}, {Sollerman}, {Soumagnac}, {Tachibana}, {Taddia}, {Walters}, {Wang}, {Yao}, \& {Zhang}}]{Ho19}
{Ho}, A. Y.~Q., {Goldstein}, D.~A., {Schulze}, S., {et~al.} 2019, \apj, 887, 169, \dodoi{10.3847/1538-4357/ab55ec}

\bibitem[{{Ho} {et~al.}(2020{\natexlab{a}}){Ho}, {Kulkarni}, {Perley}, {Cenko}, {Corsi}, {Schulze}, {Lunnan}, {Sollerman}, {Gal-Yam}, {Anand}, {Barbarino}, {Bellm}, {Bruch}, {Burns}, {De}, {Dekany}, {Delacroix}, {Duev}, {Frederiks}, {Fremling}, {Goldstein}, {Golkhou}, {Graham}, {Hale}, {Kasliwal}, {Kupfer}, {Laher}, {Martikainen}, {Masci}, {Neill}, {Ridnaia}, {Rusholme}, {Savchenko}, {Shupe}, {Soumagnac}, {Strotjohann}, {Svinkin}, {Taggart}, {Tartaglia}, {Yan}, \& {Zolkower}}]{Ho20}
{Ho}, A. Y.~Q., {Kulkarni}, S.~R., {Perley}, D.~A., {et~al.} 2020{\natexlab{a}}, \apj, 902, 86, \dodoi{10.3847/1538-4357/aba630}

\bibitem[{{Ho} {et~al.}(2020{\natexlab{b}}){Ho}, {Perley}, {Kulkarni}, {Dong}, {De}, {Chandra}, {Andreoni}, {Bellm}, {Burdge}, {Coughlin}, {Dekany}, {Feeney}, {Frederiks}, {Fremling}, {Golkhou}, {Graham}, {Hale}, {Helou}, {Horesh}, {Kasliwal}, {Laher}, {Masci}, {Miller}, {Porter}, {Ridnaia}, {Rusholme}, {Shupe}, {Soumagnac}, \& {Svinkin}}]{Ho20b}
{Ho}, A. Y.~Q., {Perley}, D.~A., {Kulkarni}, S.~R., {et~al.} 2020{\natexlab{b}}, \apj, 895, 49, \dodoi{10.3847/1538-4357/ab8bcf}

\bibitem[{{Ho} {et~al.}(2022{\natexlab{c}}){Ho}, {Margalit}, {Bremer}, {Perley}, {Yao}, {Dobie}, {Kaplan}, {O'Brien}, {Petitpas}, \& {Zic}}]{Ho22}
{Ho}, A. Y.~Q., {Margalit}, B., {Bremer}, M., {et~al.} 2022{\natexlab{c}}, \apj, 932, 116, \dodoi{10.3847/1538-4357/ac4e97}

\bibitem[{{Hugo} {et~al.}(2022){Hugo}, {Perkins}, {Merry}, {Mauch}, \& {Smirnov}}]{Hugo_2022}
{Hugo}, B.~V., {Perkins}, S., {Merry}, B., {Mauch}, T., \& {Smirnov}, O.~M. 2022, in Astronomical Society of the Pacific Conference Series, Vol. 532, Astronomical Society of the Pacific Conference Series, ed. J.~E. {Ruiz}, F.~{Pierfedereci}, \& P.~{Teuben}, 541, \dodoi{10.48550/arXiv.2206.09179}

\bibitem[{{Izzo}(2025)}]{GCN38912}
{Izzo}, L. 2025, GRB Coordinates Network, 38912, 1

\bibitem[{{Izzo} {et~al.}(2020){Izzo}, {Auchettl}, {Hjorth}, {De Colle}, {Gall}, {Angus}, {Raimundo}, \& {Ramirez-Ruiz}}]{Izzo20}
{Izzo}, L., {Auchettl}, K., {Hjorth}, J., {et~al.} 2020, \aap, 639, L11, \dodoi{10.1051/0004-6361/202038152}

\bibitem[{{Izzo} {et~al.}(2019){Izzo}, {de Ugarte Postigo}, {Maeda}, {Th{\"o}ne}, {Kann}, {Della Valle}, {Sagues Carracedo}, {Micha{\l}owski}, {Schady}, {Schmidl}, {Selsing}, {Starling}, {Suzuki}, {Bensch}, {Bolmer}, {Campana}, {Cano}, {Covino}, {Fynbo}, {Hartmann}, {Heintz}, {Hjorth}, {Japelj}, {Kami{\'n}ski}, {Kaper}, {Kouveliotou}, {Kru{\.Z}y{\'n}ski}, {Kwiatkowski}, {Leloudas}, {Levan}, {Malesani}, {Micha{\l}owski}, {Piranomonte}, {Pugliese}, {Rossi}, {S{\'a}nchez-Ram{\'\i}rez}, {Schulze}, {Steeghs}, {Tanvir}, {Ulaczyk}, {Vergani}, \& {Wiersema}}]{Izzo19}
{Izzo}, L., {de Ugarte Postigo}, A., {Maeda}, K., {et~al.} 2019, \nat, 565, 324, \dodoi{10.1038/s41586-018-0826-3}

\bibitem[{Jonas(2018)}]{Jonas2018}
Jonas, J. 2018, in Proceedings of MeerKAT Science: On the Pathway to the SKA {\textemdash} PoS(MeerKAT2016), Vol. 277, 001, \dodoi{10.22323/1.277.0001}

\bibitem[{{Jonker} {et~al.}(2013){Jonker}, {Glennie}, {Heida}, {Maccarone}, {Hodgkin}, {Nelemans}, {Miller-Jones}, {Torres}, \& {Fender}}]{Jonker2013}
{Jonker}, P.~G., {Glennie}, A., {Heida}, M., {et~al.} 2013, \apj, 779, 14, \dodoi{10.1088/0004-637X/779/1/14}

\bibitem[{{Kraft} {et~al.}(1991){Kraft}, {Burrows}, \& {Nousek}}]{Kraft91}
{Kraft}, R.~P., {Burrows}, D.~N., \& {Nousek}, J.~A. 1991, \apj, 374, 344, \dodoi{10.1086/170124}

\bibitem[{{Kumar} {et~al.}(2025){Kumar}, {Maund}, {Sun}, {Li}, {Wang}, \& {Wiersema}}]{GCN38907}
{Kumar}, A., {Maund}, J.~R., {Sun}, N.~C., {et~al.} 2025, GRB Coordinates Network, 38907, 1

\bibitem[{{Levan} {et~al.}(2025{\natexlab{a}}){Levan}, {Cotter}, {Malesani}, {Martin-Carrillo}, \& {Jonker}}]{GCN38909}
{Levan}, A.~J., {Cotter}, L., {Malesani}, D.~B., {Martin-Carrillo}, A., \& {Jonker}, P.~G. 2025{\natexlab{a}}, GRB Coordinates Network, 38909, 1

\bibitem[{{Levan} {et~al.}(2025{\natexlab{b}}){Levan}, {Rastinejad}, {Malesani}, {Fong}, {Tanvir}, {Jonker}, \& {Eyles-Ferris}}]{GCN38987}
{Levan}, A.~J., {Rastinejad}, J.~C., {Malesani}, D.~B., {et~al.} 2025{\natexlab{b}}, GRB Coordinates Network, 38987, 1

\bibitem[{{Levan} {et~al.}(2014){Levan}, {Tanvir}, {Starling}, {Wiersema}, {Page}, {Perley}, {Schulze}, {Wynn}, {Chornock}, {Hjorth}, {Cenko}, {Fruchter}, {O'Brien}, {Brown}, {Tunnicliffe}, {Malesani}, {Jakobsson}, {Watson}, {Berger}, {Bersier}, {Cobb}, {Covino}, {Cucchiara}, {de Ugarte Postigo}, {Fox}, {Gal-Yam}, {Goldoni}, {Gorosabel}, {Kaper}, {Kr{\"u}hler}, {Karjalainen}, {Osborne}, {Pian}, {S{\'a}nchez-Ram{\'\i}rez}, {Schmidt}, {Skillen}, {Tagliaferri}, {Th{\"o}ne}, {Vaduvescu}, {Wijers}, \& {Zauderer}}]{Levan14}
{Levan}, A.~J., {Tanvir}, N.~R., {Starling}, R.~L.~C., {et~al.} 2014, \apj, 781, 13, \dodoi{10.1088/0004-637X/781/1/13}

\bibitem[{{Levan} {et~al.}(2024){Levan}, {Jonker}, {Saccardi}, {Bj{\o}rn Malesani}, {Tanvir}, {Izzo}, {Heintz}, {Mata S{\'a}nchez}, {Quirola-V{\'a}squez}, {Torres}, {Vergani}, {Schulze}, {Rossi}, {D'Avanzo}, {Gompertz}, {Martin-Carrillo}, {de Ugarte Postigo}, {Schneider}, {Yuan}, {Ling}, {Zhang}, {Mao}, {Liu}, {Sun}, {Xu}, {Zhu}, {Ag{\"u}{\'\i} Fern{\'a}ndez}, {Amati}, {Bauer}, {Campana}, {Carotenuto}, {Chrimes}, {van Dalen}, {D'Elia}, {Della Valle}, {De Pasquale}, {Dhillon}, {Galbany}, {Gaspari}, {Gianfagna}, {Gomboc}, {Habeeb}, {van Hoof}, {Hu}, {Jakobsson}, {Julakanti}, {Korth}, {Kouveliotou}, {Laskar}, {Littlefair}, {Maiorano}, {Mao}, {Melandri}, {Miller}, {Mukherjee}, {Oates}, {O'Brien}, {Palmerio}, {Parviainen}, {Pieterse}, {Piranomonte}, {Piro}, {Pugliese}, {Ravasio}, {Rayson}, {Salvaterra}, {S{\'a}nchez-Ram{\'\i}rez}, {Sarin}, {Shilling}, {Starling}, {Tagliaferri}, {Linesh Thakur}, {Th{\"o}ne}, {Wiersema}, {Worssam}, \& {Zafar}}]{Levan24}
{Levan}, A.~J., {Jonker}, P.~G., {Saccardi}, A., {et~al.} 2024, arXiv e-prints, arXiv:2404.16350, \dodoi{10.48550/arXiv.2404.16350}

\bibitem[{{Li} {et~al.}(2025{\natexlab{a}}){Li}, {Chen}, {Chatterjee}, {Hua}, {Liu}, \& {Einstein Probe Team}}]{GCN38861}
{Li}, R.~Z., {Chen}, X.~L., {Chatterjee}, K., {et~al.} 2025{\natexlab{a}}, GRB Coordinates Network, 38861, 1

\bibitem[{{Li} {et~al.}(2025{\natexlab{b}}){Li}, {Chen}, {Chatterjee}, {Hua}, {Liu}, \& {Einstein Probe Team}}]{GCN38888}
---. 2025{\natexlab{b}}, GRB Coordinates Network, 38888, 1

\bibitem[{{Li} {et~al.}(2025{\natexlab{c}}){Li}, {Mao}, {Sun}, {Liu}, {Zhang}, \& {Einstein Probe Team}}]{GCN39037}
{Li}, R.~Z., {Mao}, J., {Sun}, H., {et~al.} 2025{\natexlab{c}}, GRB Coordinates Network, 39037, 1

\bibitem[{{Li} {et~al.}(2025{\natexlab{d}}){Li}, {Zhu}, {Zou}, {Geng}, {Liu}, {Wang}, {Li}, {Xu}, {Sun}, {Wang}, {Yu}, {Zhang}, {Wu}, {Yang}, {Filippenko}, {Liu}, {Yuan}, {Aguado}, {An}, {An}, {Buckley}, {Castro-Tirado}, {Fu}, {Fynbo}, {Howell}, {Hu}, {Jiang}, {Kumar}, {Mao}, {Maund}, {Liu}, {Mockler}, {Moskvitin}, {Andrews}, {Bom}, {Brink}, {Chatterjee}, {Chen}, {Cheng}, {Cooke}, {Dai}, {Du}, {Erasmus}, {Fang}, {Farah}, {Goranskij}, {Gritsevich}, {Gu}, {Guo}, {Hsiao}, {Hu}, {Hua}, {Jacobson-Gal{\'a}n}, {Jia}, {Jin}, {Kasliwal}, {Kilpatrick}, {Kumar}, {Lei}, {Li}, {Li}, {Li}, {Ling}, {Liu}, {Liu}, {Liu}, {L{\'o}pez-Oramas}, {Maslennikova}, {McCully}, {Monageng}, {Newsone}, {Padilla Gonzalez}, {Pan}, {Peng}, {Pignata}, {Poidevin}, {Potter}, {P{\'e}rez-Fournon}, {Santana-Silva}, {Santos}, {Song}, {Song}, {Spiridonova}, {Sun}, {Sun}, {Terreran}, {Wang}, {Wang}, {Wang}, {Wang}, {Wu}, {Xiang}, {Xiao}, {Xu}, {Xue}, {Yan}, {Yang}, {Yu}, {Zhang}, {Zhang}, {Zhang}, {Zhang}, {Zhang}, {Zheng}, \& {Zou}}]{Li25}
{Li}, W.~X., {Zhu}, Z.~P., {Zou}, X.~Z., {et~al.} 2025{\natexlab{d}}, arXiv e-prints, arXiv:2504.17034, \dodoi{10.48550/arXiv.2504.17034}

\bibitem[{{Liang} {et~al.}(2007){Liang}, {Zhang}, {Virgili}, \& {Dai}}]{Liang07}
{Liang}, E., {Zhang}, B., {Virgili}, F., \& {Dai}, Z.~G. 2007, \apj, 662, 1111, \dodoi{10.1086/517959}

\bibitem[{{Liu} {et~al.}(2025{\natexlab{a}}){Liu}, {An}, {Sun}, {Tinyanont}, {Anutarawiramkul}, {Butpan}, {Jiang}, {Fu}, {Zhu}, {Xu}, {Fan}, {Li}, \& {Wang}}]{GCN39583}
{Liu}, X., {An}, J., {Sun}, N.~C., {et~al.} 2025{\natexlab{a}}, GRB Coordinates Network, 39583, 1

\bibitem[{{Liu} {et~al.}(2025{\natexlab{b}}){Liu}, {Sun}, {Xu}, {Svinkin}, {Delaunay}, {Tanvir}, {Gao}, {Zhang}, {Chen}, {Wu}, {Zhang}, {Yuan}, {An}, {Bruni}, {Frederiks}, {Ghirlanda}, {Hu}, {Li}, {Li}, {Li}, {Malesani}, {Piro}, {Raman}, {Ricci}, {Troja}, {Vergani}, {Wu}, {Yang}, {Zhang}, {Zhu}, {de Ugarte Postigo}, {Demin}, {Dobie}, {Fan}, {Fu}, {Fynbo}, {Geng}, {Gianfagna}, {Hu}, {Huang}, {Jiang}, {Jonker}, {Julakanti}, {Kennea}, {Kokomov}, {Kuulkers}, {Lei}, {Leung}, {Levan}, {Li}, {Li}, {Littlefair}, {Liu}, {Lysenko}, {Ma}, {Martin-Carrillo}, {O'Brien}, {Parsotan}, {Quirola-V{\'a}squez}, {Ridnaia}, {Ronchini}, {Rossi}, {Mata-S{\'a}nchez}, {Schneider}, {Shen}, {Thakur}, {Tohuvavohu}, {Torres}, {Tsvetkova}, {Ulanov}, {Wei}, {Xiao}, {Yin}, {Bai}, {Burwitz}, {Cai}, {Chen}, {Chen}, {Chen}, {Chen}, {Chen}, {Chen}, {Cheng}, {Cordier}, {Cui}, {Cui}, {Dai}, {Dai}, {Eder}, {Eyles-Ferris}, {Fan}, {Feldman}, {Feng}, {Feng}, {Friedrich}, {Gao}, {Gonzalez}, {Guan}, {Han}, {Han}, {Hou}, {Hu}, {Hu}, {Huang}, {Huo},
  {Hutchinson}, {Ji}, {Jia}, {Jia}, {Jiang}, {Jin}, {Jin}, {Jin}, {Keereman}, {Lerman}, {Li}, {Li}, {Li}, {Li}, {Li}, {Lian}, {Liang}, {Ling}, {Liu}, {Liu}, {Liu}, {Liu}, {Liu}, {Lu}, {L{\"u}}, {Luo}, {Ma}, {Ma}, {Mao}, {Mao}, {McHugh}, {Meidinger}, {Nandra}, {Osborne}, {Pan}, {Pan}, {Ravasio}, {Rau}, {Rea}, {Rehman}, {Sanders}, {Santovincenzo}, {Song}, {Su}, {Sun}, {Sun}, {Sun}, {Tan}, {Tang}, {Tao}, {Tong}, {Wang}, {Wang}, {Wang}, {Wang}, {Wang}, {Wang}, {Wang}, {Wang}, {Wang}, {Wei}, {Willingale}, {Xiong}, {Xu}, {Xu}, {Xu}, {Xu}, {Xu}, {Xue}, {Xue}, {Yan}, {Yang}, {Yang}, {Yang}, {Yang}, {Yu}, {Zhang}, {Zhang}, {Zhang}, {Zhang}, {Zhang}, {Zhang}, {Zhang}, {Zhang}, {Zhang}, {Zhao}, {Zhao}, {Zhao}, {Zhao}, {Zhou}, {Zhou}, {Zhu}, {Zhu}, \& {Zuo}}]{Liu25}
{Liu}, Y., {Sun}, H., {Xu}, D., {et~al.} 2025{\natexlab{b}}, Nature Astronomy, \dodoi{10.1038/s41550-024-02449-8}

\bibitem[{{MacFadyen} \& {Woosley}(1999)}]{MacFadyen99}
{MacFadyen}, A.~I., \& {Woosley}, S.~E. 1999, \apj, 524, 262, \dodoi{10.1086/307790}

\bibitem[{{Malesani} {et~al.}(2009){Malesani}, {Fynbo}, {Hjorth}, {Leloudas}, {Sollerman}, {Stritzinger}, {Vreeswijk}, {Watson}, {Gorosabel}, {Micha{\l}owski}, {Th{\"o}ne}, {Augusteijn}, {Bersier}, {Jakobsson}, {Jaunsen}, {Ledoux}, {Levan}, {Milvang-Jensen}, {Rol}, {Tanvir}, {Wiersema}, {Xu}, {Albert}, {Bayliss}, {Gall}, {Grove}, {Koester}, {Leitet}, {Pursimo}, \& {Skillen}}]{Malesani09}
{Malesani}, D., {Fynbo}, J.~P.~U., {Hjorth}, J., {et~al.} 2009, \apjl, 692, L84, \dodoi{10.1088/0004-637X/692/2/L84}

\bibitem[{{Malesani} {et~al.}(2025){Malesani}, {Levan}, {van Hoof}, {Jonker}, \& {Xu}}]{GCN38902}
{Malesani}, D.~B., {Levan}, A.~J., {van Hoof}, A., {Jonker}, P.~G., \& {Xu}, D. 2025, GRB Coordinates Network, 38902, 1

\bibitem[{{Margalit}(2022)}]{margalit22}
{Margalit}, B. 2022, \apj, 933, 238, \dodoi{10.3847/1538-4357/ac771a}

\bibitem[{{Margutti} {et~al.}(2019){Margutti}, {Metzger}, {Chornock}, {Vurm}, {Roth}, {Grefenstette}, {Savchenko}, {Cartier}, {Steiner}, {Terreran}, {Margalit}, {Migliori}, {Milisavljevic}, {Alexander}, {Bietenholz}, {Blanchard}, {Bozzo}, {Brethauer}, {Chilingarian}, {Coppejans}, {Ducci}, {Ferrigno}, {Fong}, {G{\"o}tz}, {Guidorzi}, {Hajela}, {Hurley}, {Kuulkers}, {Laurent}, {Mereghetti}, {Nicholl}, {Patnaude}, {Ubertini}, {Banovetz}, {Bartel}, {Berger}, {Coughlin}, {Eftekhari}, {Frederiks}, {Kozlova}, {Laskar}, {Svinkin}, {Drout}, {MacFadyen}, \& {Paterson}}]{Margutti19}
{Margutti}, R., {Metzger}, B.~D., {Chornock}, R., {et~al.} 2019, \apj, 872, 18, \dodoi{10.3847/1538-4357/aafa01}

\bibitem[{{Masci} {et~al.}(2019){Masci}, {Laher}, {Rusholme}, {Shupe}, {Groom}, {Surace}, {Jackson}, {Monkewitz}, {Beck}, {Flynn}, {Terek}, {Landry}, {Hacopians}, {Desai}, {Howell}, {Brooke}, {Imel}, {Wachter}, {Ye}, {Lin}, {Cenko}, {Cunningham}, {Rebbapragada}, {Bue}, {Miller}, {Mahabal}, {Bellm}, {Patterson}, {Juri{\'c}}, {Golkhou}, {Ofek}, {Walters}, {Graham}, {Kasliwal}, {Dekany}, {Kupfer}, {Burdge}, {Cannella}, {Barlow}, {Van Sistine}, {Giomi}, {Fremling}, {Blagorodnova}, {Levitan}, {Riddle}, {Smith}, {Helou}, {Prince}, \& {Kulkarni}}]{Masci19}
{Masci}, F.~J., {Laher}, R.~R., {Rusholme}, B., {et~al.} 2019, \pasp, 131, 018003, \dodoi{10.1088/1538-3873/aae8ac}

\bibitem[{{Masci} {et~al.}(2023){Masci}, {Laher}, {Rusholme}, {Shupe}, {Paladini}, {Groom}, {Wold}, {Miller}, \& {Drake}}]{Masci23}
---. 2023, arXiv e-prints, arXiv:2305.16279, \dodoi{10.48550/arXiv.2305.16279}

\bibitem[{{Matthews} {et~al.}(2023){Matthews}, {Margutti}, {Metzger}, {Milisavljevic}, {Migliori}, {Laskar}, {Brethauer}, {Berger}, {Chornock}, {Drout}, \& {Ramirez-Ruiz}}]{Matthews23}
{Matthews}, D., {Margutti}, R., {Metzger}, B.~D., {et~al.} 2023, Research Notes of the American Astronomical Society, 7, 126, \dodoi{10.3847/2515-5172/acdde1}

\bibitem[{{Matzner} \& {McKee}(1999)}]{matznermckee99}
{Matzner}, C.~D., \& {McKee}, C.~F. 1999, \apjl, 526, L109, \dodoi{10.1086/312376}

\bibitem[{{Mazzali} {et~al.}(2008){Mazzali}, {Valenti}, {Della Valle}, {Chincarini}, {Sauer}, {Benetti}, {Pian}, {Piran}, {D'Elia}, {Elias-Rosa}, {Margutti}, {Pasotti}, {Antonelli}, {Bufano}, {Campana}, {Cappellaro}, {Covino}, {D'Avanzo}, {Fiore}, {Fugazza}, {Gilmozzi}, {Hunter}, {Maguire}, {Maiorano}, {Marziani}, {Masetti}, {Mirabel}, {Navasardyan}, {Nomoto}, {Palazzi}, {Pastorello}, {Panagia}, {Pellizza}, {Sari}, {Smartt}, {Tagliaferri}, {Tanaka}, {Taubenberger}, {Tominaga}, {Trundle}, \& {Turatto}}]{Mazzali08}
{Mazzali}, P.~A., {Valenti}, S., {Della Valle}, M., {et~al.} 2008, Science, 321, 1185, \dodoi{10.1126/science.1158088}

\bibitem[{{Mirabal} {et~al.}(2006){Mirabal}, {Halpern}, {An}, {Thorstensen}, \& {Terndrup}}]{Mirabal06}
{Mirabal}, N., {Halpern}, J.~P., {An}, D., {Thorstensen}, J.~R., \& {Terndrup}, D.~M. 2006, \apjl, 643, L99, \dodoi{10.1086/505177}

\bibitem[{{Modjaz} {et~al.}(2016){Modjaz}, {Liu}, {Bianco}, \& {Graur}}]{Modjaz16}
{Modjaz}, M., {Liu}, Y.~Q., {Bianco}, F.~B., \& {Graur}, O. 2016, \apj, 832, 108, \dodoi{10.3847/0004-637X/832/2/108}

\bibitem[{{Modjaz} {et~al.}(2009){Modjaz}, {Li}, {Butler}, {Chornock}, {Perley}, {Blondin}, {Bloom}, {Filippenko}, {Kirshner}, {Kocevski}, {Poznanski}, {Hicken}, {Foley}, {Stringfellow}, {Berlind}, {Barrado y Navascues}, {Blake}, {Bouy}, {Brown}, {Challis}, {Chen}, {de Vries}, {Dufour}, {Falco}, {Friedman}, {Ganeshalingam}, {Garnavich}, {Holden}, {Illingworth}, {Lee}, {Liebert}, {Marion}, {Olivier}, {Prochaska}, {Silverman}, {Smith}, {Starr}, {Steele}, {Stockton}, {Williams}, \& {Wood-Vasey}}]{Modjaz09}
{Modjaz}, M., {Li}, W., {Butler}, N., {et~al.} 2009, \apj, 702, 226, \dodoi{10.1088/0004-637X/702/1/226}

\bibitem[{{Moskvitin} {et~al.}(2025){Moskvitin}, {Spiridonova}, \& {GRB follow-up Team}}]{GCN38925}
{Moskvitin}, A.~S., {Spiridonova}, O.~I., \& {GRB follow-up Team}. 2025, GRB Coordinates Network, 38925, 1

\bibitem[{{Nakar}(2015)}]{Nakar15}
{Nakar}, E. 2015, \apj, 807, 172, \dodoi{10.1088/0004-637X/807/2/172}

\bibitem[{{Nakar} \& {Piran}(2017)}]{Nakar2017}
{Nakar}, E., \& {Piran}, T. 2017, \apj, 834, 28, \dodoi{10.3847/1538-4357/834/1/28}

\bibitem[{{Nasa High Energy Astrophysics Science Archive Research Center (Heasarc)}(2014)}]{heasoft}
{Nasa High Energy Astrophysics Science Archive Research Center (Heasarc)}. 2014, {HEAsoft: Unified Release of FTOOLS and XANADU}, Astrophysics Source Code Library, record ascl:1408.004

\bibitem[{{Nava} {et~al.}(2012){Nava}, {Salvaterra}, {Ghirlanda}, {Ghisellini}, {Campana}, {Covino}, {Cusumano}, {D'Avanzo}, {D'Elia}, {Fugazza}, {Melandri}, {Sbarufatti}, {Vergani}, \& {Tagliaferri}}]{Nava12}
{Nava}, L., {Salvaterra}, R., {Ghirlanda}, G., {et~al.} 2012, \mnras, 421, 1256, \dodoi{10.1111/j.1365-2966.2011.20394.x}

\bibitem[{{Nicholl} {et~al.}(2023){Nicholl}, {Srivastav}, {Fulton}, {Gomez}, {Huber}, {Oates}, {Ramsden}, {Rhodes}, {Smartt}, {Smith}, {Aamer}, {Anderson}, {Bauer}, {Berger}, {de Boer}, {Chambers}, {Charalampopoulos}, {Chen}, {Fender}, {Fraser}, {Gao}, {Green}, {Galbany}, {Gompertz}, {Gromadzki}, {Guti{\'e}rrez}, {Howell}, {Inserra}, {Jonker}, {Kopsacheili}, {Lowe}, {Magnier}, {McCully}, {McGee}, {Moore}, {M{\"u}ller-Bravo}, {Newsome}, {Gonzalez}, {Pellegrino}, {Pessi}, {Pursiainen}, {Rest}, {Ridley}, {Shappee}, {Sheng}, {Smith}, {Terreran}, {Tucker}, {Vink{\'o}}, {Wainscoat}, {Wiseman}, \& {Young}}]{Nicholl23}
{Nicholl}, M., {Srivastav}, S., {Fulton}, M.~D., {et~al.} 2023, \apjl, 954, L28, \dodoi{10.3847/2041-8213/acf0ba}

\bibitem[{{O'Connor} {et~al.}(2025){O'Connor}, {Pasham}, {Andreoni}, {Hare}, {Beniamini}, {Troja}, {Ricci}, {Dobie}, {Chakraborty}, {Ng}, {Klingler}, {Karambelkar}, {Rose}, {Schulze}, {Ryan}, {Dichiara}, {Monageng}, {Buckley}, {Hu}, {Srinivasaragavan}, {Bruni}, {Cabrera}, {Cenko}, {van Eerten}, {Freeburn}, {Hammerstein}, {Kasliwal}, {Kouveliotou}, {Kunnumkai}, {Leung}, {Lien}, {Palmese}, \& {Sakamoto}}]{OConnor25}
{O'Connor}, B., {Pasham}, D., {Andreoni}, I., {et~al.} 2025, \apjl, 979, L30, \dodoi{10.3847/2041-8213/ada7f5}

\bibitem[{{Offringa} {et~al.}(2014){Offringa}, {McKinley}, {Hurley-Walker}, {Briggs}, {Wayth}, {Kaplan}, {Bell}, {Feng}, {Neben}, {Hughes}, {Rhee}, {Murphy}, {Bhat}, {Bernardi}, {Bowman}, {Cappallo}, {Corey}, {Deshpande}, {Emrich}, {Ewall-Wice}, {Gaensler}, {Goeke}, {Greenhill}, {Hazelton}, {Hindson}, {Johnston-Hollitt}, {Jacobs}, {Kasper}, {Kratzenberg}, {Lenc}, {Lonsdale}, {Lynch}, {McWhirter}, {Mitchell}, {Morales}, {Morgan}, {Kudryavtseva}, {Oberoi}, {Ord}, {Pindor}, {Procopio}, {Prabu}, {Riding}, {Roshi}, {Shankar}, {Srivani}, {Subrahmanyan}, {Tingay}, {Waterson}, {Webster}, {Whitney}, {Williams}, \& {Williams}}]{Offringa_wsclean}
{Offringa}, A.~R., {McKinley}, B., {Hurley-Walker}, N., {et~al.} 2014, \mnras, 444, 606, \dodoi{10.1093/mnras/stu1368}

\bibitem[{{Olivares E.} {et~al.}(2012){Olivares E.}, {Greiner}, {Schady}, {Rau}, {Klose}, {Kr{\"u}hler}, {Afonso}, {Updike}, {Nardini}, {Filgas}, {Nicuesa Guelbenzu}, {Clemens}, {Elliott}, {Kann}, {Rossi}, \& {Sudilovsky}}]{Oliveras12}
{Olivares E.}, F., {Greiner}, J., {Schady}, P., {et~al.} 2012, \aap, 539, A76, \dodoi{10.1051/0004-6361/201117929}

\bibitem[{{Page} {et~al.}(2025){Page}, {Evans}, {DeLaunay}, \& {Swift XRT Team}}]{GCN39584}
{Page}, K.~L., {Evans}, P.~A., {DeLaunay}, J., \& {Swift XRT Team}. 2025, GRB Coordinates Network, 39584, 1

\bibitem[{{Patel} {et~al.}(2023){Patel}, {Gompertz}, {O'Brien}, {Lamb}, {Starling}, {Evans}, {Amati}, {Levan}, {Nicholl}, {Ackley}, {Dyer}, {Lyman}, {Ulaczyk}, {Steeghs}, {Galloway}, {Dhillon}, {Ramsay}, {Noysena}, {Kotak}, {Breton}, {Nuttall}, {Pall{\'e}}, \& {Pollacco}}]{Patel23}
{Patel}, M., {Gompertz}, B.~P., {O'Brien}, P.~T., {et~al.} 2023, \mnras, 523, 4923, \dodoi{10.1093/mnras/stad1703}

\bibitem[{{Perley} {et~al.}(2019){Perley}, {Mazzali}, {Yan}, {Cenko}, {Gezari}, {Taggart}, {Blagorodnova}, {Fremling}, {Mockler}, {Singh}, {Tominaga}, {Tanaka}, {Watson}, {Ahumada}, {Anupama}, {Ashall}, {Becerra}, {Bersier}, {Bhalerao}, {Bloom}, {Butler}, {Copperwheat}, {Coughlin}, {De}, {Drake}, {Duev}, {Frederick}, {Gonz{\'a}lez}, {Goobar}, {Heida}, {Ho}, {Horst}, {Hung}, {Itoh}, {Jencson}, {Kasliwal}, {Kawai}, {Khanam}, {Kulkarni}, {Kumar}, {Kumar}, {Kutyrev}, {Lee}, {Maeda}, {Mahabal}, {Murata}, {Neill}, {Ngeow}, {Penprase}, {Pian}, {Quimby}, {Ramirez-Ruiz}, {Richer}, {Rom{\'a}n-Z{\'u}{\~n}iga}, {Sahu}, {Srivastav}, {Socia}, {Sollerman}, {Tachibana}, {Taddia}, {Tinyanont}, {Troja}, {Ward}, {Wee}, \& {Yu}}]{Perley19}
{Perley}, D.~A., {Mazzali}, P.~A., {Yan}, L., {et~al.} 2019, \mnras, 484, 1031, \dodoi{10.1093/mnras/sty3420}

\bibitem[{{Perley} {et~al.}(2021){Perley}, {Ho}, {Yao}, {Fremling}, {Anderson}, {Schulze}, {Kumar}, {Anupama}, {Barway}, {Bellm}, {Bhalerao}, {Chen}, {Duev}, {Galbany}, {Graham}, {Gromadzki}, {Guti{\'e}rrez}, {Ihanec}, {Inserra}, {Kasliwal}, {Kool}, {Kulkarni}, {Laher}, {Masci}, {Neill}, {Nicholl}, {Pursiainen}, {van Roestel}, {Sharma}, {Sollerman}, {Walters}, \& {Wiseman}}]{Perley21}
{Perley}, D.~A., {Ho}, A. Y.~Q., {Yao}, Y., {et~al.} 2021, \mnras, 508, 5138, \dodoi{10.1093/mnras/stab2785}

\bibitem[{{Piro} {et~al.}(2021){Piro}, {Haynie}, \& {Yao}}]{piro21}
{Piro}, A.~L., {Haynie}, A., \& {Yao}, Y. 2021, \apj, 909, 209, \dodoi{10.3847/1538-4357/abe2b1}

\bibitem[{{Piro} \& {Kollmeier}(2018)}]{Piro2018}
{Piro}, A.~L., \& {Kollmeier}, J.~A. 2018, \apj, 855, 103, \dodoi{10.3847/1538-4357/aaaab3}

\bibitem[{{Planck Collaboration} {et~al.}(2020){Planck Collaboration}, {Aghanim}, {Akrami}, {Ashdown}, {Aumont}, {Baccigalupi}, {Ballardini}, {Banday}, {Barreiro}, {Bartolo}, {Basak}, {Battye}, {Benabed}, {Bernard}, {Bersanelli}, {Bielewicz}, {Bock}, {Bond}, {Borrill}, {Bouchet}, {Boulanger}, {Bucher}, {Burigana}, {Butler}, {Calabrese}, {Cardoso}, {Carron}, {Challinor}, {Chiang}, {Chluba}, {Colombo}, {Combet}, {Contreras}, {Crill}, {Cuttaia}, {de Bernardis}, {de Zotti}, {Delabrouille}, {Delouis}, {Di Valentino}, {Diego}, {Dor{\'e}}, {Douspis}, {Ducout}, {Dupac}, {Dusini}, {Efstathiou}, {Elsner}, {En{\ss}lin}, {Eriksen}, {Fantaye}, {Farhang}, {Fergusson}, {Fernandez-Cobos}, {Finelli}, {Forastieri}, {Frailis}, {Fraisse}, {Franceschi}, {Frolov}, {Galeotta}, {Galli}, {Ganga}, {G{\'e}nova-Santos}, {Gerbino}, {Ghosh}, {Gonz{\'a}lez-Nuevo}, {G{\'o}rski}, {Gratton}, {Gruppuso}, {Gudmundsson}, {Hamann}, {Handley}, {Hansen}, {Herranz}, {Hildebrandt}, {Hivon}, {Huang}, {Jaffe}, {Jones}, {Karakci}, {Keih{\"a}nen},
  {Keskitalo}, {Kiiveri}, {Kim}, {Kisner}, {Knox}, {Krachmalnicoff}, {Kunz}, {Kurki-Suonio}, {Lagache}, {Lamarre}, {Lasenby}, {Lattanzi}, {Lawrence}, {Le Jeune}, {Lemos}, {Lesgourgues}, {Levrier}, {Lewis}, {Liguori}, {Lilje}, {Lilley}, {Lindholm}, {L{\'o}pez-Caniego}, {Lubin}, {Ma}, {Mac{\'\i}as-P{\'e}rez}, {Maggio}, {Maino}, {Mandolesi}, {Mangilli}, {Marcos-Caballero}, {Maris}, {Martin}, {Martinelli}, {Mart{\'\i}nez-Gonz{\'a}lez}, {Matarrese}, {Mauri}, {McEwen}, {Meinhold}, {Melchiorri}, {Mennella}, {Migliaccio}, {Millea}, {Mitra}, {Miville-Desch{\^e}nes}, {Molinari}, {Montier}, {Morgante}, {Moss}, {Natoli}, {N{\o}rgaard-Nielsen}, {Pagano}, {Paoletti}, {Partridge}, {Patanchon}, {Peiris}, {Perrotta}, {Pettorino}, {Piacentini}, {Polastri}, {Polenta}, {Puget}, {Rachen}, {Reinecke}, {Remazeilles}, {Renzi}, {Rocha}, {Rosset}, {Roudier}, {Rubi{\~n}o-Mart{\'\i}n}, {Ruiz-Granados}, {Salvati}, {Sandri}, {Savelainen}, {Scott}, {Shellard}, {Sirignano}, {Sirri}, {Spencer}, {Sunyaev}, {Suur-Uski}, {Tauber}, {Tavagnacco},
  {Tenti}, {Toffolatti}, {Tomasi}, {Trombetti}, {Valenziano}, {Valiviita}, {Van Tent}, {Vibert}, {Vielva}, {Villa}, {Vittorio}, {Wandelt}, {Wehus}, {White}, {White}, {Zacchei}, \& {Zonca}}]{Planck20}
{Planck Collaboration}, {Aghanim}, N., {Akrami}, Y., {et~al.} 2020, \aap, 641, A6, \dodoi{10.1051/0004-6361/201833910}

\bibitem[{{Prentice} {et~al.}(2018){Prentice}, {Maguire}, {Smartt}, {Magee}, {Schady}, {Sim}, {Chen}, {Clark}, {Colin}, {Fulton}, {McBrien}, {O'Neill}, {Smith}, {Ashall}, {Chambers}, {Denneau}, {Flewelling}, {Heinze}, {Holoien}, {Huber}, {Kochanek}, {Mazzali}, {Prieto}, {Rest}, {Shappee}, {Stalder}, {Stanek}, {Stritzinger}, {Thompson}, \& {Tonry}}]{Prentice18}
{Prentice}, S.~J., {Maguire}, K., {Smartt}, S.~J., {et~al.} 2018, \apjl, 865, L3, \dodoi{10.3847/2041-8213/aadd90}

\bibitem[{{Prochaska} {et~al.}(2020){Prochaska}, {Hennawi}, {Westfall}, {Cooke}, {Wang}, {Hsyu}, {Davies}, {Farina}, \& {Pelliccia}}]{Prochaska20}
{Prochaska}, J., {Hennawi}, J., {Westfall}, K., {et~al.} 2020, The Journal of Open Source Software, 5, 2308, \dodoi{10.21105/joss.02308}

\bibitem[{{Quirola-V{\'a}squez} {et~al.}(2022){Quirola-V{\'a}squez}, {Bauer}, {Jonker}, {Brandt}, {Yang}, {Levan}, {Xue}, {Eappachen}, {Zheng}, \& {Luo}}]{QuirolaVasquez22}
{Quirola-V{\'a}squez}, J., {Bauer}, F.~E., {Jonker}, P.~G., {et~al.} 2022, \aap, 663, A168, \dodoi{10.1051/0004-6361/202243047}

\bibitem[{{Quirola-V{\'a}squez} {et~al.}(2023){Quirola-V{\'a}squez}, {Bauer}, {Jonker}, {Brandt}, {Yang}, {Levan}, {Xue}, {Eappachen}, {Camacho}, {Ravasio}, {Zheng}, \& {Luo}}]{QuirolaVasquez23}
---. 2023, \aap, 675, A44, \dodoi{10.1051/0004-6361/202345912}

\bibitem[{{Rastinejad} {et~al.}(2025){Rastinejad}, {Levan}, {Jonker}, {Kilpatrick}, {Fryer}, {Sarin}, {Gompertz}, {Liu}, {Eyles-Ferris}, {Fong}, {Burns}, {Gillanders}, {Mandel}, {Malesani}, {O'Brien}, {Tanvir}, {Ackley}, {Aryan}, {Bauer}, {Bloemen}, {de Boer}, {Bom}, {Chacon}, {Chambers}, {Chen}, {Chrimes}, {van Dalen}, {D'Elia}, {De Pasquale}, {Fulton}, {Groot}, {Gupta}, {Hartmann}, {van Hoof}, {Huber}, {Izzo}, {Jacobson-Galan}, {Jakobsson}, {Kong}, {Laskar}, {Lowe}, {Magnier}, {Maiorano}, {Martin-Carrillo}, {Mas-Ribas}, {Mata Sanchez}, {Nicholl}, {Nixon}, {Oates}, {Paek}, {Palmerio}, {Paris}, {Pieterse}, {Pugliese}, {Quirola Vasquez}, {van Roestel}, {Rossi}, {Salvaterra}, {Schneider}, {Smartt}, {Smith}, {Smith}, {Srivastav}, {Torres}, {Ventura}, {Vreeswijk}, {Wainscoat}, {Yang}, \& {Yang}}]{Rastinejad25}
{Rastinejad}, J.~C., {Levan}, A.~J., {Jonker}, P.~G., {et~al.} 2025, arXiv e-prints, arXiv:2504.08889, \dodoi{10.48550/arXiv.2504.08889}

\bibitem[{{Ravasio} {et~al.}(2025){Ravasio}, {Burns}, {Wilson-Hodge}, {Jonker}, \& {Fermi-GBM Team}}]{GCN39146}
{Ravasio}, M.~E., {Burns}, E., {Wilson-Hodge}, C., {Jonker}, P.~G., \& {Fermi-GBM Team}. 2025, GRB Coordinates Network, 39146, 1

\bibitem[{{Rho} {et~al.}(2021){Rho}, {Evans}, {Geballe}, {Banerjee}, {Hoeflich}, {Shahbandeh}, {Valenti}, {Yoon}, {Jin}, {Williamson}, {Modjaz}, {Hiramatsu}, {Howell}, {Pellegrino}, {Vink{\'o}}, {Cartier}, {Burke}, {McCully}, {An}, {Cha}, {Pritchard}, {Wang}, {Andrews}, {Galbany}, {Van Dyk}, {Graham}, {Blinnikov}, {Joshi}, {P{\'a}l}, {Kriskovics}, {Ordasi}, {Szakats}, {Vida}, {Chen}, {Li}, {Zhang}, \& {Yan}}]{Rho21}
{Rho}, J., {Evans}, A., {Geballe}, T.~R., {et~al.} 2021, \apj, 908, 232, \dodoi{10.3847/1538-4357/abd850}

\bibitem[{{Saccardi} {et~al.}(2025){Saccardi}, {Zhu}, {Schneider}, {Xu}, {Izzo}, {Levan}, {Malesani}, {Martin-Carrillo}, {Tanvir}, {Vergani}, \& {Stargate Collaboration}}]{GCN39585}
{Saccardi}, A., {Zhu}, Z.~P., {Schneider}, B., {et~al.} 2025, GRB Coordinates Network, 39585, 1

\bibitem[{{Sarin} {et~al.}(2024){Sarin}, {H{\"u}bner}, {Omand}, {Setzer}, {Schulze}, {Adhikari}, {Sagu{\'e}s-Carracedo}, {Galaudage}, {Wallace}, {Lamb}, \& {Lin}}]{sarin_redback}
{Sarin}, N., {H{\"u}bner}, M., {Omand}, C. M.~B., {et~al.} 2024, \mnras, 531, 1203, \dodoi{10.1093/mnras/stae1238}

\bibitem[{{Schlafly} \& {Finkbeiner}(2011)}]{Schlafly11}
{Schlafly}, E.~F., \& {Finkbeiner}, D.~P. 2011, \apj, 737, 103, \dodoi{10.1088/0004-637X/737/2/103}

\bibitem[{{Schroeder} {et~al.}(2025){Schroeder}, {Ho}, \& {Perley}}]{GCN38970}
{Schroeder}, G., {Ho}, A., \& {Perley}, D. 2025, GRB Coordinates Network, 38970, 1

\bibitem[{{Senno} {et~al.}(2016){Senno}, {Murase}, \& {M{\'e}sz{\'a}ros}}]{Senno16}
{Senno}, N., {Murase}, K., \& {M{\'e}sz{\'a}ros}, P. 2016, \prd, 93, 083003, \dodoi{10.1103/PhysRevD.93.083003}

\bibitem[{{Shilling} \& {Swift UVOT Team}(2025)}]{GCN39587}
{Shilling}, S.~P.~R., \& {Swift UVOT Team}. 2025, GRB Coordinates Network, 39587, 1

\bibitem[{{Shingles} {et~al.}(2021){Shingles}, {Smith}, {Young}, {Smartt}, {Tonry}, {Denneau}, {Heinze}, {Weiland}, {Flewelling}, {Stalder}, {Clocchiatti}, {F{\"o}rster}, {Pignata}, {Rest}, {Anderson}, {Stubbs}, \& {Erasmus}}]{Shingles21}
{Shingles}, L., {Smith}, K.~W., {Young}, D.~R., {et~al.} 2021, Transient Name Server AstroNote, 7, 1

\bibitem[{{Smith} {et~al.}(2020){Smith}, {Smartt}, {Young}, {Tonry}, {Denneau}, {Flewelling}, {Heinze}, {Weiland}, {Stalder}, {Rest}, {Stubbs}, {Anderson}, {Chen}, {Clark}, {Do}, {F{\"o}rster}, {Fulton}, {Gillanders}, {McBrien}, {O'Neill}, {Srivastav}, \& {Wright}}]{Smith20}
{Smith}, K.~W., {Smartt}, S.~J., {Young}, D.~R., {et~al.} 2020, \pasp, 132, 085002, \dodoi{10.1088/1538-3873/ab936e}

\bibitem[{{Soderberg} {et~al.}(2006){Soderberg}, {Kulkarni}, {Nakar}, {Berger}, {Cameron}, {Fox}, {Frail}, {Gal-Yam}, {Sari}, {Cenko}, {Kasliwal}, {Chevalier}, {Piran}, {Price}, {Schmidt}, {Pooley}, {Moon}, {Penprase}, {Ofek}, {Rau}, {Gehrels}, {Nousek}, {Burrows}, {Persson}, \& {McCarthy}}]{Soderberg06}
{Soderberg}, A.~M., {Kulkarni}, S.~R., {Nakar}, E., {et~al.} 2006, \nat, 442, 1014, \dodoi{10.1038/nature05087}

\bibitem[{{Soderberg} {et~al.}(2008){Soderberg}, {Berger}, {Page}, {Schady}, {Parrent}, {Pooley}, {Wang}, {Ofek}, {Cucchiara}, {Rau}, {Waxman}, {Simon}, {Bock}, {Milne}, {Page}, {Barentine}, {Barthelmy}, {Beardmore}, {Bietenholz}, {Brown}, {Burrows}, {Burrows}, {Byrngelson}, {Cenko}, {Chandra}, {Cummings}, {Fox}, {Gal-Yam}, {Gehrels}, {Immler}, {Kasliwal}, {Kong}, {Krimm}, {Kulkarni}, {Maccarone}, {M{\'e}sz{\'a}ros}, {Nakar}, {O'Brien}, {Overzier}, {de Pasquale}, {Racusin}, {Rea}, \& {York}}]{Soderberg08}
{Soderberg}, A.~M., {Berger}, E., {Page}, K.~L., {et~al.} 2008, \nat, 453, 469, \dodoi{10.1038/nature06997}

\bibitem[{{Sollerman} {et~al.}(2006){Sollerman}, {Jaunsen}, {Fynbo}, {Hjorth}, {Jakobsson}, {Stritzinger}, {F{\'e}ron}, {Laursen}, {Ovaldsen}, {Selj}, {Th{\"o}ne}, {Xu}, {Davis}, {Gorosabel}, {Watson}, {Duro}, {Ilyin}, {Jensen}, {Lysfjord}, {Marquart}, {Nielsen}, {N{\"a}r{\"a}nen}, {Schwarz}, {Walch}, {Wold}, \& {{\"O}stlin}}]{Sollerman06}
{Sollerman}, J., {Jaunsen}, A.~O., {Fynbo}, J.~P.~U., {et~al.} 2006, \aap, 454, 503, \dodoi{10.1051/0004-6361:20065226}

\bibitem[{{Song} {et~al.}(2025){Song}, {Li}, {Wang}, {Mao}, {Lu}, \& {Bai}}]{GCN38972}
{Song}, F.~F., {Li}, R.~Z., {Wang}, B.~T., {et~al.} 2025, GRB Coordinates Network, 38972, 1

\bibitem[{{Srinivasaragavan} {et~al.}(2025){Srinivasaragavan}, {Hamidani}, {Schroeder}, {Sarin}, {Ho}, {Piro}, {Cenko}, {Anand}, {Sollerman}, {Perley}, {Maeda}, {O'Connor}, {Kuncarayakti}, {Miller}, {Ahumada}, {Vail}, {Duffell}, {Ghosh Dastidar}, {Andreoni}, {Bochenek}, {Brennan}, {Carney}, {Chen}, {Freeburn}, {Gal-Yam}, {Jacobson-Gal{\'a}n}, {Kasliwal}, {Li}, {Li}, {Sravan}, \& {Warshofsky}}]{Srinivasaragavan25}
{Srinivasaragavan}, G.~P., {Hamidani}, H., {Schroeder}, G., {et~al.} 2025, arXiv e-prints, arXiv:2504.17516, \dodoi{10.48550/arXiv.2504.17516}

\bibitem[{{Srivastav} {et~al.}(2025){Srivastav}, {Chen}, {Gillanders}, {Rhodes}, {Smartt}, {Huber}, {Aryan}, {Yang}, {Beri}, {Cooper}, {Nicholl}, {Smith}, {Stevance}, {Carotenuto}, {Chambers}, {Aamer}, {Angus}, {Fulton}, {Moore}, {Smith}, {Young}, {de Boer}, {Gao}, {Lin}, {Lowe}, {Magnier}, {Minguez}, {Pan}, \& {Wainscoat}}]{Srivastav25}
{Srivastav}, S., {Chen}, T.~W., {Gillanders}, J.~H., {et~al.} 2025, \apjl, 978, L21, \dodoi{10.3847/2041-8213/ad9c75}

\bibitem[{{Steele} {et~al.}(2004){Steele}, {Smith}, {Rees}, {Baker}, {Bates}, {Bode}, {Bowman}, {Carter}, {Etherton}, {Ford}, {Fraser}, {Gomboc}, {Lett}, {Mansfield}, {Marchant}, {Medrano-Cerda}, {Mottram}, {Raback}, {Scott}, {Tomlinson}, \& {Zamanov}}]{Steele04}
{Steele}, I.~A., {Smith}, R.~J., {Rees}, P.~C., {et~al.} 2004, in Society of Photo-Optical Instrumentation Engineers (SPIE) Conference Series, Vol. 5489, Ground-based Telescopes, ed. J.~M. {Oschmann}, Jr., 679--692, \dodoi{10.1117/12.551456}

\bibitem[{{Sun} {et~al.}(2024){Sun}, {Li}, {Liu}, {Gao}, {Wang}, {Yuan}, {Zhang}, {Filippenko}, {Xu}, {An}, {Ai}, {Brink}, {Liu}, {Liu}, {Wang}, {Wu}, {Wu}, {Yang}, {Zhang}, {Zheng}, {Ahumada}, {Dai}, {Delaunay}, {Elias-Rosa}, {Benetti}, {Fu}, {Howell}, {Huang}, {Kasliwal}, {Karambelkar}, {Stein}, {Lei}, {Lian}, {Peng}, {Ridnaia}, {Svinkin}, {Wang}, {Wang}, {Wei}, {An}, {Andrews}, {Bai}, {Dai}, {Ehgamberdiev}, {Fan}, {Farah}, {Feng}, {Fynbo}, {Guo}, {Guo}, {Hu}, {Hu}, {Jiang}, {Jin}, {Li}, {Li}, {Li}, {Liang}, {Ling}, {Liu}, {Mao}, {McCully}, {Mirzaqulov}, {Newsome}, {Padilla Gonzalez}, {Pan}, {Terreran}, {Tinyanont}, {Wang}, {Wang}, {Wen}, {Xiang}, {Xue}, {Yang}, {Zhu}, {Cai}, {Castro-Tirado}, {Chen}, {Chen}, {Chen}, {Chen}, {Chen}, {Chen}, {Chen}, {Cheng}, {Cordier}, {Cui}, {Cui}, {Dai}, {Fan}, {Feng}, {Guan}, {Han}, {Hou}, {Hu}, {Huang}, {Huo}, {Jia}, {Jia}, {Jiang}, {Jin}, {Jin}, {Kuulkers}, {Li}, {Li}, {Li}, {Li}, {Li}, {Li}, {Li}, {Liu}, {Liu}, {Liu}, {Liu}, {Lu}, {Luo}, {Ma}, {Mao}, {Nandra},
  {O'Brien}, {Pan}, {Rau}, {Rea}, {Sanders}, {Song}, {Sun}, {Sun}, {Tan}, {Tang}, {Tao}, {Wang}, {Wang}, {Wang}, {Wang}, {Wang}, {Wang}, {Xiong}, {Xu}, {Xu}, {Xu}, {Xu}, {Xu}, {Xue}, {Xue}, {Yan}, {Yang}, {Yang}, {Yang}, {Zhang}, {Zhang}, {Zhang}, {Zhang}, {Zhang}, {Zhang}, {Zhang}, {Zhang}, {Zhang}, {Zhang}, {Zhao}, {Zhao}, {Zhao}, {Zhao}, {Zhou}, {Zhu}, \& {Zhu}}]{Sun24}
{Sun}, H., {Li}, W.~X., {Liu}, L.~D., {et~al.} 2024, arXiv e-prints, arXiv:2410.02315, \dodoi{10.48550/arXiv.2410.02315}

\bibitem[{{Tan} {et~al.}(2001){Tan}, {Matzner}, \& {McKee}}]{2001ApJ...551..946T}
{Tan}, J.~C., {Matzner}, C.~D., \& {McKee}, C.~F. 2001, \apj, 551, 946, \dodoi{10.1086/320245}

\bibitem[{{Th{\"o}ne} {et~al.}(2011){Th{\"o}ne}, {de Ugarte Postigo}, {Fryer}, {Page}, {Gorosabel}, {Aloy}, {Perley}, {Kouveliotou}, {Janka}, {Mimica}, {Racusin}, {Krimm}, {Cummings}, {Oates}, {Holland}, {Siegel}, {de Pasquale}, {Sonbas}, {Im}, {Park}, {Kann}, {Guziy}, {Hern{\'a}ndez-Garc{\'\i}a}, {Llorente}, {Bundy}, {Choi}, {Jeong}, {Korhonen}, {Kub{\`a}nek}, {Lim}, {Moskvitin}, {Mu{\~n}oz-Darias}, {Pak}, \& {Parrish}}]{Thone11}
{Th{\"o}ne}, C.~C., {de Ugarte Postigo}, A., {Fryer}, C.~L., {et~al.} 2011, \nat, 480, 72, \dodoi{10.1038/nature10611}

\bibitem[{{Toma} {et~al.}(2007){Toma}, {Ioka}, {Sakamoto}, \& {Nakamura}}]{Toma07}
{Toma}, K., {Ioka}, K., {Sakamoto}, T., \& {Nakamura}, T. 2007, \apj, 659, 1420, \dodoi{10.1086/512481}

\bibitem[{{Tonry} {et~al.}(2018){Tonry}, {Denneau}, {Heinze}, {Stalder}, {Smith}, {Smartt}, {Stubbs}, {Weiland}, \& {Rest}}]{Tonry18}
{Tonry}, J.~L., {Denneau}, L., {Heinze}, A.~N., {et~al.} 2018, \pasp, 130, 064505, \dodoi{10.1088/1538-3873/aabadf}

\bibitem[{{van Dalen} {et~al.}(2025){van Dalen}, {Levan}, {Jonker}, {Malesani}, {Izzo}, {Sarin}, {Quirola-V{\'a}squez}, {S{\'a}nchez}, {de Ugarte Postigo}, {van Hoof}, {Torres}, {Schulze}, {Littlefair}, {Chrimes}, {Ravasio}, {Bauer}, {Martin-Carrillo}, {Fraser}, {van der Horst}, {Jakobsson}, {O'Brien}, {De Pasquale}, {Pugliese}, {Sollerman}, {Tanvir}, {Zafar}, {Anderson}, {Galbany}, {Gal-Yam}, {Gromadzki}, {M{\"u}ller-Bravo}, {Ragosta}, \& {Terwel}}]{vanDalen25}
{van Dalen}, J. N.~D., {Levan}, A.~J., {Jonker}, P.~G., {et~al.} 2025, \apjl, 982, L47, \dodoi{10.3847/2041-8213/adbc7e}

\bibitem[{{Vernet} {et~al.}(2011){Vernet}, {Dekker}, {D'Odorico}, {Kaper}, {Kjaergaard}, {Hammer}, {Randich}, {Zerbi}, {Groot}, {Hjorth}, {Guinouard}, {Navarro}, {Adolfse}, {Albers}, {Amans}, {Andersen}, {Andersen}, {Binetruy}, {Bristow}, {Castillo}, {Chemla}, {Christensen}, {Conconi}, {Conzelmann}, {Dam}, {de Caprio}, {de Ugarte Postigo}, {Delabre}, {di Marcantonio}, {Downing}, {Elswijk}, {Finger}, {Fischer}, {Flores}, {Fran{\c{c}}ois}, {Goldoni}, {Guglielmi}, {Haigron}, {Hanenburg}, {Hendriks}, {Horrobin}, {Horville}, {Jessen}, {Kerber}, {Kern}, {Kiekebusch}, {Kleszcz}, {Klougart}, {Kragt}, {Larsen}, {Lizon}, {Lucuix}, {Mainieri}, {Manuputy}, {Martayan}, {Mason}, {Mazzoleni}, {Michaelsen}, {Modigliani}, {Moehler}, {M{\o}ller}, {Norup S{\o}rensen}, {N{\o}rregaard}, {P{\'e}roux}, {Patat}, {Pena}, {Pragt}, {Reinero}, {Rigal}, {Riva}, {Roelfsema}, {Royer}, {Sacco}, {Santin}, {Schoenmaker}, {Spano}, {Sweers}, {Ter Horst}, {Tintori}, {Tromp}, {van Dael}, {van der Vliet}, {Venema}, {Vidali}, {Vinther}, {Vola},
  {Winters}, {Wistisen}, {Wulterkens}, \& {Zacchei}}]{Vernet11}
{Vernet}, J., {Dekker}, H., {D'Odorico}, S., {et~al.} 2011, \aap, 536, A105, \dodoi{10.1051/0004-6361/201117752}

\bibitem[{{Virgili} {et~al.}(2009){Virgili}, {Liang}, \& {Zhang}}]{Virgili09}
{Virgili}, F.~J., {Liang}, E.-W., \& {Zhang}, B. 2009, \mnras, 392, 91, \dodoi{10.1111/j.1365-2966.2008.14063.x}

\bibitem[{{Wang} {et~al.}(2024){Wang}, {Dastidar}, {Giannios}, \& {Duffell}}]{Wang2024}
{Wang}, H., {Dastidar}, R.~G., {Giannios}, D., \& {Duffell}, P.~C. 2024, \apjs, 273, 17, \dodoi{10.3847/1538-4365/ad4d9d}

\bibitem[{{Wang} {et~al.}(2019){Wang}, {Wang}, {Cano}, {Wang}, {Liu}, {Dai}, {Deng}, {Yu}, {Li}, {Song}, {Qiu}, \& {Wei}}]{Wang19}
{Wang}, L.~J., {Wang}, X.~F., {Cano}, Z., {et~al.} 2019, \mnras, 489, 1110, \dodoi{10.1093/mnras/stz2184}

\bibitem[{{Whitesides} {et~al.}(2017){Whitesides}, {Lunnan}, {Kasliwal}, {Perley}, {Corsi}, {Cenko}, {Blagorodnova}, {Cao}, {Cook}, {Doran}, {Frederiks}, {Fremling}, {Hurley}, {Karamehmetoglu}, {Kulkarni}, {Leloudas}, {Masci}, {Nugent}, {Ritter}, {Rubin}, {Savchenko}, {Sollerman}, {Svinkin}, {Taddia}, {Vreeswijk}, \& {Wozniak}}]{Whitesides17}
{Whitesides}, L., {Lunnan}, R., {Kasliwal}, M.~M., {et~al.} 2017, \apj, 851, 107, \dodoi{10.3847/1538-4357/aa99de}

\bibitem[{{Woosley}(1993)}]{Woosley93}
{Woosley}, S.~E. 1993, \apj, 405, 273, \dodoi{10.1086/172359}

\bibitem[{{Woosley} \& {Bloom}(2006)}]{Woosley06}
{Woosley}, S.~E., \& {Bloom}, J.~S. 2006, \araa, 44, 507, \dodoi{10.1146/annurev.astro.43.072103.150558}

\bibitem[{{Woosley} {et~al.}(1999){Woosley}, {Eastman}, \& {Schmidt}}]{Woosley99}
{Woosley}, S.~E., {Eastman}, R.~G., \& {Schmidt}, B.~P. 1999, \apj, 516, 788, \dodoi{10.1086/307131}

\bibitem[{{Xu} {et~al.}(2025){Xu}, {Zhu}, {Liu}, {Fynbo}, {Zou}, {Kumar}, {Liu}, {Song}, {Li}, {Mao}, {Liu}, {An}, {Li}, {Wang}, {Geng}, {Wu}, {Sun}, {Yuan}, \& {Zhang}}]{GCN38984}
{Xu}, D., {Zhu}, Z.~P., {Liu}, X., {et~al.} 2025, GRB Coordinates Network, 38984, 1

\bibitem[{{Yao} {et~al.}(2022){Yao}, {Ho}, {Medvedev}, {Nayana}, {Perley}, {Kulkarni}, {Chandra}, {Sazonov}, {Gilfanov}, {Khorunzhev}, {Khatami}, \& {Sunyaev}}]{Yao22}
{Yao}, Y., {Ho}, A. Y.~Q., {Medvedev}, P., {et~al.} 2022, \apj, 934, 104, \dodoi{10.3847/1538-4357/ac7a41}

\bibitem[{{Yin} {et~al.}(2024){Yin}, {Zhang}, {Yang}, {Sun}, {Zhang}, {Shao}, {Hu}, {Zhu}, {Xu}, {An}, {Gao}, {Wu}, {Zhang}, {Castro-Tirado}, {Pandey}, {Rau}, {Lei}, {Xie}, {Ghirlanda}, {Piro}, {O'Brien}, {Troja}, {Jonker}, {Yu}, {An}, {Chen}, {Chen}, {Dong}, {Eyles-Ferris}, {Fan}, {Fu}, {Fynbo}, {Gao}, {Huang}, {Jiang}, {Jiang}, {Julakanti}, {Kuulkers}, {Lao}, {Li}, {Ling}, {Liu}, {Liu}, {Mou}, {Pan}, {Wei}, {Wu}, {Yadav}, {Yang}, {Yuan}, \& {Zhang}}]{Yin24}
{Yin}, Y.-H.~I., {Zhang}, B.-B., {Yang}, J., {et~al.} 2024, \apjl, 975, L27, \dodoi{10.3847/2041-8213/ad8652}

\bibitem[{{Yuan} {et~al.}(2022){Yuan}, {Zhang}, {Chen}, \& {Ling}}]{Yuan22}
{Yuan}, W., {Zhang}, C., {Chen}, Y., \& {Ling}, Z. 2022, in Handbook of X-ray and Gamma-ray Astrophysics, ed. C.~{Bambi} \& A.~{Sangangelo}, 86, \dodoi{10.1007/978-981-16-4544-0_151-1}

\bibitem[{{Zackay} {et~al.}(2016){Zackay}, {Ofek}, \& {Gal-Yam}}]{Zackay16}
{Zackay}, B., {Ofek}, E.~O., \& {Gal-Yam}, A. 2016, \apj, 830, 27, \dodoi{10.3847/0004-637X/830/1/27}

\bibitem[{{Zhang} {et~al.}(2025){Zhang}, {Yuan}, {Ling}, {Chen}, {Rea}, {Rau}, {Cai}, {Cheng}, {Coti Zelati}, {Dai}, {Hu}, {Jia}, {Jin}, {Li}, {O'Brien}, {Shen}, {Shu}, {Sun}, {Sun}, {Wang}, {Yang}, {Zhang}, {Zhang}, {Zhang}, {Zhang}, {An}, {Buckley}, {Coleiro}, {Cordier}, {Dou}, {Eyles-Ferris}, {Fan}, {Feng}, {Fu}, {Fynbo}, {Galbany}, {Jha}, {Jiang}, {Kong}, {Kuulkers}, {Lei}, {Li}, {Liu}, {Liu}, {Liu}, {Liu}, {Liu}, {Maitra}, {Marino}, {Monageng}, {Nandra}, {Sanders}, {Soria}, {Tao}, {Wang}, {Wang}, {Wang}, {Wang}, {Wu}, {Wu}, {Xu}, {Xu}, {Xue}, {Xue}, {Zhang}, {Zhu}, {Zou}, {Bao}, {Chen}, {Chen}, {Chen}, {Chen}, {Chen}, {Chen}, {Cui}, {Cui}, {Dai}, {Fan}, {Guan}, {Han}, {Hou}, {Hu}, {Huang}, {Huo}, {Jia}, {Jiang}, {Jin}, {Li}, {Li}, {Li}, {Li}, {Li}, {Li}, {Lian}, {Liu}, {Liu}, {Liu}, {Lu}, {Luo}, {Ma}, {Mao}, {Pan}, {Pan}, {Song}, {Sun}, {Tan}, {Tang}, {Tao}, {Wang}, {Wang}, {Wang}, {Wang}, {Wang}, {Wang}, {Wu}, {Xu}, {Xu}, {Xu}, {Xu}, {Xu}, {Xue}, {Xue}, {Yan}, {Yang}, {Yang}, {Yang}, {Zhang}, {Zhang},
  {Zhang}, {Zhang}, {Zhang}, {Zhang}, {Zhao}, {Zhao}, {Zhao}, {Zhao}, {Zhou}, {Zhou}, {Zhu}, \& {Zhu}}]{Zhang25}
{Zhang}, W., {Yuan}, W., {Ling}, Z., {et~al.} 2025, Science China Physics, Mechanics, and Astronomy, 68, 219511, \dodoi{10.1007/s11433-024-2524-4}

\bibitem[{{Zheng} {et~al.}(2025){Zheng}, {Zhu}, {Lu}, \& {Zhang}}]{Zheng25}
{Zheng}, J.-H., {Zhu}, J.-P., {Lu}, W., \& {Zhang}, B. 2025, arXiv e-prints, arXiv:2503.24266, \dodoi{10.48550/arXiv.2503.24266}

\bibitem[{{Zhu} {et~al.}(2025{\natexlab{a}}){Zhu}, {Liu}, {Fu}, {Jiang}, {An}, \& {Xu}}]{GCN38885}
{Zhu}, Z.~P., {Liu}, X., {Fu}, S.~Y., {et~al.} 2025{\natexlab{a}}, GRB Coordinates Network, 38885, 1

\bibitem[{{Zhu} {et~al.}(2025{\natexlab{b}}){Zhu}, {Corcoran}, {Levan}, {Izzo}, {Eyles-Ferris}, {Fynbo}, {Leloudas}, {Malesani}, {Saccardi}, {Thakur}, {Xu}, {Gompertz}, {Jonker}, \& {Stargate Collaboration}}]{GCN38908}
{Zhu}, Z.~P., {Corcoran}, G., {Levan}, A.~J., {et~al.} 2025{\natexlab{b}}, GRB Coordinates Network, 38908, 1

\bibitem[{{Zou} {et~al.}(2025){Zou}, {Liu}, {Kumar}, {Chen}, {Du}, {Wang}, {Lagioia}, {Fang}, {Pan}, {Han}, {Zhang}, {Xin}, {Wu}, {Liu}, {Liu}, \& {Mephisto Team}}]{GCN38914}
{Zou}, X., {Liu}, C., {Kumar}, B., {et~al.} 2025, GRB Coordinates Network, 38914, 1

\end{thebibliography}
\bibliographystyle{aasjournal}



\end{document}